\RequirePackage{fix-cm}
\documentclass[twocolumn]{svjour3}    
\makeatletter
\let\cl@chapter\undefined
\makeatletter   
\let\proof\relax
\let\endproof\relax

\usepackage{xr}
\usepackage{xr-hyper}
\usepackage{hyperref}       

\usepackage{hyperref}
\smartqed  
\usepackage{graphicx}
\usepackage[utf8]{inputenc} 
\usepackage[T1]{fontenc}    
\usepackage[autonum,tchypersetup]{tchdr}
\usepackage[sort&compress]{natbib} 
\usepackage{fancyhdr}
\usepackage{mathptmx}     
\crefname{appendix}{Supp.}{Supps.}
\crefname{subappendix}{Supp.}{Supps.}

\begin{document}

\title{The computational asymptotics of Gaussian variational inference and the Laplace approximation}

\author{Zuheng Xu         \and
      Trevor Campbell 
}


\institute{Zuheng Xu  \at
              Department of Statistics, The University of British Columbia, Vancouver, Canada\\
              \email{zuheng.xu@stat.ubc.ca}           
           \and
           Trevor Campbell \at
           Department of Statistics, The University of British Columbia, Vancouver, Canada\\
           \email{trevor@stat.ubc.ca}
}

\date{Received: date / Accepted: date}

\maketitle

\begin{abstract}
Gaussian variational inference and the Laplace approximation are popular alternatives 
to Markov chain Monte Carlo that formulate Bayesian posterior inference
as an optimization problem, 
enabling the use of simple and scalable stochastic optimization algorithms. 
However, a key limitation of both methods is that  
the solution to the optimization problem is typically not tractable to compute; 
even in simple settings the problem is nonconvex.
Thus, recently developed statistical guarantees---which all involve 
the (data) asymptotic properties of the global optimum---are 
not reliably obtained in practice.
In this work, we provide two major contributions: a theoretical analysis of the
asymptotic convexity properties of variational inference with a Gaussian family
and the maximum a posteriori (MAP) problem required by the Laplace approximation; 
and two algorithms---\emph{consistent Laplace approximation} (CLA) and
\emph{consistent stochastic variational inference} (CSVI)---that exploit these 
properties to find the optimal approximation in the asymptotic regime. 
Both CLA and CSVI involve a tractable initialization procedure that finds
the local basin of the optimum, and CSVI further includes 
a scaled gradient descent
algorithm that provably stays locally confined to that basin.  
Experiments on nonconvex synthetic and real-data examples show that compared
with standard variational and Laplace approximations, 
both CSVI and CLA improve the likelihood of obtaining the global optimum
of their respective optimization problems. 

\end{abstract}

\section{Introduction}
\label{sec: intro}

Bayesian statistical models are powerful tools for learning from data, 
with the ability to encode complex hierarchical dependence and domain 
expertise, as well as coherently quantify uncertainty in latent parameters.
For many modern Bayesian models, exact computation of the
posterior is intractable \citep[Section 2.1]{blei2017variational} and
statisticians must resort to approximate inference algorithms. 
Currently, the most popular type of Bayesian inference algorithm in statistics is
Markov Chain Monte Carlo (MCMC) \citep{hastings1970monte,gelfand1990sampling,robert2013monte}, 
which provides approximate samples from the posterior distribution and is supported by a 
comprehensive literature of theoretical guarantees \citep{roberts2004general,meyn2012markov}. 
An alternative---variational 
inference \citep{jordan1998introduction,wainwright2008graphical,blei2017variational}---approximates the intractable
posterior with a distribution chosen from a pre-specified family, e.g., the family
of Gaussian distributions parametrized by mean and covariance.
The approximating distribution is chosen by minimizing a discrepancy
(e.g., Kullback-Leibler (KL) \citep[Section 2.8]{murphy2012machine} or 
R\'enyi divergence \citep{van2014renyi}) to the posterior distribution over the family.
Another alternative---the Laplace approximation \citep{shun1995laplace,hall2011asymptotic}---involves
first finding the maximum of the posterior density, and then fitting a Gaussian using a second-order Taylor expansion.
Both methods 
convert Bayesian inference into an optimization problem,
enabling the use of simple, scalable stochastic optimization algorithms \citep{robbins1951stochastic,Bottou2004}
that require only a subsample of the data at each iteration and avoid computation
on the entire dataset. 

But despite their scalability, both variational inference and the Laplace
approximation have key limitations. First, one is forced to use an approximation of the posterior from a preselected
parametric family of distributions. In particular, the Laplace approximation involves the family
of Gaussians, while in variational inference the choice of family is left to the practitioner. 
In either case, it is in general difficult to know how limited the family is
before actually optimizing; and not only that, it is also often difficult to
estimate the approximation error once the optimization is complete \citep{huggins2020validated}.
For example, if one uses a mean-field Gaussian family with a diagonal covariance, 
the resulting posterior approximation will typically
underestimate posterior variances and cannot capture its covariances \citep[Section 21.2.2]{murphy2012machine},
two quantities of particular interest to statisticians.
The second key limitation is that even if the Laplace approximation or optimal variational approximation 
are known to have low error, 
the optimization problems required by both methods are typically nonconvex,
and the global optimum cannot be found reliably.

The key to addressing the first limitation 
is to understand the optimal approximation error within the chosen family.
Aside from nonparametric mixtures \citep{guo2016boosting,miller2017variational,locatello18bbvi,campbell2019universal}---which 
can be designed to achieve arbitrary approximation quality---available results in the finite-data setting are quite limited.
For example, \citet{han2019statistical} provides a non-asymptotic
analysis of optimal mean-field variational approximations, but extending these results to more general distribution families is not
straightforward. 
On the other hand, multiple threads of research have explored the statistical properties
of parametric posterior approximations in the data-asymptotic regime by taking advantage of the
limiting behavior of the Bayesian posterior.
The Laplace approximation is well studied in
statistical literature in this regard
\citep{shun1995laplace,hall2011asymptotic,miller2021asymptotic,bassett2019maximum,barber2016laplace},
while research on variational inference is ongoing.
\citet{wang2019frequentist} exploits the asymptotic normality of the posterior
distribution in a parametric Bayesian setting to show that the variational KL minimizer
asymptotically converges to the KL minimizer to
the limiting normal posterior distribution.
 \citet{alquier2020concentration} analyze the rate of convergence of
the variational approximation to a fractional posterior---a posterior with a
tempered likelihood---in a high dimensional setting where the posterior itself may
not have the ideal asymptotic behavior.
\citet{zhang2020convergence} studies the contraction rate of the variational
distribution for non-parametric Bayesian inference and provides general
conditions on the Bayesian model that characterizes the rate.
\citet{yang2020alpha} and \citet{jaiswal2019asymptotic} build a framework for
analyzing the statistical properties of $\alpha$-R\'enyi variational inference,
and provide sufficient conditions that guarantee an optimal convergence rate of
the obtained point estimate. 
But while the literature has built a comprehensive understanding of the asymptotic
guarantees of both the Laplace and optimal variational approximations,
the nonconvexity of the optimization problems involved
makes these guarantees difficult to obtain reliably in practice.  
In fact, \cref{prop:vbfails} of the present work demonstrates that 
both the Laplace approximation and Gaussian variational inference
involve nonconvex optimization, 
even in simple cases with ideal asymptotic posterior behaviour.

In this work, we address the nonconvexity of Gaussian variational inference and
the maximum a posteriori (MAP) problem in
the data-asymptotic regime when the posterior distribution admits asymptotic
normality. Rather than focusing on the statistical
properties of the optimum,
 we investigate  
the asymptotic properties of the optimization problems themselves (\cref{sec:asymptotics}),
and use these to design procedures (\cref{sec:consistentvi}) which
involve only tractable optimization 
and hence make theoretical results regarding global optima applicable. 
In particular, we develop
\textit{consistent stochastic variational inference} (CSVI) and
\textit{consistent Laplace approximation} (CLA), two 
algorithms for Gaussian posterior approximation. CSVI is guaranteed to find the optimal
variational approximation, and CLA the maximum a posteriori (MAP), 
with probability that converges to $1$ in the limit of
observed data. The first key innovation in both CSVI and CLA is that we initialize
the optimization at the mode of a \emph{smoothed posterior}---the posterior distribution convolved with
Gaussian noise. We prove that, with enough data, the smoothed MAP falls in a local region in which the 
optimization problem is locally convex and contains the global optimum, and that 
finding the smoothed MAP is a convex optimization problem and hence tractable (\cref{sec:smoothedmap}).
The second innovation, which pertains only to CSVI, is a gradient scaling 
during stochastic optimization (\cref{sec:scaledsgd}) that ensures that the optimization
remains inside the aforementioned local region and converges to the global optimum. 
Experiments on synthetic examples in \cref{sec:simulation} show
that CSVI and CLA provide numerically stable and asymptotically consistent posterior
approximations.

\section{Variational and Laplace posterior approximations} \label{sec:gaussianvi}

In the setting of Bayesian inference considered in this paper, we are given a
sequence of posterior distributions $\Pi_n$, $n\in\nats$ each with full support
on $\reals^d$.  The index $n$ represents the amount of observed data; denote
$\Pi_0$ to be the prior.  We also assume that each posterior $\Pi_n$ has
density $\pi_n$ with respect to the Lebesgue measure.

\subsection{Gaussian variational inference}
Gaussian variational inference aims to find 
a Gaussian approximation to the posterior distribution by 
solving the optimization problem
\[
\begin{aligned}
\argmin_{\mu \in\reals^d, \Sigma\in\reals^{d\times d}} \quad &\kl{\distNorm(\mu, \Sigma)}{\Pi_n} 
\quad \text{s.t.} \quad  \Sigma \succ 0, 
\end{aligned}\label{eq:basicgaussianvi}
\]
where the Kullback-Leibler divergence \citep[Section 2.8]{murphy2012machine} is defined as
\[
\kl{Q}{P} \defined \int \log \der{Q}{P} \dee Q
\]
for any pair of probability distributions $P, Q$ such that $Q \ll P$, and
$\der{Q}{P}$ is the Radon-Nikodym derivative of $Q$ with respect to $P$
\citep[Section 3.2]{folland1999real}. 
We use the standard reparametrization of
$\Sigma$ using the Cholesky factorization $\Sigma = n^{-1}LL^T$ 
to arrive at the common formulation of Gaussian variational inference \citep{kucukelbir2017automatic}
that is the focus of the present work:
\[
\begin{aligned}
\mu_n^\star, L_n^\star &= \argmin_{\mu \in\reals^d, L\in\reals^{d\times d}} \!  - n^{-1}\log \det L + F_n(\mu, L)\\
& \text{s.t.} \quad  L  \text{ lower triangular with positive diagonal},
\end{aligned}\label{eq:gaussianvi}
\]
where
\[
f_n(x) &\defined -n^{-1}\log \pi_n(x)\label{eq:fn}\\
F_n(\mu, L) &\defined \EE\left[f_n(\mu+n^{-1/2}LZ)\right], \quad Z\dist\distNorm(0,I)\label{eq:Fn}.
\]
Denote the optimal Gaussian distribution 
\[
    \distNorm_\text{VI,n} \defined \distNorm(\mu^\star_n, n^{-1}L_n^\star L_n^{\star T}).    
\]
Intuitively, this optimization problem encodes a tradeoff between
maximizing the expected posterior density under the variational approximation---which tries
to make $L$ small and move $\mu$ close to the maximum point of $\pi_n$---and 
maximizing the entropy of the variational approximation---which prevents 
$L$ from becoming too small.
It crucially does not depend on the (typically unknown) normalization of $\pi_n$, 
which appears as an additive constant in \cref{eq:gaussianvi}; it is common to drop this
constant and instead equivalently maximize the \emph{expectation lower bound (ELBO)} \citep{blei2017variational}.
Note that there are a number 
of unconstrained parametrizations of the covariance matrix variable $\Sigma$ \citep{Pinheiro96}.
We select the (unique) positive-diagonal Cholesky factor $L$ as it makes the optimization problem
\cref{eq:gaussianvi} more amenable to both theoretical analysis and computational optimization.

One typically attempts to solve \cref{eq:gaussianvi} using an iterative local descent optimization algorithm.
As the expectation is intractable in most scenarios, 
this involves stochastic optimization \citep{hoffman2013stochastic,Ranganath14,Kingma14,kucukelbir2017automatic}. 
In particular, assuming 
one can interchange expectation and differentiation (see \cref{sec:assumptions} for details),
the quantities
\[
\begin{aligned}
\hat\nabla_{\mu,n}(\mu, L, Z) \!&\defined\! \nabla f_n(\mu + n^{-1/2}LZ)\\
\hat\nabla_{L,n}(\mu, L, Z)\! &\defined \!-n^{-1}(\diag L)^{-1} \\
& \quad + n^{-1/2}\ltri\nabla f_n(\mu + n^{-1/2}LZ) Z^T,
\end{aligned}\label{eq:stochasticgradients}
\]
are unbiased estimates of the $\mu$- and $L$-gradients of the objective in
\cref{eq:gaussianvi} given $Z\dist\distNorm(0, I)$, where the 
functions $\diag : \reals^{d\times d}\to\reals^{d\times d}$ and $\ltri: \reals^{d\times d}\to\reals^{d\times d}$ set the off-diagonal and upper triangular elements of
their arguments to 0, respectively.
These unbiased gradient estimates may be used in a wide variety of stochastic optimization algorithms \citep{robbins1951stochastic,Bottou2004}
applied to \cref{eq:gaussianvi}.  In this paper, we will focus on projected stochastic gradient
descent (SGD) \citep[Section 3.]{bubeck2015convex} due to its simplicity; we expect that the mathematical theory in this work
extends to other related methods. 
In general, Gaussian variational inference is a nonconvex optimization problem
and standard iterative methods such as SGD are not guaranteed to produce a
sequence of iterates that converge to $\mu_n^\star, L_n^\star$.  

\subsection{Laplace approximation}
The Laplace approximation \citep[Section 4.4]{bishop2006pattern} 
constructs a Gaussian approximation to the posterior 
centered at the maximum a posteriori (MAP) point $\theta_n^\star$, 
and with covariance
based on a second-order Taylor expansion of $\log \pi_n$ at the MAP,
i.e., 
\[
  \distNorm_{\text{Lap}, n} &\defined \distNorm(\theta^\star_{n}, \Sigma_{\text{Lap},n})\label{eq:laplace}\\
  \theta^\star_n &= \argmin_{\theta\in \reals^d} f_n(\theta)\label{eq:map}\\
 \Sigma_{\text{Lap}, n} &\defined \left( - \nabla^2 \log\pi_{n}(\theta^\star_n)\right)^{-1}.
\]
Under certain regularity conditions, the total variation error 
of the Laplace approximation
diminishes as the number of samples increases
\citep{kass1990validity,miller2021asymptotic,schillings2020convergence};
we present the precise statement in \cref{thm_deterbvm}.
However, 
since $\log\pi_n$ is typically not concave, obtaining the MAP point---and hence computing the
Laplace approximation---is generally intractable.

\section{Consistent variational and Laplace approximations}\label{sec:consistentvi}
In this section, we provide two methods---\emph{Consistent Stochastic Variational Inference (CSVI)} and \emph{Consistent Laplace (CLA)}---that 
\emph{asymptotically solve} the Gaussian variational inference and MAP problems in \cref{eq:gaussianvi,eq:map}  
in the sense that the probability that the iterates converge to the global
optimum converges to $1$ in the asymptotic limit of observed data (see \cref{def:asympsolvable}).
The development of the algorithms and results in this section depend heavily on  
asymptotic convexity and smoothness analysis later in \cref{sec:asymptotics}.
While the interested reader can refer to that section for a precise treatment,
the key intuitive points in the development are that, very informally:
(1) $f_n$ is generally nonconvex, even asymptotically in well-behaved models,
and this renders both the MAP and variational inference problems nonconvex;
however, 
(2) both optimization problems are asymptotically locally strongly convex around 
a fixed location $\theta_0$,
and $\mu^\star_n$ and $\theta^\star_{n}$ both converge to $\theta_0$;
and (3) the posterior density smoothed by a Gaussian kernel is asymptotically log concave, and the 
smoothed MAP $\htheta^\star_n$ also converges to $\theta_0$.
Therefore in CSVI and CLA, we first use the smoothed posterior MAP $\htheta^\star_n$ as a tractable initialization
near the global optimum,
and carefully scale gradient steps while optimizing so as to remain in the local basin around the optimum.
In the remainder of this section, we provide
the details of the smoothed MAP optimization, CSVI, and CLA; 
theoretical details are deferred to \cref{sec:asymptotics}.

Similar two-stage designs involving initialization followed
by local optimization have been employed for nonconvex problems 
in statistics, e.g., \citet{balakrishnan2017statistical}.
However, this work assumes the existence of the initialization---which generally
requires intractable nonconvex optimization itself---and requires a model-specific theoretical
analysis to tune the local optimization method.
In contrast, our work provides practical, model-agnostic, and asymptotically tractable 
optimization methods.

\begin{figure*}[t!]
	\centering
	\begin{subfigure}{\textwidth}
	\includegraphics[width = \linewidth]{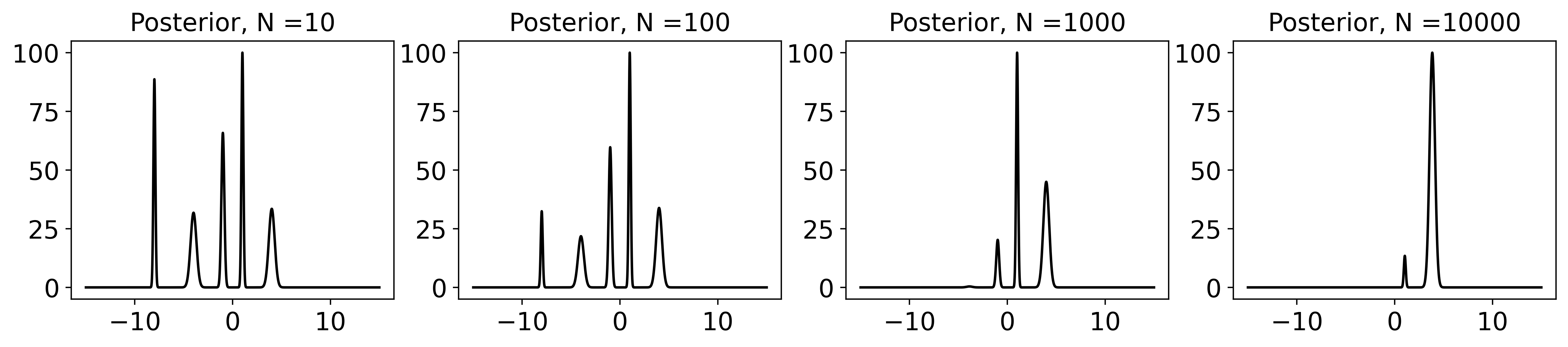}
	\end{subfigure}
	\begin{subfigure}{\textwidth}
	\includegraphics[width = \linewidth]{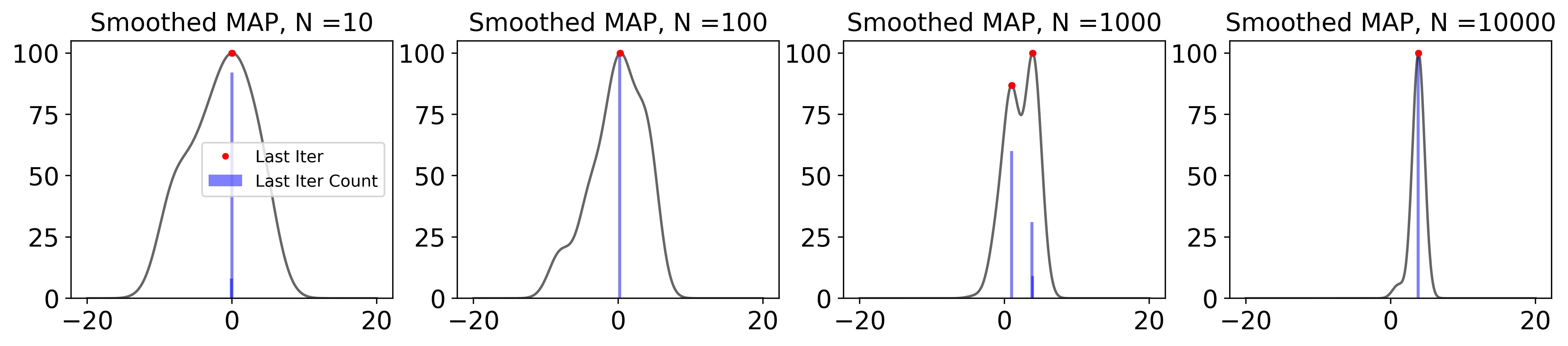}
	\end{subfigure}
	\caption{\textbf{Top:} A synthetic Bayesian model with posterior density for
increasing dataset sizes ($n = 10, 100, 1000, 10000$). The posterior density
displayed in the figure are normalized, fixing the maximum at $100$.
The details of this synthetic model and experimental setup 
can be found in \cref{supp:smoothexample}.
\textbf{Bottom:} The smoothed Bayesian posterior densities corresponding to the 
top row. Black curves show the smoothed posterior (with smoothing
constant of $\alpha_n = 10 n^{-0.3}$), the blue histogram shows the counts
(over 100 trials) of the output of the smoothed MAP initialization, and the red
dots denote the last iteration of the smoothed MAP optimization.}
	\label{fig:posts}
\end{figure*}
 
\subsection{Smoothed MAP initialization}\label{sec:initial_smoothmap}

Given the $n^\text{th}$ posterior distribution $\Pi_n$, we define the \emph{smoothed posterior} $\hPi_n$
with smoothing variance $\alpha_n$ to be the $\theta$-marginal of the  generative process
\[
W \dist \Pi_n, \quad \theta \dist \distNorm(W, \alpha_n I).	
\]
The probability density function $\hpi_n$ of $\hPi_n$ is given by the
convolution
of $\pi_n$ with a multivariate normal density,
\[ 
\hat{\pi}_n(\theta) 
=& \EE\left[\frac{1}{(2\pi)^{d/2} \alpha_n^{d/2}} \exp\left( -\frac{1}{2\alpha_n}\|\theta - W \|^2 \right)\right]. \label{eq:smootheddensity}
\]
Given these definitions, the \emph{smoothed MAP problem} is the MAP inference problem for the smoothed posterior distribution, i.e.,
\[ \label{eq:smoothedmap}
\htheta^\star_{n} =&
\argmin_{\theta \in \reals^d } \,\, - \log \EE\left[\exp\left(-\frac{1}{2\alpha_n}\| \theta - W\|^2\right)\right].
\]

\begin{figure}[t!]
	\centering
	\includegraphics[width = \linewidth]{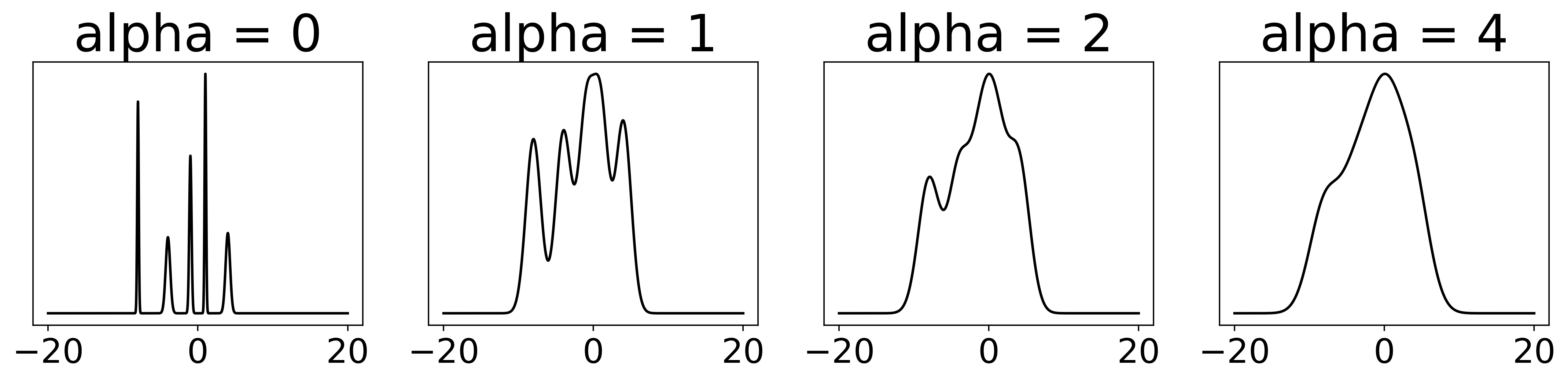}
	\caption{Plots of the smoothed posterior density $\hpi_n$ with increasing smoothing variance $\alpha$. Note that $\alpha = 0$ denotes the original posterior.}
	\label{fig:smoothing}
\end{figure}

Gaussian smoothing is commonly used in image and signal processing
\citep{forsyth2002computer,nixon2012feature,haddad1991class,lindeberg1990scale},
and has previously been applied to reduce the presence of spurious local
optima in nonconvex optimization problems, making them easier to solve with
local gradient-based methods \citep{addis2005local,mobahi2013optimization}.
This effect is demonstrated in \cref{fig:posts}, 
where we construct a synthetic Bayesian model
where the posterior is asymptotically normal but has multiple modes even given
a large sample size. The details of this synthetic model and the setting of
the experiment can be found in \cref{supp:smoothexample}. 
The variance $\alpha_n$ controls the degree of smoothing; larger values
create a smoother density $\hpi_n$, at the cost of making $\hpi_n$ a poorer
approximation of the original function $\pi_n$. 
\cref{fig:smoothing}
demonstrates how increasing $\alpha_n$ increases the smoothing effect,
resulting in fewer and flatter local optima in the objective.
In practice with a fixed finite data set, one needs to tune the smoothing constant $\alpha_n$.
However, we show later in \cref{thm:convexsmoothedMAP} that if $\alpha_n$
satisfies $n \alpha_n^3 \to \infty$, then eventually 
\cref{eq:smoothedmap} becomes a convex problem,
and $\htheta^\star_n$ converges in probability to the original posterior mode $\theta^\star_n$.

We use SGD to solve the smoothed MAP problem. 
 By change of variables and reparametrization,
the gradient of the smoothed MAP objective function in \cref{eq:smoothedmap} is
\[\label{eq:smoothgrd}
\nabla (-\log \hat{\pi}_n(\theta) )
=&\, \alpha_n^{-1/2} \frac{
\EE\left[W \pi_n\left(\theta-\alpha_n^{1/2} W\right)\right]
}{
\EE\left[\pi_n\left(\theta- \alpha_n^{1/2}W\right)\right]
},
\]
where $W \sim \distNorm\left(0, I\right)$. 
Note that the unknown
normalization constant in $\pi_n$ cancels in the numerator and denominator.
We obtain stochastic estimates of the gradient using a Monte Carlo
approximation of the numerator and denominator using the same samples, i.e.,
self-normalized importance sampling \citep[p.~95]{robert2013monte}. It is
known that the variance of this gradient estimate may be quite large or even
infinite; although techniques such as truncation \citep{ionides2008truncated}
and smoothing \citep{vehtari2015pareto} exist to address it, we leave this
issue as an open problem for future work. The resulting SGD procedure with
explicit gradient estimates are shown in \cref{alg:smoothedmap}.

\begin{algorithm}[t!]
\caption{Smoothed MAP optimization} \label{alg:smoothedmap}
\begin{algorithmic}
\Procedure{SmoothedMAP}{$\pi_n$, $\alpha_n$, $K$, $S$, $(\gamma_k)_{k\in\nats}$}
\State $\theta \gets 0$
\For{$k=1, \dots, K$}
	\State Sample $(Z_s)_{s = 1}^S \distiid \distNorm(0,I)$
	\State $\hg \!\gets \alpha_n^{-\frac{1}{2}}\!\left(\!\sum_{s = 1}^S \!Z_s \! \pi_n(\theta-\alpha_n^{\frac{1}{2}}Z_s)\!\right)\!\Big/\!\left(\!\sum_{s = 1}^S\pi_n(\theta- \alpha_n^{\frac{1}{2}}Z_s)\!\right) $ 
	\State $\theta \gets \theta - \gamma_k \hg$ 
\EndFor
\State \Return $\theta$
\EndProcedure
\end{algorithmic}
\end{algorithm}

\subsection{Consistent Laplace approximation (CLA)}\label{sec:csl}

The consistent Laplace approximation (CLA) involves first 
initializing $\theta$ to the smoothed MAP $\htheta^\star_n$ (which we estimate with \cref{alg:smoothedmap}),
and then proceeding as in the standard
Laplace approximation: we use gradient descent to find the posterior
mode, and then construct $\distNorm_{\text{Lap},n}$ per \cref{eq:laplace}.
The only concern is that the iterates produced by gradient descent
need to stay confined to the local basin of $f_n$ around the optimum point;
we can ensure this by using a small enough step size, 
or by using a line-search to set the step size adaptively
(\citealp[Page~466]{boyd2004convex}; \citealp{armijo1966minimization,bertsekas2000gradient}).
We present CLA with backtracking line search in
\cref{alg:csl}.

\begin{algorithm}[t!]
\caption{Consistent Laplace approximation} \label{alg:csl}
\begin{algorithmic}
  \Procedure{CLA}{$f_n$, $\beta \in (0, 1)$, $K$}
 \State $\theta \gets$\texttt{SmoothedMAP} (\cref{alg:smoothedmap}) 
\For{$k=1, \dots, K$}
  \State $t \gets 1$
  \While {$f_n(\theta - t \nabla f_n(\theta)) > f_n(\theta) -\frac{t}{2} \|\nabla f_n(\theta)\|^2$}
	\State $t \gets \beta t$
	\EndWhile 
  \State $\theta \gets \theta - t \nabla f_n(\theta)$ 
\EndFor
\State\Return $\theta$, $\frac{1}{n}\left(\nabla^2 f_n(\theta)\right)^{-1}$
\EndProcedure
\end{algorithmic}
\end{algorithm}

We provide a convergence result for CLA  in \cref{thm:mapasymp}. 
In particular, under mild regularity conditions on the Bayesian
model, \cref{thm:mapasymp} shows 
that CLA asymptotically solves the MAP optimization problem \cref{eq:map},
and hence enables reliable computation of the Laplace approximation.
\cref{def:asympsolvable} clarifies what it means for an algorithm
to solve an optimization problem in the data-asymptotic limit.
We use the usual notation $o_{P_n}$ and $O_{P_n}$ to denote
stochastic order \citep[Section 2.2]{van2000asymptotic}. 
\bnumdefn\label{def:asympsolvable}
An iterative algorithm \emph{asymptotically solves} a (random) sequence of
optimization problems indexed by $n\in\nats$,
each with a single global optimum point $x_n^\star\in\reals^d$,
if the sequence of iterates $(x_{k,n})_{k\in\nats}$ produced by the algorithm satisfies
\[
  \lim_{n\to\infty} \Pr\left( \|x_{k, n}- x_n^\star\|^2 = o_{P_n}(1) \right)  = 1,
\]
where $\Pr$ denotes the law of the sequence of optimization problems
and $P_n$ denotes the law of the iterates of the
$n^\text{th}$ optimization problem.
Further, we say that it asymptotically solves the 
sequence of problems \emph{at a rate $t_k$}, if
\[
  \lim_{n\to\infty} \Pr\left( \|x_{k, n} - x_n^\star\|^2 = O_{P_n}(t_k) \right)
  = 1, \quad \lim_{k \to \infty} t_k = 0.
\]
\enumdefn

\bnthm \label{thm:mapasymp}
Suppose \cref{assump:regularity} holds. Then there exist $r, \ell > 0$ and $0\leq \eta < 1$
such that if we initialize $\theta$ such that $\|\theta - \htheta^\star_n\| \leq \frac{r}{4(\ell + 1)}$,
then CLA asymptotically solves the MAP problem \cref{eq:map} at a rate of $\eta^k$.
\enthm
The constant $r$ in the statement of \cref{thm:mapasymp} intuitively represents the radius
of the local convex basin around the optimum point of $f_n$, 
and $\ell$ is the local Lipschitz smoothness constant, as defined in
\cref{lem:asymplocalconv}. In practice, since the constants $r$ and $\ell$ are not known,
one should run the smoothed MAP optimization \cref{alg:smoothedmap} 
until convergence to $\htheta^\star_n$ 
according to some diagnostic, e.g., a small gradient norm.

\subsection{Consistent stochastic variational inference (CSVI)}\label{sec:scaledsgd}

Consistent stochastic variational inference (CSVI) begins by initializing $L=I$
and $\mu$ to the smoothed MAP $\htheta^\star_n$ (which we estimate with \cref{alg:smoothedmap}), 
and then
optimizes the variational objective \cref{eq:gaussianvi} using 
projected stochastic
gradient descent (SGD) \citep[Section 3]{bubeck2015convex}.
Much like in CLA, the remaining concern is that the iterates of SGD
stay within the local region around the optimum $\mu^\star_n, L^\star_n$
in which the variational objective is strongly convex.
The major issue is that the regularization term 
$-n^{-1}\log\det L$ in the objective of \cref{eq:gaussianvi}
is not Lipschitz smooth, which both makes theoretical guarantees on convergence difficult
to obtain and in practice results in instability in $L$
during optimization. We address this issue by applying a novel
scaling matrix to the gradient steps; in particular,
define the scaled $L$ gradient matrix $\tilde\nabla_{L,n}(\mu, L, Z)\in\reals^{d\times d}$ via
\[
& \left[\tilde\nabla_{L,n}(\mu, L, Z)\right]_{ij} \\
 &= 
\left\{\begin{array}{ll}
\left[\hat\nabla_{L,n}(\mu, L, Z)\right]_{ij} & j \neq i \\
\frac{1}{1+(nL_{ii})^{-1}}\left[\hat\nabla_{L,n}(\mu, L, Z)\right]_{ii} & j = i, L_{ii} > 0\\
-1  & j = i, L_{ii} = 0.
\end{array}\right. \label{eq:scaledLgradient}
\]
This scaling prevents the gradient of $L$ from diverging when diagonal elements of $L\to 0$,
and also creates a well-defined gradient for $L$ at the boundary of the feasible region.
Then given a sequence of step sizes $(\gamma_k)_{k\in\nats}$, $\gamma_k \geq 0$,
Monte Carlo samples $(Z_k)_{k\in\nats} \distiid \distNorm(0, I)$,
and initialization $\mu_0 = \htheta^\star_n, L_0 = I$,
the standard stochastic gradient update applied to the Gaussian variational inference 
problem is
\[
\mu_{k+1} &\gets \mu_k - \gamma_k \hat\nabla_{\mu,n}(\mu_k, L_k, Z_k)\\
L_{k+1} &\gets L_k - \gamma_k \tilde\nabla_{L,n}(\mu_k, L_k, Z_k).
\]
After making this scaled gradient update, we ensure that the diagonal of $L$
remains nonnegative by employing a simple projection step after each update: we
set any negative diagonal entry in the current iterate $L_k$ to 0. CSVI based 
on SGD with these two simple modifications is presented in \cref{alg:csvi}. The convergence
of CSVI in the sense of \cref{def:asympsolvable} 
is provided by \cref{thm:csvi}.
\bnthm \label{thm:csvi} 
Suppose \cref{assump:regularity,assump:lsmooth} hold.
There exist constants $r, C > 0$ such that
if we initialize $\mu$ such that $\|\mu  - \htheta^\star_n\|_2^2 \leq \frac{r^2}{32}$,
and
\[
& \gamma_k = \Theta(k^{-\rho})\text{ for some }\rho\in(0.5, 1], \text{ and} \\ 
& \forall k\in\nats, \quad 0 < \gamma_k < C,  \label{eq:gammaconds}
\]
then for any $\rho' \in (0, \rho - 0.5)$, CSVI asymptotically solves Gaussian 
variational inference at rate $k^{-\rho'}$.
\enthm
Note that both constants $r$ and $C$ in the statement of \cref{thm:csvi} are not known in practice.
The constant $C$ is a step-size parameter that is typical in the analysis of stochastic 
gradient methods, and must be tuned.
The constant $r$, defined in \cref{cor:asymplocalconv}, intuitively represents the radius of the 
convex local basin around the optimal variational parameters.
In practice, 
one should run the smoothed MAP optimization \cref{alg:smoothedmap} 
until convergence to $\htheta_n^\star$ according to some diagnostic, e.g., a small gradient norm.
Further, note that one could use other stochastic gradient-based optimization schemes
in CSVI and the smoothed MAP optimization,
such as the Nesterov accelerated gradient algorithm \citep{nesterov1983method}, 
AdaGrad \citep{duchi2011adaptive}, and Adam \citep{kingma2015adam};
in our experiments we use Adam.
Although our theoretical
results do not cover those variants, we expect that they could be extended 
as long as one uses the gradient scaling on $L$ in \cref{eq:scaledLgradient}. 

\begin{algorithm}[t!]
\caption{Consistent stochastic variational inference} \label{alg:csvi}
\begin{algorithmic}
  \Procedure{CSVI}{$-\frac{1}{n}\log \pi_n$, $g$, $(\gamma_k)_{k\in\nats}$, $K$}
 \State $\mu \gets$\texttt{SmoothedMAP} (\cref{alg:smoothedmap}) 
 \State $L \gets I$
 \For{$k=1, \dots, K$}
    \State Sample $Z \sim \distNorm(0, I)$
    \State $g \gets \hat\nabla_{\mu,n}(\mu, L, Z)\,$ and $\,G \gets \hat\nabla_{L,n}(\mu, L, Z)$
    \For{$i=1, \dots, d$}
        \If{$L_{ii} > 0$}
        \State $G_{ii} \gets \frac{1}{1+(nL_{ii})^{-1}}G_{ii}$
        \Else
        \State $G_{ii} \gets -1$
        \EndIf
    \EndFor
    \State $\mu \gets \mu - \gamma_k g\,$ and $\,L \gets L - \gamma_k G$
    \For{$i=1, \dots, d$}
       \State $L_{ii} \gets \max\left\{0, L_{ii}\right\}$
    \EndFor
 \EndFor
 \State \Return $\mu, L$
\EndProcedure
\end{algorithmic}
\end{algorithm}

\section{Computational asymptotic theory} \label{sec:asymptotics}
In this section, we provide a detailed investigation of the
MAP and variational inference optimization problems, which underpins the
convergence of CLA (\cref{thm:mapasymp}) and CSVI (\cref{thm:csvi}).  We take
advantage of the theory of statistical asymptotics to show that
as we obtain more data, 
the optimum solutions of \cref{eq:map,eq:gaussianvi} each converge to a fixed value,
the objective functions become locally strongly convex around that fixed value,
and the smoothed MAP initialization lies within that local region. 

\subsection{Statistical model and assumptions}\label{sec:assumptions}

As is common in past work
\citep{shen2001rates,ghosal2000convergence,kleijn2012bernstein}, we take a
frequentist approach to analyzing Bayesian inference.
We assume that the sequence of observations are independent and identically distributed $(X_i)_{i=1}^n \distiid P_{\theta_0}$
from a distribution $P_{\theta_0}$ with parameter $\theta_0\in\reals^d$
 selected from a parametric family $\{P_{\theta} : \theta \in \reals^d\}$. 
We further assume that for each $\theta\in\reals^d$, $P_\theta$ has common support, has
 density $p_\theta$ with respect to some common base
measure, and that $p_\theta(x)$ is a Lebesgue measurable function of $\theta$ for all $x$. Finally, we assume the prior
distribution $\Pi_0$ has full support on $\reals^d$ with density $\pi_0$ with respect to the Lebesgue measure. Thus
by Bayes' rule, the posterior distribution $\Pi_n$ has density proportional to the prior density times the likelihood, i.e.,
\[
\pi_n(\theta) \propto \pi_0(\theta) \prod_{i=1}^n p_\theta(X_i).
\]
In order to develop the theory in this work, we require a set of additional
technical assumptions on $\pi_0$ and $p_\theta$ given by
\cref{assump:regularity}. These are a collection of regularity conditions that
are standard for parametric models, which 
guarantee that the \emph{maximum likelihood estimate (MLE)} $\theta_{\text{MLE},n}\defined \argmax_{\theta \in \reals^d} \sum_{i =1}^n \log p_{\theta}\left(X_i\right)$ 
is well-defined and asymptotically $\sqrt{n}$-consistent for
$\theta_0$ \citep[Theorem 5.39]{van2000asymptotic}, and that the Bayesian
posterior distribution  of $\sqrt{n}(\theta - \theta_0)$ converges in total
variation to a Gaussian distribution; this is known as the \emph{Bernstein-von Mises theorem}
\citep[Theorem 10.1]{van2000asymptotic}. 
Also, \cref{assump:regularity} is sufficient to guarantee the consistency of
MAP estimator $\theta^\star_{n}$ to $\theta_0$
\citep[Lemma 2.1]{grendar2009asymptotic}, and the asymptotic exactness of the
Laplace approximation \cref{eq:laplace} \citep[Theorem 4]{miller2021asymptotic}.
\bnassum\textbf{(Regularity Conditions)} 
\label{assump:regularity}
\benum
\item $\left\{P_\theta : \theta\in\reals^d\right\}$ is an identifiable family of distributions;
\item For all $x,\theta$, the densities $\pi_0, p_\theta$ are positive and twice continuously differentiable in $\theta$;
\item There  exists a measurable function $L(x)$  such that for $\theta, \theta'$ in a neighbourhood of $\theta_0$ and for all $x$,
\[
&|\log p_{\theta}(x) - \log p_{\theta'}(x)| \leq L(x) \|\theta - \theta'\|,\\
& \EE_{\theta_0} \left[L(X)^2 \right] <\infty;
\]
\item For all $\theta$, 
\[
H_\theta  \defined & -\EE_\theta\left[\nabla^2\log p_\theta(X)\right] \\
 =&  \EE_\theta\left[\nabla \log p_\theta(X)\nabla \log p_\theta(X)^T\right],
\]
and $H_{\theta_0} \succeq \epsilon I$ for some $\epsilon > 0$. Further, for $\theta, \theta'$ in a neighbourhood of $\theta_0$, 
\[
(\theta, \theta')\to \EE_{\theta'} \left[ -\nabla^2 \log p_\theta(X)\right]
\]
is continuous in spectral norm;
\item There exists a measurable function $g(x)$ such that for $ \theta$ in a neighbourhood of $\theta_0$ and for all $x$,
\[
\max_{i,j\in [d]} \left|\left[\nabla^2 \log p_{\theta}(x) \right]_{i,j} \right| < g(x), \quad \EE_{\theta_0}[g(X)] < \infty.
\]
\eenum 
\enassum
\bnthm[Bernstein-von Mises \& MLE consistency {(\citep[Theorems 5.39, 10.1]{van2000asymptotic}}] \label{thm_bvm} 
\phantom{p} \\Under \cref{assump:regularity},
\[\label{bvm}
\begin{aligned}
& \sqrt{n} \left( \theta_{\text{MLE},n} - \theta_0\right) \convd \distNorm\left(0, H_{\theta_0}^{-1} \right),\text{ and } \\
& \tvd{\Pi_{n}\,}{\,\mcN\left(n^{-1/2}\Delta_{n, \theta_0}+\theta_0, n^{-1}H_{\theta_0}^{-1}\right)}\stackrel{P_{\theta_0}}{\longrightarrow} 0,
\end{aligned}
\]
where
$\Delta_{n, \theta_{0}}=n^{-1/2}\sum_{i=1}^{n} H_{\theta_{0}}^{-1}\nabla \log p_\theta(X_i)$.
\enthm

\bnthm[Laplace \& MAP consistency {(\citealp[Lemma 2.1]{grendar2009asymptotic}; \citealp[Theorem 4]{miller2021asymptotic})}] \label{thm_deterbvm} 
 \phantom{p}\\Under \cref{assump:regularity},
\[\label{deterbvm}
\begin{aligned}
& \theta^\star_{n} \stackrel{P_{\theta_0}}{\to} \theta_0,\text{ and }
\tvd{\Pi_n\,}{\distNorm\left( \theta^\star_{n}, \Sigma_{\text{Lap},n} \right)} \stackrel{P_{\theta_0}}{\longrightarrow} 0.
\end{aligned}
\]
\enthm

Note that the above conditions in \cref{assump:regularity} are stronger \citep[Lemmas 7.6 and 10.6]{van2000asymptotic}
than the usual \emph{local asymptotic normality} \citep[Section 7]{van2000asymptotic} 
and \emph{testability} \citep[p.~141]{van2000asymptotic} conditions required
for asymptotic posterior concentration and Gaussianity in the 
Bayesian asymptotics literature. Many of the results in this work would still hold with these
weaker conditions, but we prefer \cref{assump:regularity} for the present work as 
these conditions are simpler to state and check.

The regularity conditions in
\cref{assump:regularity}---which essentially all pertain to a local neighborhood of $\theta_0$---are sufficient
for the analysis of the smoothed MAP optimization and Laplace approximation.
For variational inference, however, we additionally
require asymptotic control on the \emph{global} smoothness of the negative log
posterior density $f_n$. This is essentially because 
$F_n$ in the variational objective \cref{eq:gaussianvi} 
is an expectation of $f_n$ under a normal distribution;
if the tails of $f_n$ are nonconvex and grow quickly, they can
influence the convexity of $F_n$ even locally around $\theta_0$.
In this work, we impose a bound
on the second derivative, but we conjecture that bounds on higher-order
derivatives would also suffice; see \cref{subsec:cvxsmthF} for details.

\bnassum \label{assump:lsmooth}
\textbf{(Asymptotic Smoothness)}
There exists an $\ell > 0$ such that
\[
	\Pr\left(\sup_{\theta}\left\|n^{-1}\nabla^2\log\pi_n(\theta)\right\|_2 > \ell\right) &\to 0,	
\]
where the $\|\cdot\|_2$ denotes the matrix spectral norm.
\enassum

\subsection{Global optimum consistency}\label{sec:globalconsist}

The first important property of both Gaussian variational inference and the
MAP problem is that the optimum point converges to a fixed location;
this substantially simplifies the convergence analysis of both CLA and CSVI.
In particular, for the Laplace approximation, $\theta^\star_n$ converges in
probability to the data-generating parameter $\theta_0$.  For Gaussian
variational inference, $\mu^\star_n, L^\star_n$ converges in probability to
$\theta_0, L_0$, where $L_0$ is the unique positive-diagonal Cholesky factor of
the inverse Fisher information matrix $H^{-1}_{\theta_0} = L_0L_0^T$.  The
precise statement for the Laplace approximation was given earlier in
\cref{thm_deterbvm}, while \cref{thm:optimum_inside} provides the precise
statement for variational inference; the proof follows directly from a result
regarding the total variation consistency of the optimal variational
distribution \citep{wang2019frequentist} and the continuity of the
positive-diagonal Cholesky decomposition \citep[p.~295]{Schatzman02}.
\bnthm \label{thm:optimum_inside}
Under \cref{assump:regularity},
\[
\forall \epsilon > 0, \qquad \lim_{n\to \infty} 
\Pr\left(
\left\|\mu^\star_n - \theta_0\right\| < \epsilon,\,\,  
\left\|L^\star_n - L_0\right\| < \epsilon 
\right) = 1.
\]
\enthm

\subsection{Convexity of smoothed MAP and consistency of $\htheta^\star_n$}\label{sec:smoothedmap} 

Next, we analyze the properties of the smoothed MAP problem
and its optimum $\htheta^\star_n$.
Although intuitively reasonable, Gaussian smoothing typically does not
typically come with strong practical theoretical guarantees, essentially
because a good choice of the smoothing variance $\alpha_n$ is not known.
\citet{mobahi2013optimization} shows for a continuous integrable function
with quickly decaying tails (at rate $\|x\|^{-d-3}$ as $\|x\|\to\infty$), the
smoothed function is strictly convex given a large enough selection of
$\alpha_n$. \citet{addis2005local} studies the smoothing effect of a
log-concave kernel on a special type of piecewise constant function, and
proves that the smoothed function is either monotonic or unimodal. 
To the best of our knowledge, previous analyses of smoothed optimization 
do not provide guidance regarding the choice of $\alpha_n$
or bounds on the error of the 
smoothed optimum point versus the original.

In contrast to past work, we leverage the asymptotic concentration of the
statistical model as $n\to\infty$ to obtain error bounds
as well as guidance on choosing $\alpha_n$. 
In particular, \cref{thm:convexsmoothedMAP} shows that if the sequence
$\alpha_n$ is chosen to decrease slower than $n^{-1/3}$, the smoothed MAP
problem is eventually strictly convex within any arbitrary compact domain,
and that the solution of the smoothed MAP problem $\htheta^\star_n$ 
is asymptotically consistent for $\theta_0$ at a $\sqrt{\alpha_n}$ rate. 
Therefore, we can tractably estimate $\htheta_n^\star$,
and---combined with the result of the previous section---use it as an
initialization for $\theta$ in CLA and $\mu$ in CSVI that is guaranteed
to be close to $\theta^\star_n$ and $\mu^\star_n$, respectively.

\bnthm \label{thm:convexsmoothedMAP}
Suppose \cref{assump:regularity} holds and $n \alpha_n^3 \to
\infty$. Then for all $M>0$, the probability that the smoothed MAP optimization problem 
\[
\min_{\|\theta-\theta_0\|\leq M}  - \log \hat{\pi}_n(\theta)
\]
is strictly convex converges to $1$ as $n \to \infty$ under the data generating
distribution. Further, the optima for the smoothed MAP problem is asymptotically $\sqrt{\alpha_n}$-consistent, that is
\[
	\|\hat{\theta}_{n}^\star - \theta_{0}\| = O_{P_{\theta_0}}(\sqrt{\alpha_n}).
\]
\enthm

\subsection{Asymptotic local convexity and smoothness of $f_n$}\label{subsec:cvxsmthf}

Note that the statistical consistency of the 
optima $\mu^\star_n$, $L^\star_n$, $\theta^\star_n$
and smoothed MAP $\htheta^\star_n$
alone do not provide a complete analysis of the asymptotics;
in order to make use of these
results, we require that the variational inference objective \cref{eq:gaussianvi} 
and the MAP objective \cref{eq:map} are
well-behaved in some sense. 
Since we have access only to (stochastic estimates of)
the gradient of the objective function in \cref{eq:gaussianvi,eq:map}, 
and stochastic gradient descent is known
to solve optimization problems with \emph{strongly convex} and \emph{Lipschitz
smooth} objectives \citep{Bottou2004,rakhlin2012making}, 
this amounts to
analyzing the convexity and smoothness of the objective
functions.\footnote{There are many other properties one might require of a
tractable optimization problem, e.g., pseudoconvexity
\citep{crouzeix1982criteria}, quasiconvexity \citep{arrow1961quasi}, or
invexity \citep{ben1986invexity}.  We focus on convexity as it does not impose
overly stringent assumptions on our theory and has stronger implications than
each of the aforementioned conditions.} 
We begin with a generalization of the typical definitions of strong convexity
and Lipschitz smoothness found in the literature \citep{boyd2004convex} in
\cref{def_convexity}.
\bnumdefn[Convexity and Smoothness]\label{def_convexity} 
Let $g: \mcX \rightarrow \reals$ be a twice differentiable function on a convex set $\mcX\subseteq \reals^d$,
and let $D: \mcX \rightarrow \reals^{d\times d}$ be a positive definite matrix depending on $x$. Then
$g$ is \emph{$D$-strongly convex} if 
\[
\begin{aligned}
\forall x\in \mcX, \quad& \nabla^2 g(x) \succeq D(x),
\end{aligned}
\]
and $g$ is \emph{$D$-Lipschitz smooth} if 
\[
\begin{aligned}
\forall x \in \mcX, \quad & -D(x) \preceq \nabla^2 g(x) \preceq D(x).
\end{aligned}\label{eq:lipschitzsmooth}
\]
If either property holds only within a convex subset
$\mcY \subset \mcX$, we say it holds \emph{locally} within $\mcY$.
\enumdefn

In general, the MAP objective function in \cref{eq:map}---i.e., the scaled negative log 
posterior density $f_n$---need 
not be strongly convex or smooth for any particular $n$,
and as such it is difficult to make any claim regarding convergence.
This is where statistical asymptotics provides a major benefit in 
optimization: by \cref{lem:asymplocalconv}, as $n\to\infty$, the 
probability
that $f_n$ becomes locally strongly convex and Lipschitz smooth
around $\theta_0$ converges to 1 (see \cref{example_asymp_convex} for an illustration of this effect). 
The convergence of CLA (\cref{thm:mapasymp}) is then essentially a consequence of 
the fact that we initialize near $\htheta_n$, which converges to $\theta_0$ by \cref{thm:convexsmoothedMAP},
and that the local convexity and smoothness in \cref{lem:asymplocalconv}
ensures that gradient descent will contract towards $\theta^\star_n$.
\bnlem\label{lem:asymplocalconv}
Under \cref{assump:regularity}, there exist $r, \epsilon, \ell > 0$ such that
\[
 & \Pr\left( f_n\text{ is }\epsilon I\text{-strongly convex and } \right. \\
 & \qquad \qquad \left. \ell I\text{-Lipschitz smooth in the set } B_r(\theta_0) \right) \to 1,   
 \]
as $n \to \infty$, where $B_r(\theta_0) \defined \{\theta \in \reals^d: \|\theta - \theta_0\| \leq r \}$.
\enlem

\begin{figure}[t!]
\centering
\includegraphics[width = \linewidth]{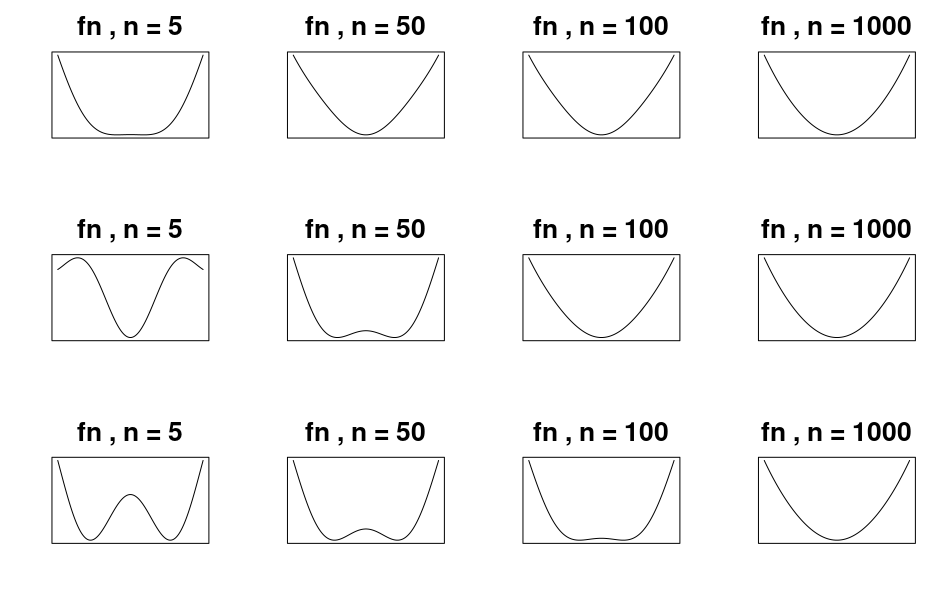}
\caption{Plots of the function $f_n(y)$ from \cref{example_asymp_convex}.
Each row of figures represents a single realization of the sequence $(f_n)_{n\in\nats}$
for increasing sample sizes 5, 50, 100, and 1000. Each column includes three repetitions of $f_n$ under a single $n$. As $n$
increases, the function $f_n(y)$ is more likely to be strongly convex and Lipschitz smooth
with constants approaching 2. }
\label{fig:asym_invex_example}
\end{figure}

\bnexa \label{example_asymp_convex}
Let $f_n(y) = y^2 + \left(\frac{1}{n}\sum_{i=1}^n X_i \right) \cos 5y$, where $X_i \sim \mcN(0,1)$.
Then
\[
\left|\frac{\dee^2\!f_n}{\dee y^2} - 2\right| = 25\left|\cos(5y)\right|\cdot\left|n^{-1}\sum_{i =1}^n X_i \right|.
\]
Therefore by the law of large numbers and the fact that $|\cos(5y)|\leq 1$, 
for any $\epsilon > 0$, the sequence $(f_n)_{n\in\nats}$ is asymptotically 
$(2-\epsilon)$-strongly convex and $(2+\epsilon)$-Lipschitz smooth.
\cref{fig:asym_invex_example} visualizes the asymptotic behaviour of $f_n$ as $n$ increases.
\enexa

\subsection{Asymptotic local convexity and smoothness of $F_n$}\label{subsec:cvxsmthF}

The variational objective function \cref{eq:gaussianvi} contains two terms:
a regularization $-n^{-1}\log \det L$, and an expectation of the negative log posterior
density under Gaussian noise, $F_n(\mu, L)$.
The regularization term is known to be convex \citep[p.73]{boyd2004convex}
and Lipschitz smooth in any compact subset of the domain (the gradient scaling 
\cref{eq:scaledLgradient} essentially makes the smoothness uniform on the whole domain;
see the proof of \cref{thm:csvi} in \cref{supp:proof}).
Therefore in this section we focus on the analysis of 
the data-dependent term, $F_n(\mu, L)$. 
The first main result is that $F_n(\mu, L)$ inherits the global 
convexity and smoothness behaviour of $f_n$.
\bnthm\label{thm:globalcvxsmth}
Suppose
$f_n$ is $D$-strongly convex (-Lipschitz smooth)
for positive definite matrix $D\in\reals^{d\times d}$.
Then $F_n$ reinterpreted as a function from $\reals^{(d+1)d}\to \reals$---by stacking $\mu$ and each
column of $L$ into a single vector---is $D'$-strongly convex (-Lipschitz smooth), where
\[
D' = \block(D, n^{-1}D, \dots, n^{-1}D) \in \reals^{(d+1)d\times (d+1)d},
\] 
and $\block$ creates a block-diagonal matrix out of its arguments.
\enthm

For example, if the posterior distribution $\Pi_n$ is a multivariate Gaussian distribution $\distNorm(\mu_n, n^{-1}\Sigma_n)$ 
with mean $\mu_n$ and covariance $n^{-1}\Sigma_n$,
then the expectation component of the Gaussian variational inference objective function becomes
\[
 F_n(\mu, L) =  n^{-1}\tr\Sigma_{n}^{-1} LL^T +\left(\mu-\mu_{n}\right)^{T} \Sigma_{n}^{-1}\left(\mu-\mu_{n}\right),	
\]
which is a jointly convex quadratic function in $\mu, L$
with Hessian matrix (for $\mu$ and columns of $L$ stacked together in a single vector) 
equal to 
\[\block(\Sigma_n^{-1}, n^{-1}\Sigma_n^{-1}, \dots, n^{-1}\Sigma_n^{-1})\in\reals^{(d+1)d\times (d+1)d}.\]
Combined with the convexity of the log determinant term $-n^{-1}\log \det L$ \citep[p.73]{boyd2004convex},
Gaussian variational inference for strongly convex and 
Lipschitz smooth negative log posterior density $-n^{-1}\log\pi_n$  
is itself strongly convex and Lipschitz smooth in any compact set contained in 
the optimization domain.

However, in a typical statistical model, the posterior is typically neither
Gaussian nor strongly convex.  But when the Bernstein-von Mises theorem
holds~\citep{van2000asymptotic}, the posterior distribution (scaled and shifted
appropriately) converges asymptotically to a Gaussian distribution.  Thus, it
may be tempting to think that the Bernstein von-Mises theorem implies that
Gaussian variational inference
should eventually become a convex optimization problem. This is unfortunately
not true, essentially because Bernstein-von Mises only implies convergence to a
Gaussian in total variation distance, but not necessarily in the log density
function or its gradients.  The second main result in this
section---\cref{prop:vbfails}---is a simple demonstration of the fact that the
Bernstein-von Mises theorem is not sufficient to guarantee the convexity of
Gaussian variational inference.
\bnprop\label{prop:vbfails}
Suppose $d=1$, $f_n$ is differentiable to the third order
for all $n$, that there exists an open interval $U\subseteq \reals$ and
$\eps > 0$ such that 
\[
\sup_{\theta \in U} \dder{f_n}{\theta}  \leq -\eps,
\]
and that there exists  $\eta > 0$ such that 
\[
\sup_{\theta\in\reals} \left| \frac{d^3f_n}{d \theta^3}\right| \leq \eta. 
\]
Then there exists a $\delta>0$ such that
\[
\sup_{\sigma < \delta, \, \mu \in U} \dder{}{\mu}\kl{\distNorm(\mu, \sigma^2)}{\Pi_n} < 0.
\]
\enprop

Although \cref{prop:vbfails} is a negative result about the global convexity of $F_n$,
it does hint at a very useful fact:  the \emph{local} convexity 
of $F_n$ matches that of $f_n$, assuming that we control the global growth of $f_n$
(e.g., in \cref{prop:vbfails}, we imposed a uniform bound on the $3^\text{rd}$ derivative).
This is due to the fact that 
$F_n(\mu, L) = \EE f_n(\mu + n^{-1/2}LZ)$, 
where $Z$ is a standard Gaussian random vector;
intuitively,
since the Gaussian distribution has very light tails,
as $n$ grows the Taylor approximation
\[
	F_n(\mu, L) \approx f_n(\mu) + \frac{1}{2}n^{-1} L^T\nabla^2 f_n(\mu) L + o(n^{-1})
\]
becomes accurate, 
assuming $f_n(x)$ does not grow too quickly as $\|x\|\to\infty$.
Therefore when $n$ is large, we expect $F_n$ to behave like $f_n$ as a 
function of $\mu$, and be roughly quadratic in $L$ with Hessian  $n^{-1}\nabla^2 f_n(\mu)$.
Thus, as long as $f_n$ is locally convex in $\mu$, we expect $F_n$ to be locally convex in $\mu, L$ as well.
The third main result of this section specifies the general link between the local convexity
behaviour of $f_n$ and $F_n$ assuming the global Lipschitz smoothness of $f_n$.
\bnthm\label{thm:strongconvexsmooth}
Suppose there exist $\epsilon, \ell, r > 0$ and $x\in\reals^d$ such that $f_n$ is 
globally $\ell I$-Lipschitz smooth
and locally $\epsilon I$-strongly convex in the set $\{y : \|y-x\|\leq r\}$.
Define 
\[
 D_n &\defined \block\left(I, n^{-1} I,  \dots, n^{-1} I\right) \in \reals^{(d+1)d\times (d+1)d} &&\\ 
 \tau_n(\mu,L) &\defined 1 - \distChiSq_{d+2}\left(n\frac{(r^2 - 2\|\mu - x\|^2)}{2\|L\|^2_F}\right), \label{eq:Ttau}
\]
where $\distChiSq_{k}$ is the CDF of a chi-square random variable with $k$ degrees of freedom.
Then $F_n$ reinterpreted as a function of $\reals^{(d+1)d} \to \reals$---by stacking $\mu$ and each column of $L$ into a
single vector---is $\ell D_n$-Lipschitz smooth; and is $(\epsilon - \tau_n(\mu, L)\cdot(\epsilon+\ell))D_n$-strongly convex when $\|\mu - x\|^2 \leq \frac{r^2}{2}$. 
\enthm

The function $\tau_n(\mu, L)$ in \cref{thm:strongconvexsmooth} 
characterizes how much the tails of $f_n$
can influence the local strong convexity of $F_n$ around the point $x$.  In
particular, as long as $\mu$ is close to $x$ and $\|L\|_F$ (which modulates the
effect of noise) is sufficiently small, then the argument of the
$\distChiSq_{d+2}$ CDF is large, so $\tau_n$ is small, 
so $(\epsilon - \tau_n(\mu, L)\cdot(\epsilon+\ell)) \approx \epsilon$; thus we recover local
strong convexity of the same magnitude as $f_n$.  A further note is that
although \cref{thm:strongconvexsmooth} requires a uniform bound on the Hessian
of $f_n$, we conjecture that a similar result would hold under the assumption
of a uniform bound on the $k^\text{th}$ derivative.  For simplicity of the
result and ease of use later on in \cref{sec:consistentvi}, we opted for the
second derivative bound.

The last result of this section---\cref{cor:asymplocalconv}---combines
\cref{thm:strongconvexsmooth,thm:optimum_inside,lem:asymplocalconv}
to provide the key asymptotic convexity/smoothness result that we use in the
development of the optimization algorithm in \cref{sec:consistentvi}.

\bncor \label{cor:asymplocalconv}
Suppose \cref{assump:regularity,assump:lsmooth} hold, and define
$D_n$ as in \cref{thm:strongconvexsmooth}. Then there exist $\epsilon, \ell, r > 0$ such that 
$F_n$ reinterpreted as a function of $\reals^{(d+1)d}\to\reals$---by stacking $\mu$ and each column 
of $L$ into a single vector---satisfies
\[
	& \Pr\left(  F_n\text{ is } \frac{\epsilon}{2}D_n \text{-strongly convex in $\mcB_{r,n}$ and} \right.\\
	&\qquad \qquad  \left.\text{ globally } \ell D_n\text{-Lipschitz smooth}\right) \to 1,	
\]
as $n \to \infty$, where
\[
\mcB_{r,n} = \left\{\mu\in\reals^d, L\in\reals^{d\times d} : \| \mu - \mu_n^\star\|^2 \leq \frac{r^2}{4}  \text{ and } \right.\\
 \left. \|L-L_n^\star\|_F^2 \!\leq 4\|I-L_n^\star\|_F^2\right\}.
\]
\encor

\section{Experiments}\label{sec:simulation}

\begin{figure*}[t!]
    \centering
    \includegraphics[width=\textwidth]{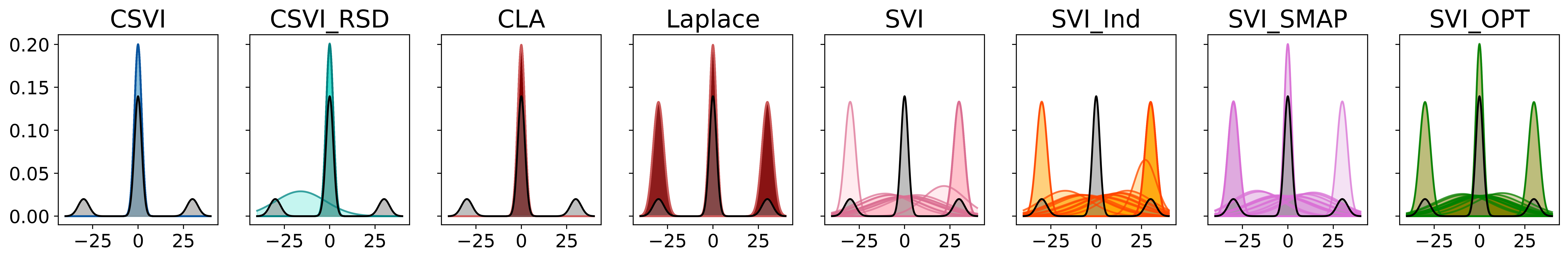}
    \caption{The result of 20 trials of Gaussian approximation with the 
    Gaussian mixture target (grey) given in \cref{eq:pigaussmix}. Each plot
    shows the target distribution and 20 Gaussian approximations obtained from one algorithm-initialization combination. 
    }
    \label{fig:mixline}
\end{figure*}

\begin{figure}[t!]
    \centering
    \includegraphics[width=\linewidth]{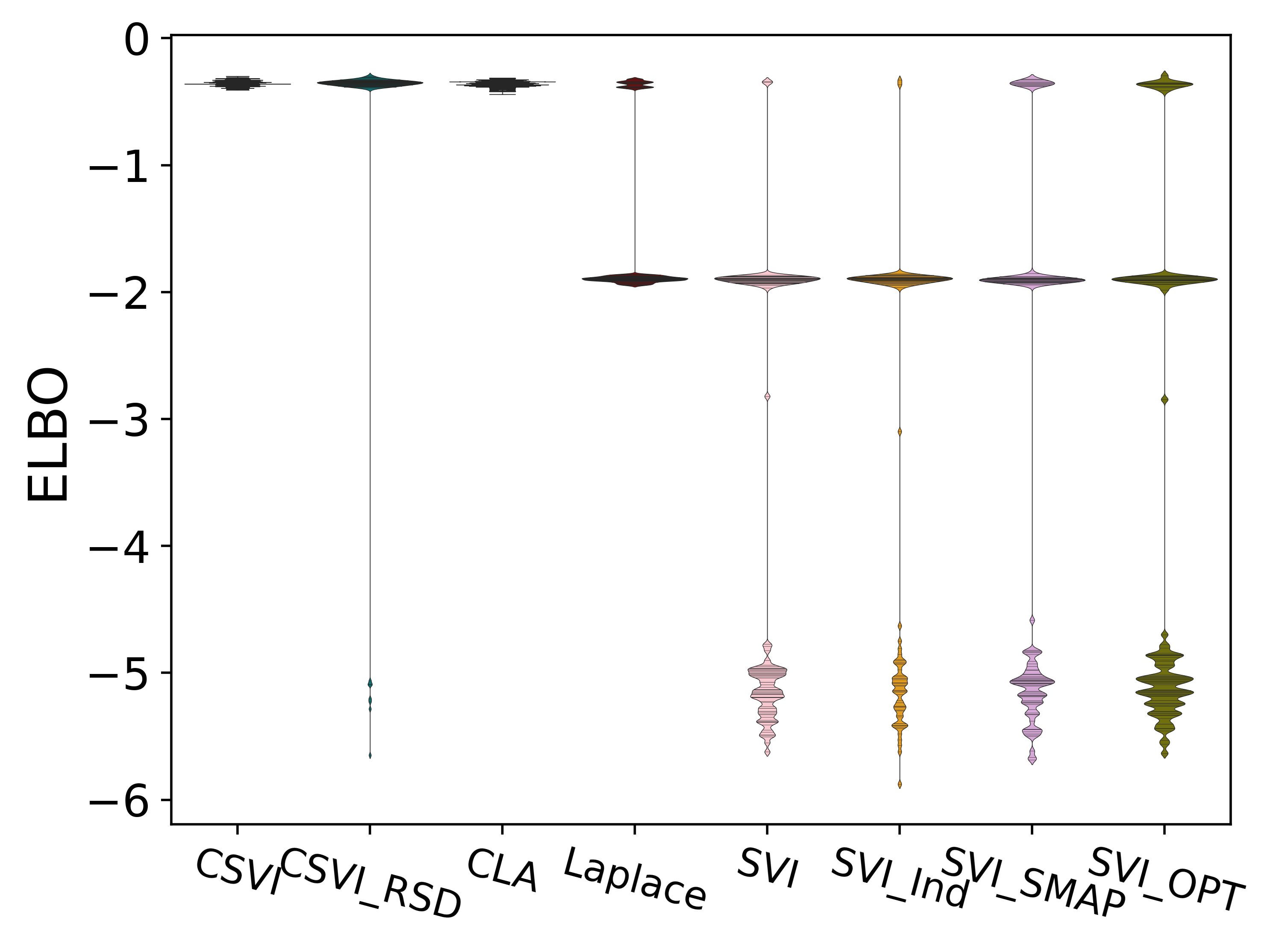}
    \caption{The violin plots of the ELBOs of running $100$ trials of Gaussian
      approximation with the 
    Gaussian mixture target given in \cref{eq:pigaussmix}.  
    The output of CSVI and CLA reliably finds the global optimum solution corresponding to the central mixture peak;
    SVI often provides solutions corresponding to local optima.}
    \label{fig:mixelbo}
\end{figure}

In this section, we compare CSVI and CLA to standard 
Gaussian stochastic variational inference 
(SVI)\footnote{Code for the experiments is available 
at \url{https://github.com/zuhengxu/Consistent-Stochastic-Variational-Inference}.} and 
the standard Laplace approximation on both synthetic and real data inference problems. 
By default we run both 
variational optimization algorithms for $100{,}000$ iterations and 
the smoothed MAP optimization for $20{,}000$ iterations.
We base the gradients for the smoothed MAP mean initialization (\cref{alg:smoothedmap}) on $100$ samples, 
and the gradients for VI algorithms on a single sample. For both CLA
and Laplace, we run backtracking line search for $20{,}000$
iterations, with $\beta = 0.5$. 

\subsection{Synthetic Gaussian mixture}

In the first experiment, we compare the reliability of CSVI and SVI, CLA and
Laplace approximation on a
simple target function under different initialization schemes, choices of smoothing constant, and
learning rates. The inferential goal is to approximate a Gaussian mixture
distribution $\Pi$,
\[
\Pi\!=\! 0.7\distNorm(0, 4)\! + \!0.15 \distNorm(-30, 9)\! + \!0.15\distNorm(30, 9). \label{eq:pigaussmix}
\]
We set $n = 1$ in this example as there is no data likelihood. We use the
smoothing constant $\alpha_n = 100$ in the implementation of CSVI,
and initialize the smoothed MAP optimization, Laplace approximation  and the mean of SVI uniformly in
the range $(-50, 50)$. The standard deviation $\sigma$ for CSVI is initialized
to be $1$ and the log standard deviation $\log \sigma$ for SVI is initialized
uniformly in the range  $(\log 0.1, \log 10)$.  Unless otherwise indicated, we
hand-tune the learning rates for both CSVI and SVI to optimize 
performance---we set $\gamma_k = 5/(1+ k)$ for CSVI, $\gamma_k = 15/(1+k)$ for
SVI. $t$ for CLA is set as $1$.

In \cref{sec:mix_init}, we compare CLA, Laplace approximation, CSVI and SVI under various initialization
schemes, aiming to dissect the contribution of each element of our methodology.
Specifically, aside from the standard CSVI and SVI methods described above, we consider
$4$ additional combinations of VI algorithm and initialization: CSVI\_RSD,
SVI\_Ind, SVI\_SMAP and SVI\_OPT. In particular, 
CSVI\_RSD differs from CSVI by initializing $\log \sigma$ uniformly in the range  $(\log 0.5, \log 10)$,
SVI\_Ind denotes SVI with $\sigma_0 = 1$, 
SVI\_SMAP is SVI using $\mu_0$ as the smoothed MAP,
 and SVI\_OPT uses the optimal initialization (smoothed MAP for mean and $\sigma = 1$) for SVI. 
The results demonstrate that
both the smoothed MAP initialization and scaled gradient estimates are
necessary to produce consistent and reliable VI approximations; and the smoothed
MAP initialization also improve the reliability of Laplace method significantly.

In \cref{sec:mix_sens}, we investigate the sensitivity of 
CSVI to the smoothing constant $\alpha_n$. The results demonstrate that
the performance of CSVI is very robust to the change of $\alpha_n$. We also compare the reliability of CSVI
and SVI  across different optimization step schedules $\gamma_k$, in which CSVI
outperforms SVI in all settings and generally favours smaller learning rate.

\subsubsection{Sensitivity to initialization} \label{sec:mix_init}

\begin{figure*}[t!]
    \centering
    \includegraphics[width=\linewidth]{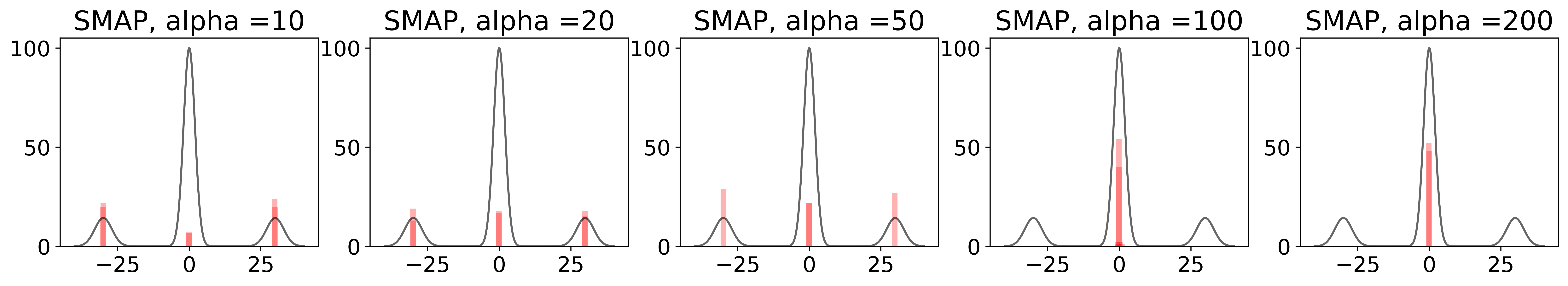}
    \caption{The result of smoothed MAP of the Gaussian mixture target \cref{eq:pigaussmix} across different values of $\alpha_n$. The black curve corresponds to the Gaussian mixture density and the red histogram shows the counts (over 100 trials) of the output of the smoothed MAP initialization. }
    \label{fig:alpha_smap}
\end{figure*}

\begin{figure}[t!]
    \centering
    \includegraphics[width=\linewidth]{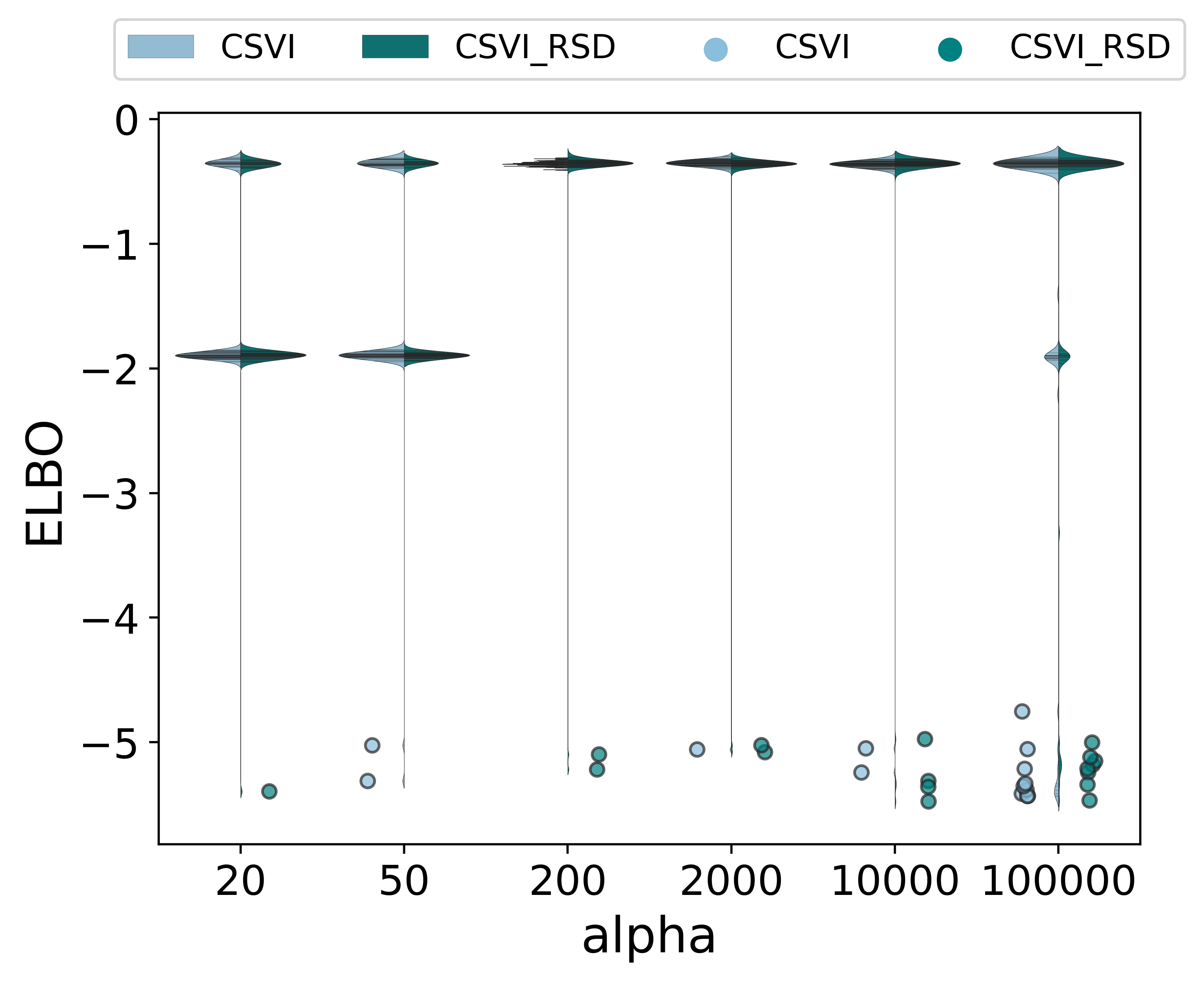}
    \caption{The result of running 100 trials of CSVI and CSVI\_RSD  with the 
    Gaussian mixture target  \cref{eq:pigaussmix} using different smoothing values $\alpha  =  20,50,200,2000,10000, 100000$. The scatter points mark ELBO outliers.} 
    \label{fig:alpha_elbo}
\end{figure}

We first demonstrate the performance of CLA, Laplace, CSVI, SVI and their $4$ variants under
different initialization schemes---CSVI\_RSD, SVI\_Ind, SVI\_SMAP and SVI\_OPT. 
We run $100$ trials for each variant.
 \cref{fig:mixline} visualizes $20$
variational approximations and Laplace approximations that are randomly selected from the $100$ trials.  Note
that the majority of the mass of the Gaussian mixture target distribution
concentrates on the central mode with mean 0 and standard deviation 2; the
optimal variational approximation has these same parameters. As shown in the
plot, CSVI reliably learns the optimal variational distribution.
CSVI\_RSD---with randomly initialized standard deviation---occasionally becomes trapped in
a local optimum that places the
mean between the central mode and the adjacent peaks with a large standard
deviation. The variants without gradient scaling (SVI, SVI\_Ind, SVI\_SMAP, and SVI\_Opt) 
are significantly more likely to find this same local optimum;
this is because the standard projected gradient has unstable
behaviour for small $\sigma$ 
 due to the log-determinant regularization term. The comparison between CLA and
 standard Laplace method reinforce  the importance of a reliable initialization
 shceme for deterministic posterior approximation; Laplace approxiamtion with
 random initilization is particularly sensitive to the local optima of posterior
 due to its lack of stochasticity.

These observations reveal two important facts. First, a
good mean initialization is important and helps recover
the global optimum. Second, the gradient scaling
described  in \cref{eq:scaledLgradient} aids the stability of the VI algorithm,
which ensures that the algorithm stays in the region around the optimum, and hence
converges to the optimal solution. 

\cref{fig:mixelbo} presents a quantitative characterization of this result. In
particular, we plot the final \emph{expectation lower bound (ELBO)}
\citep{blei2017variational} for each method, which is equivalent to the
negative KL divergence between the posterior and variational distribution up to a
constant; a larger ELBO value corresponds to a better approximation.  
We estimate the ELBO using $1000$ Monte Carlo samples.
As demonstrated in the violin plots, CSVI and CSVI\_RSD find the global
optimum significantly more reliably than SVI and its variants. Also, by
comparing the distribution of the ELBO of the $100$ trials of SVI and its
variants, we find that the influence of the initial value alone is limited. This
aligns well with our earlier theory in \cref{cor:asymplocalconv}; in order to reliably 
find the global optimum of variational inference problem, one needs \emph{both} a careful initialization and to stay
in the basin of the global optimum during optimization. However, for the Laplace
approximation, careful intialization determines its performance.

\subsubsection{Sensitivity to smoothing and learning rate} \label{sec:mix_sens}

\captionsetup[subfigure]{labelformat=empty}
\begin{figure}[t!]
    \centering 
    \includegraphics[width=0.95\linewidth]{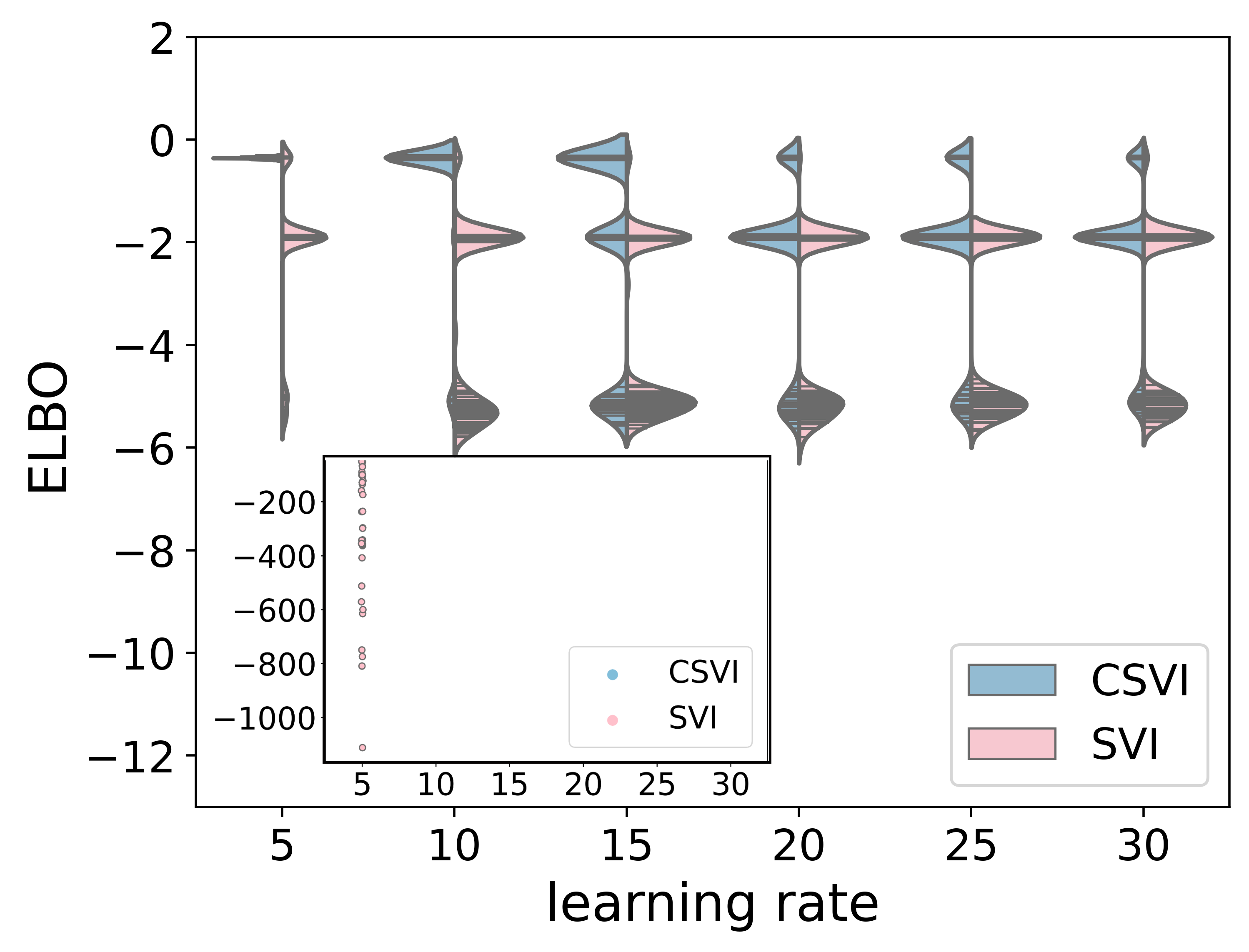}
\caption{The sensitivity results of CSVI and SVI with target distribution  \cref{eq:pigaussmix} across different $\gamma_k$. 
SVI often does not converge when $\gamma_k= 5$; the inset shows the results for these trials.} 
\label{fig:step}
\end{figure}

Next, we study the sensitivity of the VI algorithms to the choice of
the smoothing constant $\alpha_n$.  Note that the smoothing constant $\alpha_n$ is
the variance of the Gaussian smoothing kernel. A larger value corresponds to a
more aggressive smoothing effect, and hence a lower likelihood of finding a spurious
local peak.
As shown in \cref{fig:alpha_smap}, as long as $\alpha_n$ is set large enough,
the smoothed MAP initialization has a reasonable chance to locate $\mu_0$ at
the central mode of target distribution \cref{eq:pigaussmix}. As a result,
CSVI and CSVI\_RSD has a mean initialization close to that of the
optimal variational distribution.  \cref{fig:alpha_elbo} presents the
distribution of the ELBOs over $100$ trials of CSVI and CSVI\_RSD. 
This figure demonstrates that CSVI and CSVI\_RSD reliably find the optimal
ELBO for a wide range of $\alpha_n$ ranging from about 100 to 100{,}000.
In other words, the approach is not overly sensitive to the value of $\alpha_n$. 

Finally, we illustrate the sensitivity of CSVI and SVI to the optimization step
schedule. Both algorithms are run for $100$ trials across different step
schedules, i.e., $\gamma_k = C/(1+k)$ for $C=5, 10, 15, 20, 25, 30$.  
In  \cref{fig:step}, we display the spread of the ELBOs. In
general, CSVI outperforms SVI for all choices of $\gamma_k$---it is more likely to find
the optimum and the ELBO variation between trials is significantly smaller. This
confirms that CSVI is less sensitive to the choice of learning rate.
Further, SVI requires many more steps to converge than CSVI 
when $\gamma_k$ is small. As the step size gets larger,
CSVI may overshoot its original local basin and converge to a suboptimal point.

\subsection{Bayesian sparse linear regression}

In this experiment, we compare the quality of CSVI, SVI, CLA and Laplace on a Bayesian
sparse linear regression problem. As mentioned at the end of
\cref{sec:scaledsgd}, we use Adam \cite{kingma2015adam} updates in both the smoothed MAP estimation
and the variational inference algorithm
to achieve faster convergence. The detailed implementation of the
Adam version of CSVI is presented in \cref{supp:csvi_adam_alg}.

In the Bayesian sparse linear regression model, 
we are given a
set of data points $\left(x_n, y_n \right)_{n=1}^N$ with feature $x_n \in \reals^d$
and response $y_n \in \reals$, we assume that the responses
were generated from a Gaussian likelihood
\[
y_n   \mid x_n, \beta & \distind \distNorm \left( x_n^T \beta, \sigma^2 \right),
\]
and we assert that the feature coefficients
each have a ``spike and slab'' prior distribution consisting of a mixture of two
Gaussian distributions with different variances
\[
 \left( \beta_i \right)_{i = 1}^d &\distiid \frac{1}{2} \distNorm(0, \tau_1^2) + \frac{1}{2} \distNorm(0, \tau_2^2),
\]
where $\tau_1$ is set to a small value and $\tau_2$ is set to be large. Priors
of this type are commonly used to encode variable selection \citep{george1993variable}. The 
goal of the inference is to approximate the posterior distribution of $(\beta_1, \dots ,
\beta_d)$ with a full rank Gaussian distribution.

\begin{figure}[t!]
    \centering
    \includegraphics[width=\linewidth]{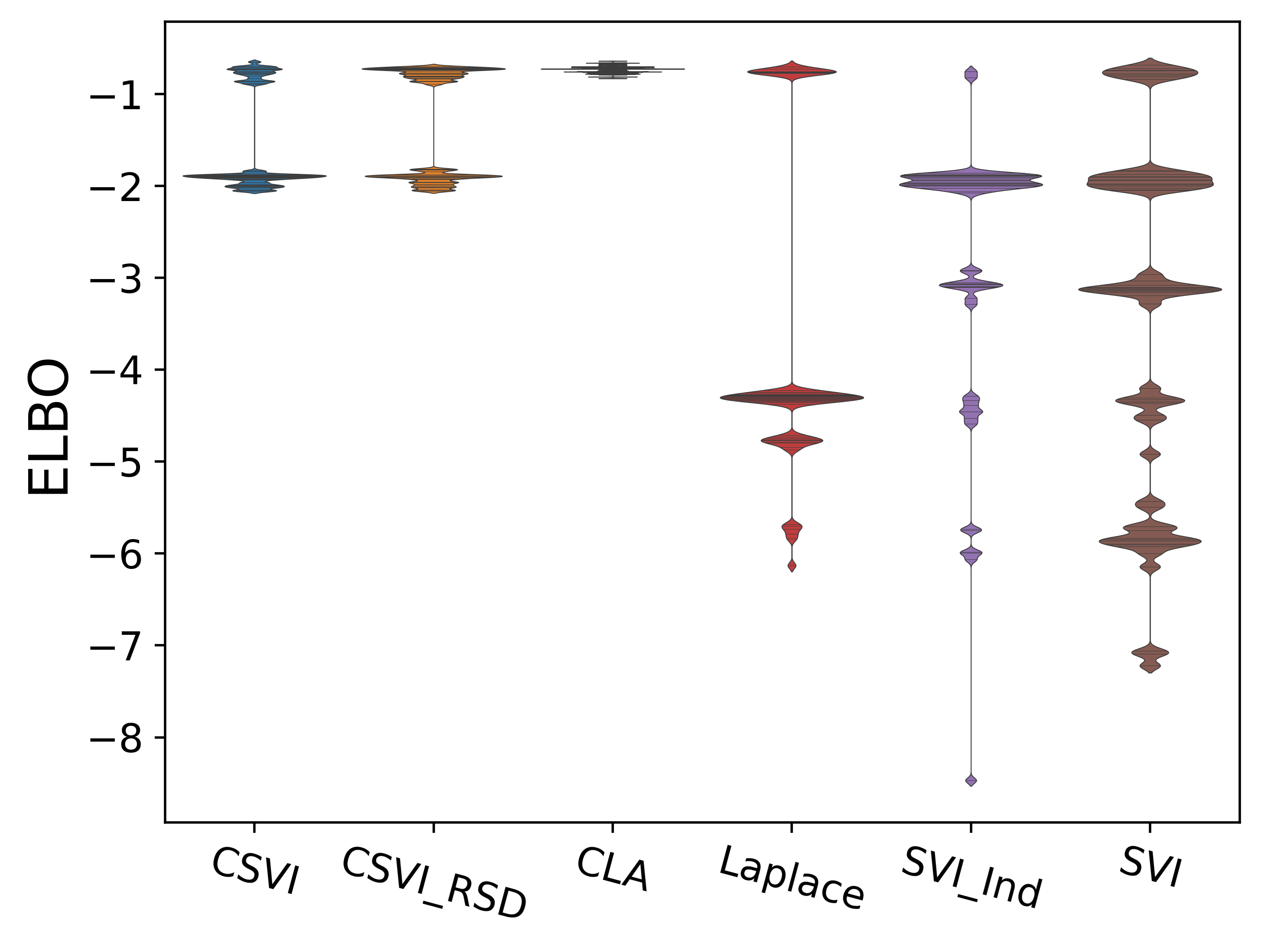}
    \caption{Sparse regression results on the synthetic dataset. The violin plots show the distribution of the ELBOs over $100$ trials.} \label{fig:sr_syn_elbo}
\end{figure}

\begin{figure*}[t!]
    \centering
    \includegraphics[width= \linewidth]{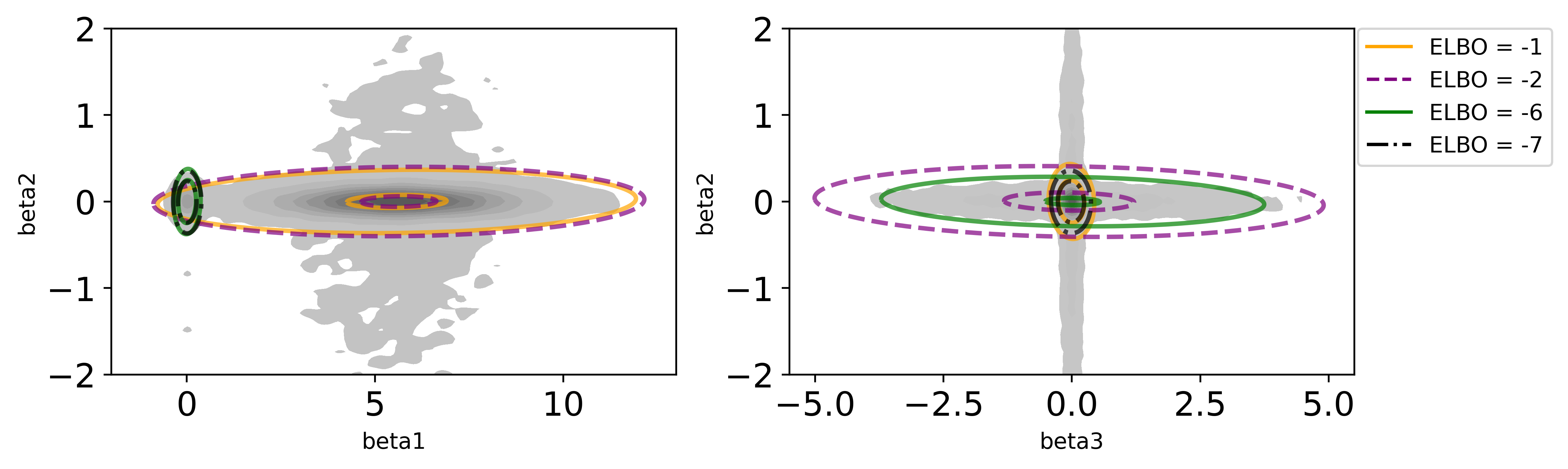}
    \caption{Visualization of different VI approximations to the posterior distribution on synthetic dataset ($\beta_1 \text{v.s.} \beta_2$; $\beta_3 \text{v.s.} \beta_2$ ).The grey area depicts the posterior density and four Gaussian approximations displayed in contour plots corresponding to different ELBOs.}
    \label{fig:sr_syn_vis}
\end{figure*}

We run $100$ trials of CSVI and SVI on two datasets---a synthetic
dataset, and a dataset of measurements of 97 men with prostate cancer\footnote{Available online at
\url{http://www.stat.cmu.edu/~ryantibs/statcomp/data/pros.dat}.}.
For the synthetic example, we set $\sigma = 5, \tau_1 = 0.1, \tau_2 = 10$ 
and generate features $x_n \in \reals^5 $ \iid from $\distNorm(0,I )$ with $N = 10$. The response $y_n$ is generated from the following process,
\[
y_n =  \begin{bmatrix}
    1 &0 &0 &0 &0
\end{bmatrix} ^T   x_n + \eps, \quad \eps \sim \distNorm(0, 0.25).
\]
We use learning rates of $\gamma_k = 0.01$ for the optimization of SMAP,
$t = 0.01$ for backtracking line search,
and set the smoothing constant $\alpha_n = 2$. Both the initial value of
SMAP optimization, MAP optimization and the mean initialization of SVI are randomly
sampled from the prior distribution.
In terms of $L_0$, we consider two different initialization schemes---the
identity matrix and a random diagonal matrix---for both CSVI and
SVI, where for the random $L_0$, $\log \left[L_0 \right]_{ii}, i\in [d]$
is uniform in the range $(\log 0.1, \log 100)$. The learning rates for
both VI algorithms are set to $0.001$. \cref{fig:sr_syn_elbo} shows the
ELBOs of VI inference on the synthetic data over $100$ trials. Compared to
SVI, CSVI  reliably learns a better approximation to the posterior and
rarely gets trapped at a local optimum.  
And CLA is much more robust to local minima compared to standard Laplace due to a good initialization.
To visualize different Gaussian approximation, 
\cref{fig:sr_syn_vis} shows contours of the optimal variational
approximation ($\text{ELBO} = -1$) and three local approximations
corresponding to $\text{ELBO} = -2, -6, -7$ respectively;
it is clear that CSVI tends to find better local optima than SVI.
Notice that in this example, the ELBO distribution of CLA is more consistent
than CSVI's. This is essentially due to the fact that CLA is a deterministic
approximation methods---the Hessian of log posterior evaluated at the mean
determines the fitted covariance, while CSVI is able to produce multiple
covariance fit given the Gaussian mean. 

\begin{figure}[t!]
   \centering
   \includegraphics[width=\linewidth]{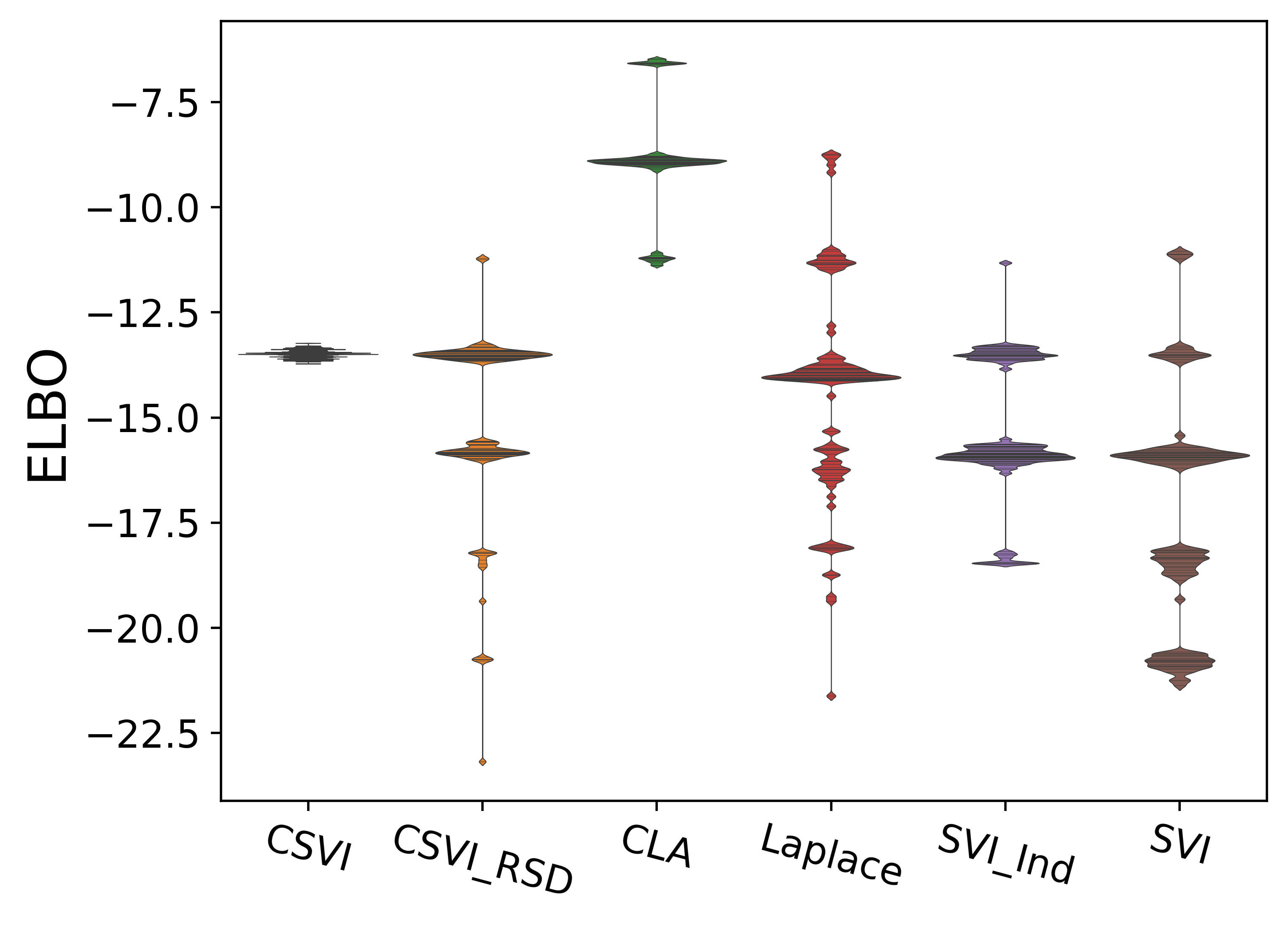}
   \caption{Sparse regression results on the prostate cancer dataset. The violin plots show the distribution of the ELBOs over $100$ trials.} \label{fig:sr_real_elbo}
\end{figure}

For the real dataset experiment, we subsample the original dataset to $N =30$ 
data points. We apply the SMAP optimization for $200{,}000$
iterations with the smoothing constant $\alpha_n = 0.03$; and the learning rate
for SMAP and VI algorithms are set as $0.002$ and $0.0002$
respectively. Other settings remain identical to the synthetic experiment.
The results in \cref{fig:sr_real_elbo} generally align with those from the
previous synthetic experiment---CSVI and CLA outperform SVI and Laplace
respetively with a more consistent and accurate Gaussian approximation.
But an interesting observation is that CLA dominates all other methods and 
CSVI fails to learn those Gaussian disitrbuiton produced by CLA. It is mainly
because that CLA fits into the dominating mode of posterior while CSVI prefers to fit
a wider Gaussian distribution that covers the whole range of the posterior. A
more detailed discussion, including several supporting visualizations 
is included in \cref{supp:sr_detail}.

\subsection{Bayesian Gaussian mixture model}

\begin{figure}[t!]
    \centering
    \includegraphics[width=\linewidth]{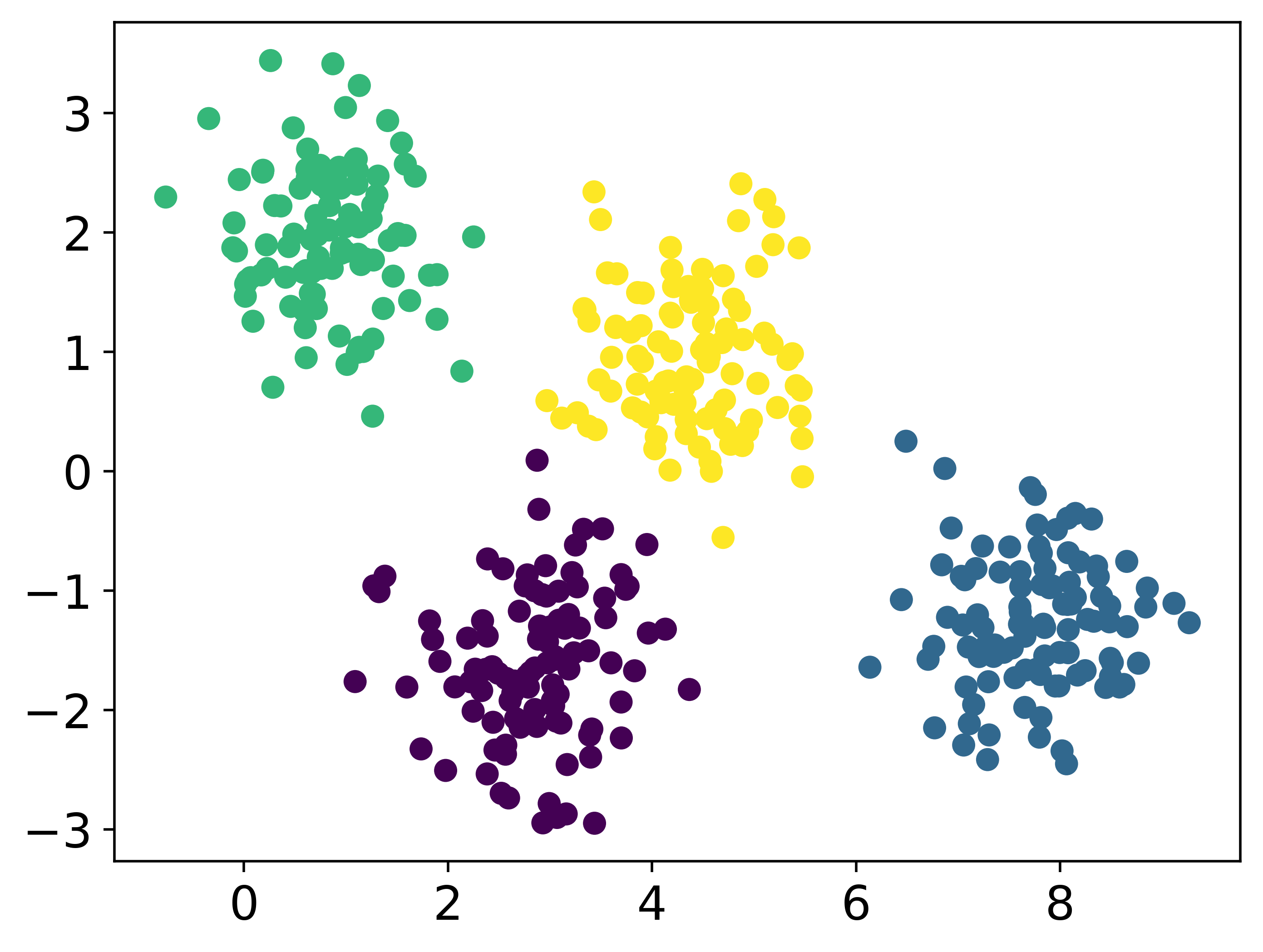}
    \caption{Synthetic dataset of GMM. } \label{fig:gmm_syn_dat}
\end{figure}

\captionsetup[subfigure]{labelformat=empty}
\begin{figure}[t!]
    \centering 
\begin{subfigure}[b]{.495\textwidth} 
    \scalebox{1}{\includegraphics[width=\textwidth]{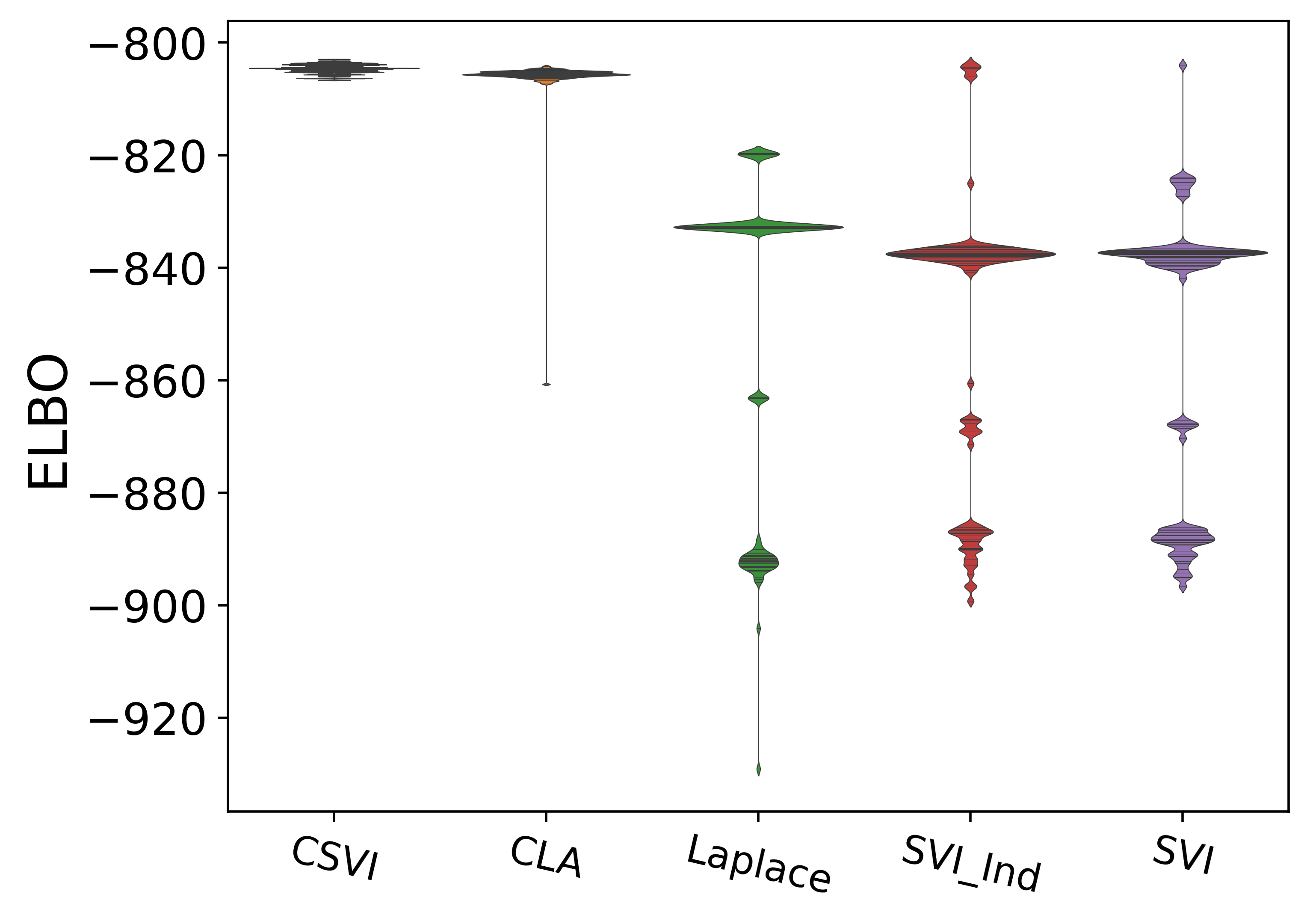}}
    \caption{(a)\label{fig:gmm_syn_elbo}}
\end{subfigure}
\hfill
\centering
\begin{subfigure}[b]{0.495\textwidth}
    \scalebox{1}{\includegraphics[width=\textwidth]{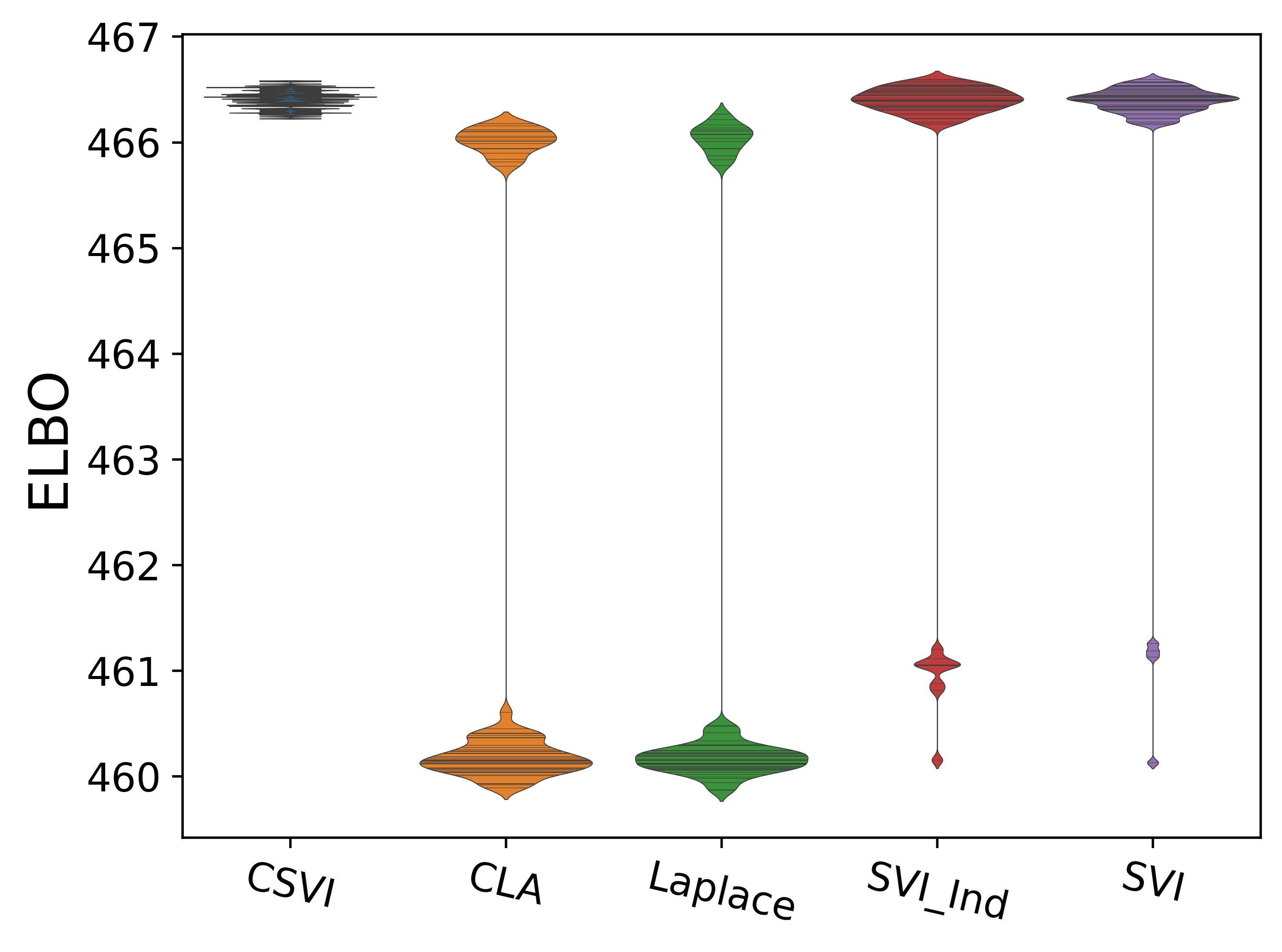}}
    \caption{(b)\label{fig:gmm_real_elbo}}
\end{subfigure}

\caption{GMM results on the synthetic dataset \cref{fig:gmm_syn_elbo} and real dataset \cref{fig:gmm_real_elbo}. The violin plots show the distribution of the ELBOs over $100$ trials. }
\label{fig:gmm_syn}
\end{figure}

 We finally compare the methods on a Bayesian Gaussian mixture model (GMM) applied to a synthetic dataset and the Shapley galaxy dataset 
 \footnote{This dataset contains the measurements of the redshifts for $4215$
 galaxies in the Shapley Concentration regions and was generously made available
 by Michael Drinkwater, University of Queensland, which can be downloaded from
 \url{https://astrostatistics.psu.edu/datasets/Shapley_galaxy.html} .}.  Similar
 to the previous experiments, we employ Adam for all optimization procedures
 involved in the inference and each experiment is repeated for $100$ trials. In
 the Bayesian Gaussian mixture model, 
we are given $N$ observations $(x_n)^N_{n=1}\subseteq \reals^d$, each following a $k$-component Gaussian mixture distribution:
\[
    x_n | \theta_{1:K}, \mu_{1:K}, \sigma_{1:K, 1:D} &\sim \sum_{k = 1}^K \theta_k\distNorm\left(\mu_k, \diag(\sigma_{k1}, \dots , \sigma_{kD})\right).
\]
The goal is to infer the posterior distribution of all the latent parameters
$(\theta_{1:k}, \mu_{1:k}, \sigma_{1:K, 1:D})$, on which we place a Dirichlet
prior on the mixture proportions, a Gaussian prior on the Gaussian means, and a
lognormal prior on the standard deviations, i.e.,
\[(\theta_1, \dots, \theta_K) &\sim \distDir(\alpha_0) \\
    \mu_k &\sim \distNorm(0, I), \quad k  = 1, 2, \dots, K\\
\sigma_{kd} &\sim \distNamed{LogNormal}(0,1), \\
& \qquad k  = 1, 2, \dots, K, d = 1, 2, \dots D.
\]
One may notice that the support of $\theta_k$ and $ \sigma_{kd}$ is
constrained; the present work is limited to posterior
distributions with full support. Therefore, we 
first apply transformations  such that the
transformed random variables have unconstrained support \cite{kucukelbir2017automatic}. The details of the
transformation can be found in \cref{supp:gmmtransformation}.

For the synthetic dataset, we generate $N = 400$ data points equally from $4$
isotropic bivariate Gaussian distributions with equal covariance $0.6^2 I$. The
mean of those Gaussian distributions are generated randomly within $[-10,
10]^2$. \cref{fig:gmm_syn_dat} displays the synthetic data points, where
samples from the four Gaussian distributions are distinguished by different
colors. We fit this dataset with $K =3$ and set $\alpha_0 = 1$---we pick a
misspecified setting to make the posterior distribution have multiple
meaningful modes and hence create local optima for GVB inference.  During
inference, we use samples from the prior distribution as 
the initial value for SMAP, Laplace and the mean for SVI. The smoothing
constant for smoothed MAP estimator is set to $1$. For the initialization of
covariance matrix, we set $L_0 = I$ for CSVI and use two different settings
for SVI; we consider both the identity matrix and random diagonal matrix,
of which the log diagonal indices are uniformly sampled in the range $(\log
0.1, \log 100)$. We use a learning rate $\gamma_k = 0.1$ for SMAP and
$\gamma_k = 0.01$ for both VI algorithms. The initial step size $t$ for the line search
of Laplace is set to $1$.
The experimental results are illustrated in \cref{fig:gmm_syn_elbo}, from which
we notice that CSVI and CLA are able to consistently find the global optimum in
almost every
trial while both SVI and Laplace tends to converge to some local optima.
A similar phenomenon is also observed in the real data experiment (\cref{fig:gmm_real_elbo}),
suggesting that CSVI is more reliable. But in this case, though CLA is
noticabily better than the standard Laplace approxiamtion, it is outperformed
by all variational inference methods. This reveals the limitation of Laplace
approximation---it is a local approximation method which totally depends on the
curvature of log posterior function; while variational inference is based on
reducing KL divergence to the target, which is a gloabl metric.
The details of the real data experiment
are deferred to \cref{supp:gmm_real_setting}.

\section{Conclusion}\label{sec:conclusion}

This work provides an extensive theoretical analysis of the computational
aspects of Laplace approximation and Guassian variational inference, and uses the theory to design a
general procedure that addresses the nonconvexity of the problem
in an asymptotic regime. We show that under mild conditions, the MAP estimation
problem and 
Gaussian variational optimization are locally asymptotically convex.
Based on this fact, we developed consistent stochastic variational inference
(CSVI), a scheme that asymptotically solves Gaussian variational inference; and
consistent Laplace approximation, a variant of Laplace approximation that
address the intractability of finding MAP value.
Both CSVI and CLA solves a smoothed MAP problem to initialize the Gaussian mean within
 the locally convex area, and then CSVI further runs a scaled
projected stochastic gradient descent to create iterates that converge to the optimum. 
The asymptotic
consistency of CSVI is mathematically justified, and experimental results
demonstrate the advantages over traditional SVI.

There are many avenues of further exploration for the present work. For
example, we limit consideration to the case of Gaussian variational families
due to their popularity; but aside from the mathematical details, nothing about
the overall strategy necessarily relied on this choice. It would be worth
examining other popular variational families, such as mean-field exponential
families \citep{xing2002generalized}.

Furthermore, the current work is limited to posterior distributions with full
support on $\reals^d$---otherwise, the KL divergence variational objective is
degenerate. It would be of interest to study whether variational inference
using a Gaussian variational family truncated to the support of the posterior
possesses the same beneficial asymptotic properties and asymptotically
consistent optimization algorithm as developed in the present work.  

Another interesting potential line of future work is to investigate other
probability measure divergences as variational objectives. For example, the
chi-square divergence \citep[p.~51]{liese1987convex,csiszar1967information}, R\'enyi
$\alpha$-divergence \citep{van2014renyi}, Stein discrepancy
\citep{stein1972bound}, and more \citep{gibbs2002choosing} have all been used
as variational objectives.  Along a similar vein, we studied the convergence
properties of only a relatively simple stochastic gradient descent algorithm;
other base algorithms with better convergence properties exist
\citep{kingma2015adam,duchi2011adaptive,nesterov1983method}, and it may be fruitful to see if they have similar
asymptotic consistency properties.

A final future direction is to investigate the asymptotic behaviour of
variational inference with respect to other measures of optimization
tractability. In particular, (local) pseudoconvexity
\citep{crouzeix1982criteria}, quasiconvexity \citep{arrow1961quasi}, and
invexity \citep{ben1986invexity,craven1985invex} are all weaker than (local)
convexity, but provide similar guarantees for stochastic optimization.  These
may be necessary to consider when examining other divergences as variational
objectives.

\begin{acknowledgements}
  The authors gratefully acknowledge 
  the support of an Natural Sciences and Engineering Research Council of Canada (NSERC)
  Discovery Grant and Discovery Launch Supplement, and UBC four year doctoral fellowship.
\end{acknowledgements}

\bibliographystyle{spbasic}      
\bibliography{sources.bib}   

\newpage
\clearpage
\appendix
\section{Details of experiments} \label{supp:expts}

\subsection{Details of the toy example for smoothed MAP}\label{supp:smoothexample}

The underlying  synthetic model for \cref{fig:posts} is as follows,
\[
& \theta \dist \frac{1}{5}\distNorm(0, 0.15^2)+ \frac{1}{5}\distNorm(1, 0.1^2) +\frac{1}{5}\distNorm(-4, 0.3^2) + \frac{1}{5}\distNorm(4, 0.3^2) \\
& \quad  + \frac{1}{5}\distNorm(-8, 0.1^2) \\
&X_i \given \theta \distiid \distNorm(\theta, 5000),
\]
where the data are truly generated from $(X_i)_{i=1}^n \distiid \distNorm(3, 10)$.
For smoothed MAP optimization, we use a smoothing constant of $\alpha_n = 10 n^{-0.3}$, and set the initial value uniformly within the range $(-50, 50)$. The learning rate for the SGD is chosen as $\gamma_k = 15/(1+ k^{0.9})$.

\subsection{Algorithm: CSVI (Adam)} \label{supp:csviadam}

\begin{algorithm}[H]
    \caption{CSVI with adaptive moment estimation} \label{supp:csvi_adam_alg}
    \begin{algorithmic}
    \Procedure{CSVI}{$f_n$, $g$, $\gamma$, $K$}
     \State $\mu_0 \gets $\texttt{SmoothedMAP} (\cref{alg:smoothedmap}), $L_0 \gets I$
     \State $\epsilon \gets 10^{-8}$, $\beta_1 \gets 0.9$, $\beta_2 \gets 0.9999$
     \State $m_0 \gets 0$, $\nu_0 \gets 0$ 
     \For{$k= 0, 1, \dots, K-1$} \Comment{All operations on vector/matrix are elementwise}
        \State Sample $Z_k \sim \distNorm(0, I)$ 
        \State $g_k\gets \left( \hat\nabla_{\mu,n}(\mu_k, L_k, Z_k), \tilde\nabla_{L,n} (\mu_k, L_k, Z_k)\right) $ \Comment{$\hat\nabla_{\mu,n},\tilde\nabla_{L,n} $ are same to \cref{alg:csvi}} 
        \State $m_{k+1}  \gets \beta_1 m_k + (1-\beta_1 )g_k$
        \State $v_{k+1}  \gets \beta_2 v_k + (1-\beta_2 )g_k^{ 2}$
        \State $\hat{m}_{k+1 } \gets m_k / (1  - \beta_1^k)$
        \State $\hv_{k+1 } \gets v_k / (1  - \beta_2^k)$
        \State $\left( \mu_{k+1}, L_{k+1}\right) \gets \left( \mu_{k}, L_{k}\right) - \gamma \hat{m}_k/ \left( \sqrt{\hv_k}  + \eps\right)  $ 
        \For{$i=1, \dots, d$}
           \State $L_{k+1,ii} \gets \max\left\{0, L_{k+1,ii}\right\}$
        \EndFor
     \EndFor
     \State \Return $\mu_K, L_K$
    \EndProcedure
    \end{algorithmic}
    \end{algorithm}

\subsection{Discussion of sparse regression experiment} \label{supp:sr_detail}

\begin{figure}[t!]
    \centering
    \includegraphics[width=\linewidth]{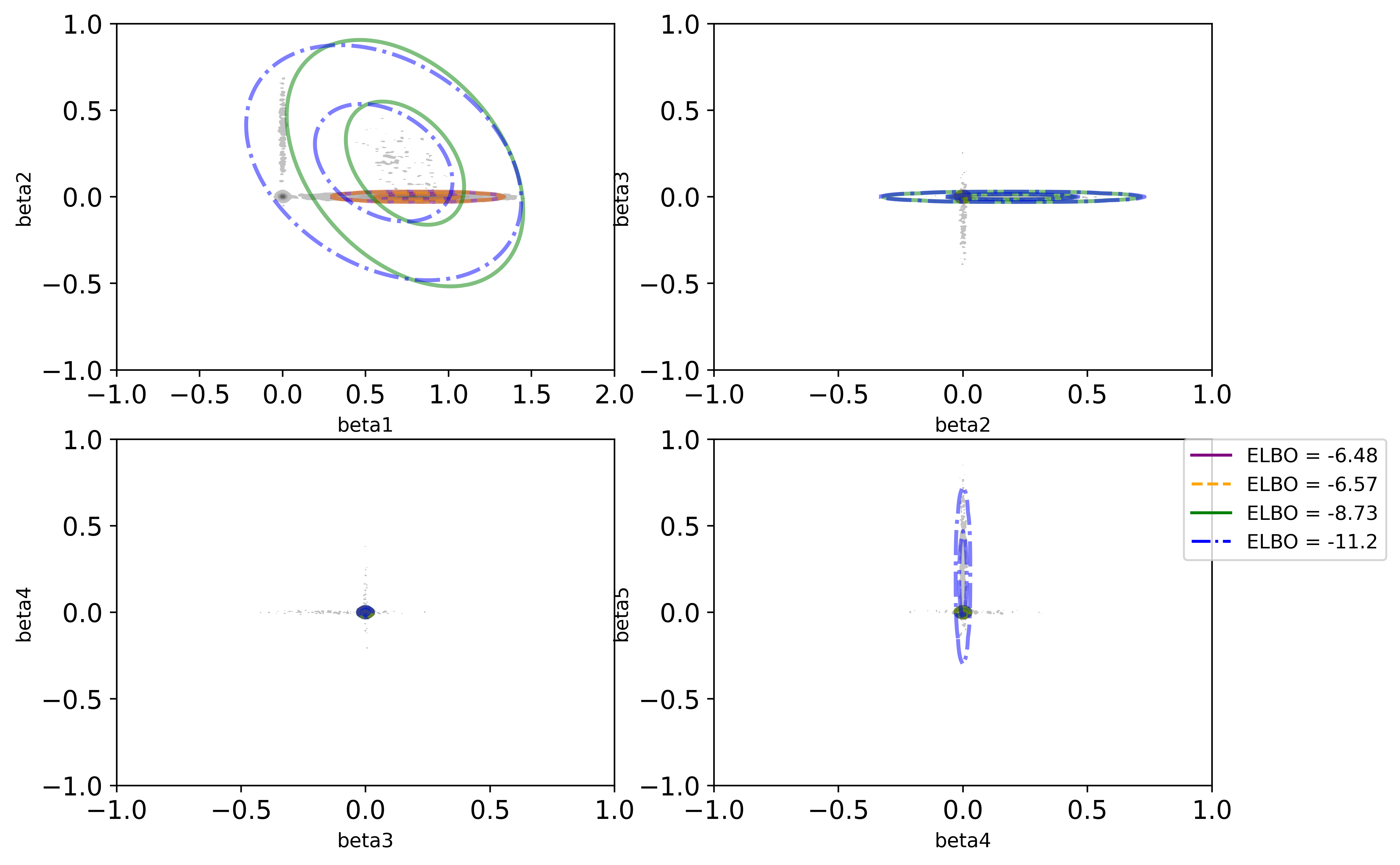}
    \caption{Visualization of different CLA approximations to the posterior
    distribution on real dataset.The grey area depicts the posterior density and four Gaussian approximations displayed in contour plots corresponding to different ELBOs.} 
    \label{fig:vis_csl}
\end{figure}

\begin{figure}[t!]
    \centering
    \includegraphics[width=\linewidth]{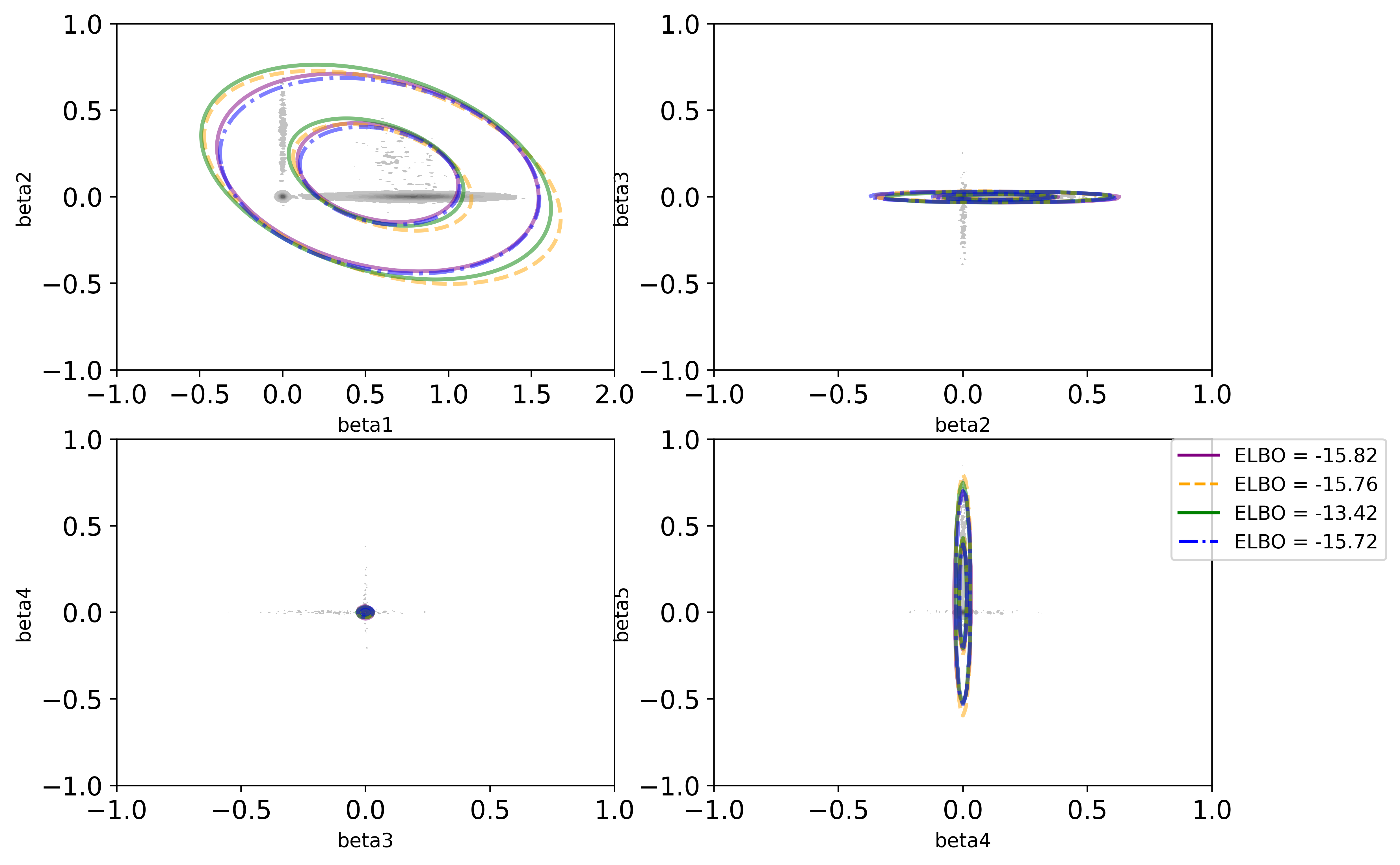}
    \caption{Visualization of different CSVI approximations to the posterior
    distribution on real dataset.The grey area depicts the posterior density and four Gaussian approximations displayed in contour plots corresponding to different ELBOs.} 
    \label{fig:vis_csvi}
\end{figure}

In this section, we provide further discussion to the result presented in
\cref{fig:sr_real_elbo}.
\cref{fig:vis_csl,fig:vis_csvi} visualizes the Gaussian approximations produced
by CLA and CSVI.
Instead of fitting a single mode, CSVI covers the range of posterior and
fit a Gaussian distribution with larger variance. Even though the performance of
CSVI is consistent across runs, it does find the local optimum instead of the
global solution. In this case, reverse KL---the
objective function of Gaussian VI---can be limited. We compare the forward KL of
these fitted Gaussians using $32000$ posterior samples obatined from Stan,
suggesting that CSVI find a solution that is better in forward KL.

\begin{figure}[t!]
    \centering
    \includegraphics[width=\linewidth]{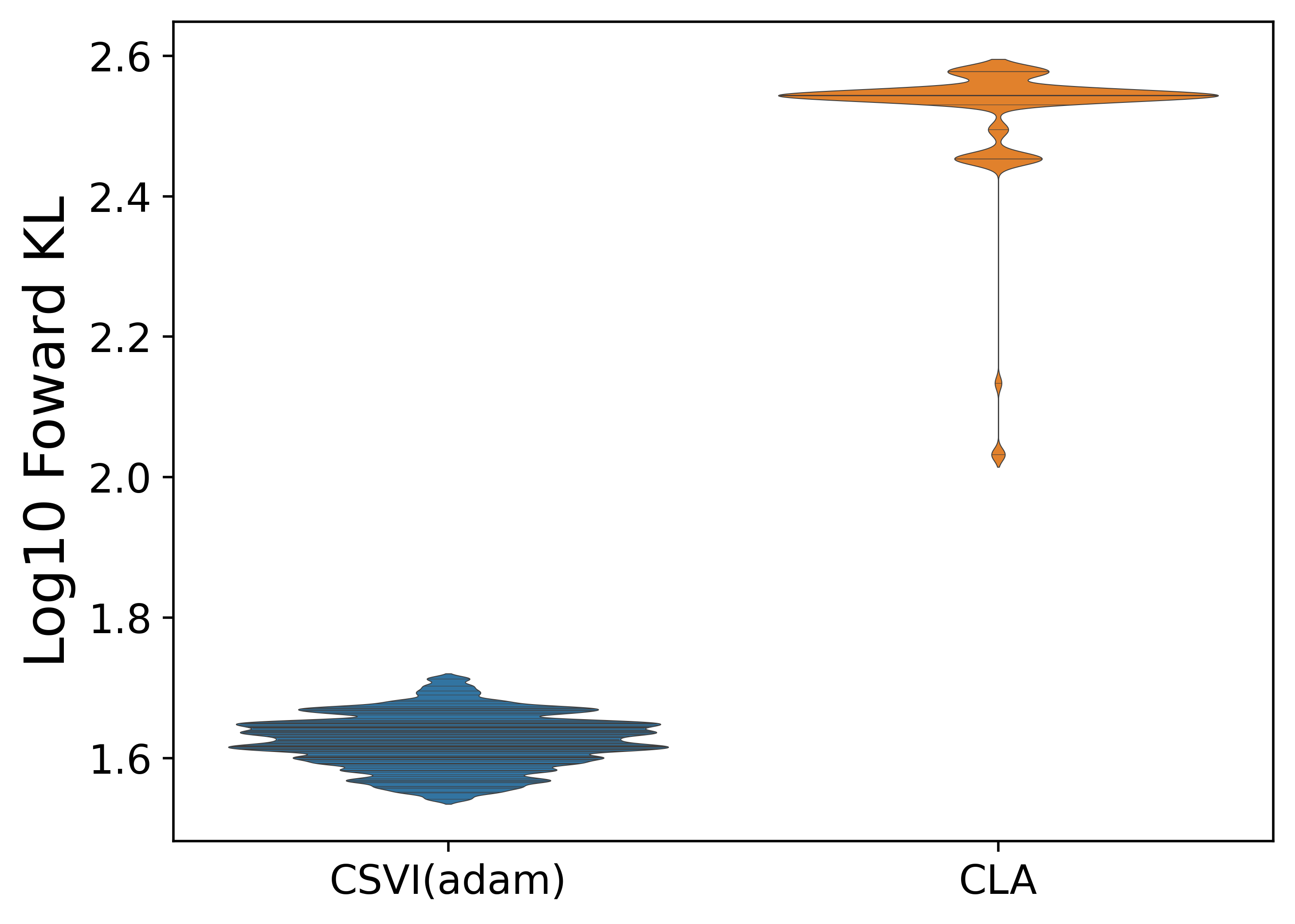}
    \caption{Comparison of forward KL of sparse regression experiment} 
    \label{fig:sr_fwd_kl}
\end{figure}

\subsection{Details of the GMM experiment}    \label{supp:gmm_detail}

\subsubsection{Variable transformations of Bayesian Gaussian mixture model} \label{supp:gmmtransformation}

To transform $(\theta_1, \dots, \theta_K) \in \Delta(K)$ and $\sigma_{kd}\in \reals_+$ into unconstrained space, we consider the change of random variables as below:
\benum
\item For $\sigma_{kd}\sim \distNamed{LogNormal}(0,1)$, we consider 
\[
     \tau_{kd} = \log(\sigma_{kd}) \sim \distNorm(0,1),   
\]
which has a full support on $\reals$.
\item For $\theta \sim \distDir(\alpha_0)$, we consider using marginalized LogGamma random variables. Notice the relationship of Gamma distribution and Dirichlet distirbution as follows, 
\[
&\left( \lambda_k\right)_{k = 1}^K \distiid \distNamed{LogGamma}(\alpha_0, 1)\\
& \left( \frac{\exp(\lambda_1)}{\sum_k \exp(\lambda_k)}, \dots, \frac{\exp(\lambda_K)}{\sum_k \exp(\lambda_k)}\right) \sim \distDir(\alpha_0),
\]
then $\lambda_k$ is supported on $\reals$.
\eenum
Therefore, instead of inferring the original parameters, we perform Gaussian variational inference on the posterior distribution of $(\lambda_{1:k}, \mu_{1:k}, \tau_{11:kd})$.

\subsubsection{Detailed settings for real data experiment} \label{supp:gmm_real_setting}

We subsample the Shapley galaxy dataset to $N = 500$ and the goal is to cluster the distribution of galaxies in the Shapley Concentration region. In our experiment, we fix the number of component $K = 3$ and set $\alpha_0 = 1$. During inference, we initialize the SMAP with random samples from the prior distribution, which are also used as the mean initialization for SVI. In SMAP, we perform the smoothed MAP estimation to a tempered posterior distribution $\pi_n^\kappa$ with $\kappa = 1/2$; and set the smoothing constant $\alpha_n = 3$. The learning rate for SMAP and VI algorithms are chosen as $0.005$ and $0.001$ respectively. And similar to the synthetic experiment, CSVI and SVI\_Ind use the identity matrix for $L_0$ and SVI use random diagonal matrix for $L_0$, whose log diagonal indices are uniformly in the range $(\log 0.1, \log 100)$.

\section{Proofs} \label{supp:proof}

\subsection{Proof of \cref{thm:optimum_inside}}

\bprf
We consider the KL cost for the scaled and shifted posterior distribution. Let
$\tPi_n$ be the Bayesian posterior distribution of $\sqrt{n} (\theta - \theta_0)$. 
The KL divergence measures the difference between the distributions of
two random variables and is invariant when an invertible transformation is
applied to both random variables
\citep[Theorem 1]{qiao2010study}. Note that $\tPi_n$ is shifted and
scaled from $\Pi_n$,  and that this linear transformation is invertible, so
\[
\kl{\distNorm(\mu, \Sigma)}{\Pi_n} = \kl{\distNorm\left(\sqrt{n}(\mu - \theta_0), n \Sigma\right)}{\tPi_n}.
\]
Let
$\tmu_n^\star, \tSigma_n^\star$ be the parameters of the optimal Gaussian variational approximation
to $\tPi_n$, i.e.,
\[
	\tmu_n^\star, \tSigma_n^\star = \argmin_{\mu \in\reals^d, \Sigma \in\reals^{d\times d}}  \kl{\distNorm(\mu, \Sigma)}{\tPi_n} \quad \text{s.t.} \quad  \Sigma \succ 0,
\]
and let
\[
	\tilde{\distNorm_n}^\star \defined \distNorm\left(\tmu_n^\star, \tSigma_n^\star\right) = \distNorm\left(\sqrt{n}(\mu_n^\star - \theta_0), L_n^\star L_n^{\star T} \right).	
\]
\citet[Corollary 7]{wang2019frequentist} shows that 
under \cref{assump:regularity}, 
\[
\tvd{\tilde{\distNorm_n}^\star}{\distNorm\left( \Delta_{n, \theta_{0}}, H_{\theta_0}^{-1}\right)}  \stackrel{P_{\theta_0}}{\to} 0.	    
\]	
Convergence in total variation implies
weak convergence, which then implies pointwise convergence of the characteristic
function. Denote $\tphi^\star_n(t)$ and $\phi_n(t)$ to be the characteristic functions of
$\tilde{\distNorm_n^\star}$ and $\distNorm\left( \Delta_{n, \theta_{0}}, H_{\theta_0}^{-1}\right)$. Therefore 
\[
\forall t\in\reals^d, \; \frac{\phi^\star_n(t)}{\phi_n(t)} &= \exp\!\left( i ( \sqrt{n}(\mu^\star_n - \theta_0)-\Delta_{n, \theta_{0}})^T t - \frac{1}{2}
t^T \left(L_n^\star L_n^{\star T} - H_{\theta_0}^{-1} \right) t \right)\\
& \stackrel{P_{\theta_0}}{\longrightarrow} 1,
\]
 which implies	
\[
 \mu_n^\star \stackrel{P_{\theta_0}}{\rightarrow} \frac{1}{\sqrt{n}}\Delta_{n, \theta_{0}} + \theta_0,
\quad\text{and}\quad
  L_n^\star L_n^{\star T} \stackrel{P_{\theta_0}}{\rightarrow} H^{-1}_{0} = L_0 L_0^T.    
\]
Under \cref{assump:regularity}, \citet[Theorem 8.14]{van2000asymptotic} states that
\[
\|\Delta_{n, \theta_{0}} - \sqrt{n}( \theta_{\text{MLE},n} -   \theta_0)\| \stackrel{P_{\theta_0}}{ \to}	0,
\]
and $\theta_{\text{MLE},n} \overset{P_{\theta_0}}{\to} \theta_0$ according to \cref{bvm}, yielding $\mu_n^\star \stackrel{P_{\theta_0}}{ \to } \theta_0$.

Finally since the Cholesky decomposition defines a continuous mapping from the
set of positive definite Hermitian matrices to the set of lower triangular
matrices with positive diagonals (both sets are equipped with the spectral norm)
\citep[p.~295]{Schatzman02}, we have
\[
L_n^\star \stackrel{P_{\theta_0}}{ \to } L_0.
\]
\eprf

\subsection{Proof of \cref{thm:globalcvxsmth}}

\bprf
We provide a proof of the result for strong convexity; the result for Lipschitz smoothness 
follows the exact same proof technique. Note that if $D'$ does not depend on $x$, $F_n(x)$ is $D'$-strongly convex if and only if $F_n(x) - \frac{1}{2}x^TD'x$
is convex. We use this equivalent characterization of strong convexity in this proof.

Note that for $Z\dist\distNorm(0, I)$,
\[
\EE\left[\frac{1}{2}(\mu+n^{-1/2}LZ)^TD(\mu+n^{-1/2}LZ)\right] = \frac{1}{2}\mu^TD\mu + \frac{1}{2}\tr L^T(n^{-1}D)L.
\]
Define $\lambda \in [0,1]$, vectors $\mu, \mu' \in \reals^d$,
 positive-diagonal lower triangular matrices $L, L'\in\reals^{d\times d}$,
and vectors $x, x'\in\reals^{(d+1)d}$ by stacking $\mu$ and the columns of $L$ 
and likewise $\mu'$ and the columns of $L'$. Define $x(\lambda) = \lambda x + (1-\lambda)x'$,
$\mu(\lambda) = \lambda \mu + (1-\lambda)\mu'$, and $L(\lambda) = \lambda L + (1-\lambda) L'$. Then
\[
&F_n(x(\lambda)) - \frac{1}{2}x(\lambda)^T\diag(D, n^{-1}D, \dots, n^{-1}D)x(\lambda) \\
=&F_n(\mu(\lambda), L(\lambda)) - \left(\frac{1}{2}\mu(\lambda)^TD\mu(\lambda) + \frac{1}{2}\tr L(\lambda)^T(n^{-1}D)L(\lambda)\right)\\
=& \EE\left[n^{-1}\log\pi_n(\mu(\lambda)+n^{-1/2}L(\lambda)Z) - \frac{1}{2}(\mu(\lambda) +n^{-1/2}L(\lambda)Z)^TD(\mu(\lambda)\right.\\
& \left. + n^{-1/2}L(\lambda)Z)\right].
\]
By the $D$-strong convexity of $n^{-1}\log \pi_n$,
\[
\leq& \lambda \left(F_n(\mu, L) - \frac{1}{2}\mu^TD\mu - \frac{1}{2}\tr L^T(n^{-1}D)L\right)\\
&+ (1-\lambda) \left(F_n(\mu', L') - \frac{1}{2}\mu'^TD\mu' - \frac{1}{2}\tr L'^T(n^{-1}D)L'\right)\\
=& \lambda \left(F_n(x) - \frac{1}{2}x^T\diag(D, n^{-1}D, \dots, n^{-1}D)x\right)\\
&+(1-\lambda) \left(F_n(x') - \frac{1}{2}x'^T\diag(D, n^{-1}D, \dots, n^{-1}D)x'\right).
\]
\eprf

\subsection{Proof of \cref{prop:vbfails}}

\bprf
Note that by reparameterization, 
\[
\argmin_{\mu} \kl{\distNorm(\mu, \sigma^2)}{\Pi_n} = \argmin_{\mu}   \EE\left[ -n^{-1}\log\pi_n (\mu + \sigma Z)\right],
\]
where $Z \sim \distNorm(0,1)$. Using a Taylor expansion, 
\[
& - \EE\left[ \dder{}{\mu} \left(-n^{-1}\log\pi_n (\mu + \sigma Z)\right) \right]\\
& = \EE \left[ -n^{-1} \log\pi_n^{(2)} (\mu )  - n^{-1}\log\pi_n^{(3)} (\mu') \cdot\sigma Z  \right],
\]
for some $\mu'$ between $\mu$ and $\mu+\sigma Z$. By the uniform bound on the third derivative
and local bound on the second derivative, for any $\mu \in U$,
\[
\EE \left[ -n^{-1} \log\pi_n^{(2)} (\mu )  - n^{-1}\log\pi_n^{(3)} (\mu') \cdot\sigma Z  \right]
&\leq -\epsilon  + \eta \sigma\EE \left| Z \right|\\
&\leq -\epsilon  + \eta \sigma.
\]
The result follows for any $0<\eps < \epsilon / \eta$.
\eprf

\subsection{Proof of \cref{thm:strongconvexsmooth}}

\bprf

Note that we can split $L$ into columns and express $LZ$ as
\[
LZ = \sum_{i=1}^d L_i  Z_i,
\]
where $L_i\in\reals^p$ is the $i^\text{th}$ column of $L$, and $(Z_i)_{i=1}^d\distiid\distNorm(0,1)$.
Denoting $\nabla^2 f_n \defined \nabla^2 f_n(\mu+LZ)$ for brevity,
 the $2^\text{nd}$ derivatives in both $\mu$ and $L$ are
\[
\nabla^2_{\mu\mu} F_n &= \EE\left[\nabla^2 f_n \right]\\
\nabla^2_{L_iL_j} F_n &= n^{-1}\EE\left[Z_i Z_j \nabla^2 f_n \right]\\
\nabla^2_{\mu L_i} F_n &= n^{-1/2}\EE\left[Z_i\nabla^2 f_n\right]
\]
where we can pass the gradient and Hessian through the expectation by dominated
convergence because $Z$ has a normal distribution and $f_n$ has $\ell$-Lipschitz gradients.
Stacking these together in block matrices yields the overall Hessian,
\[
A &= \left[\begin{array}{cccc}
I &
n^{-1/2} Z_1 I&
\dots &
n^{-1/2} Z_d I
\end{array}\right] \in \reals^{d \times d(d+1)}\\
\nabla^2 F_n &= \EE\left[A^T \nabla^2 f_n A\right] \in \reals^{d(d+1)\times d(d+1)}.
\]
Since $f_n$ has $\ell$-Lipschitz gradients, for all $x\in\reals^d$, $-\ell I \preceq \nabla^2 f_n(x) \preceq \ell I$.
Applying the upper bound and evaluating the expectation yields the Hessian upper bound (and the same technique
yields the corresponding lower bound):
\[
\nabla^2 F_n &= \EE\left[A^T \nabla^2 f_n A\right]\\
& \preceq \ell \EE\left[A^TA\right]\\
&= \ell\left[\begin{array}{cccc}
I & 0 & 0 & 0\\
0 & n^{-1}I & 0 & 0\\
0 & 0 & \ddots & 0\\
0 & 0 & 0 & n^{-1} I
\end{array}\right] = \ell D_n.
\]

To demonstrate local strong convexity, 
we split the expectation into two parts: one where $n^{-1/2} LZ$ is small enough to 
guarantee that $\|\mu+n^{-1/2} LZ - x\|^2 \leq r^2$,
and the complement. Define
\[
r_n^2(\mu, L) \defined  n\frac{(r^2 - 2\|\mu - x\|^2_2)}{2\|L\|^2_F}.
\]
Note that when $\|Z\|^2 \leq r_n^2(\mu, L) $,
\[
\left\|\mu + \frac{1}{\sqrt{n}}LZ - x \right\|_2^2 &\leq 2 \|\mu - x\|^2 + 2n^{-1}\|LZ\|^2\\
& \leq 2 \|\mu - x\|^2 + 2n^{-1}\|L\|_F^2\|Z\|^2\\
& \leq r^2.
\]
Then we may write
\[
\nabla^2 F_n &= \EE\left[\ind\left[\|Z\|^2 \leq r_n^2(\mu, L) \right] A^T\nabla^2 f_n A\right] \\
& \qquad + \EE\left[\ind\left[\|Z\|^2 > r_n^2(\mu, L) \right] A^T\nabla^2 f_n A\right].
\]
Since $f_n$ has $\ell$-Lipschitz gradients and is locally $\epsilon$-strongly convex,
\[
\nabla^2 F_n &\succeq  \epsilon \cdot \EE\left[\ind\left[\|Z\|^2 \leq r_n^2(\mu, L) \right] A^T A\right]
- \ell\cdot \EE\left[\ind\left[\|Z\|^2 > r_n^2(\mu, L) \right] A^T A\right].
\] 
Note that $A^TA$ has entries $1$ and $n^{-1} Z_i^2$ along the diagonal,
as well as $n^{-1} Z_iZ_j$, $i\neq j$ and $n^{-1/2} Z_i$ on the off-diagonals.
By symmetry, since $Z$ is an isotropic Gaussian, censoring 
by $\ind\left[\|Z\|^2 \leq \dots\right]$ or $\ind\left[\|Z\|^2 > \dots\right]$
maintains that the off-diagonal expectations are 0.
Therefore the quantity
$\EE\left[\ind\left[\|Z\|^2 \leq r_n^2(\mu, L) \right]A^TA \right]$ 
is diagonal with coefficients $1 - \alpha_n(\mu,L)$ and $n^{-1} \beta_n(\mu,L)$, 
and
$\EE\left[\ind\left[\|Z\|^2 > r_n^2(\mu, L) \right]A^TA \right]$ 
is diagonal with coefficients $\alpha_n(\mu,L)$ and $n^{-1} \tau_n(\mu,L)$ where
\[
	\alpha_n(\mu,L) &= \Pr\left(\|Z\|^2 > r_n^2(\mu, L) \right)\\
 \beta_n(\mu,L) &= \EE\left[Z_1^2 \ind\left[\|Z\|^2 \leq r_n^2(\mu, L) \right]\right] = d^{-1}\EE\left[\|Z\|_2^2 \ind\left[\|Z\|^2 \leq r_n^2(\mu, L) \right]\right]\\
 \tau_n(\mu,L) &= \EE\left[Z_1^2 \ind\left[\|Z\|^2 > r_n^2(\mu, L) \right]\right] = d^{-1}\EE\left[\|Z\|_2^2 \ind\left[\|Z\|^2 > r_n^2(\mu, L) \right]\right].
\]
Note that $\|Z\|^2 \dist \distChiSq_d$; so $\alpha_n(\mu,L) = 1-\distChiSq_d(r_n^2(\mu, L) )$ and
\[
\tau_n(\mu,L) &= \int_{r_n^2(\mu, L) }^{\infty}\ind\left[x\geq 0\right]\frac{1}{2^{(d+2)/2}\Gamma((d+2)/2)} x^{\frac{d+2}{2}-1}e^{-x/2}\dee x \\
& = 1- \distChiSq_{d+2}(r_n^2(\mu, L) )\\
\beta_n(\mu,L) &= 1-\tau_n(\mu,L).
\]
Therefore, 
\[
&\nabla^2 F_n \\
&  \succeq 
 \diag\!\left((\epsilon(1-\alpha_n(\mu,L))\!\! - \!\!\ell \alpha_n(\mu,L))I, (\epsilon n^{-1}(1\!-\!\tau_n(\mu,L)) \!\!- \!\!\ell n^{-1}\tau_n(\mu,L))I,  \right. \\
 & \qquad \qquad \qquad \qquad \qquad \left. \dots, (\epsilon n^{-1}(1\!-\!\tau_n(\mu,L)) - \ell n^{-1}\tau_n(\mu,L))I\right) \\
&= \epsilon D_n
- (\epsilon + \ell)\diag\left(\alpha_n(\mu,L)I, n^{-1}\tau_n(\mu,L)I, \dots, n^{-1}\tau_n(\mu,L)I\right) \\
&\succeq \epsilon D_n 
- (\epsilon + \ell)\diag\left(\tau_n(\mu,L)I, n^{-1}\tau_n(\mu,L)I, \dots, n^{-1}\tau_n(\mu,L)I\right) \\
&= D_n\left( \epsilon - \tau_n(\mu,L) \cdot(\epsilon + \ell) \right).
\]

\eprf

\subsection{Proof of \cref{lem:asymplocalconv}}

\bprf
Given \cref{assump:regularity}, we know $f_n$ is twice continuously differentiable. Thus, using the second order characterization of strong convexity, it is equivalent to show the existence of $r,\eps >0$ such that 
\[
\sP\left( \forall \theta \in B_r(\theta_0), \quad \nabla^2 f_n(\theta) \succeq \eps I \right) \to 1,
\]
as $n \to \infty$. Note that by Weyl's inequality 
\[\label{eq:weylonfn}
\nabla^2 f_n(\theta) =& 	\nabla^2 f_n(\theta) - H_\theta + H_\theta\\
\succeq& \lambda_{\min}\left( \nabla^2 f_n(\theta) - H_\theta \right) I + \lambda_{\min}(H_\theta) I.
\]
Condition $4$ of \cref{assump:regularity} guarantees that $H_{\theta_0} \succeq \eps I$ and that there exists a $\kappa > 0$ such that $H_\theta$ is continuous in $B_{\kappa}(\theta_0)$. Hence there exists $0<\kappa' \leq \kappa$, such that   $\forall \theta \in B_{\kappa'}(\theta_0),\; H_\theta \succeq \frac{\eps}{2}I$. 

We then consider $\lambda_{\min}\left( \nabla^2 f_n(\theta) - H_\theta \right)$. We aim to find a $0 <r \leq \kappa'$ such that $|\lambda_{\min}\left( \nabla^2 f_n(\theta) - H_\theta \right)|$ is sufficiently small.
Note that for any fixed $r > 0$,
\[
& \sup_{\theta \in B_r(\theta_0)} \left|\lambda_{\min}\left( \nabla^2 f_n(\theta) - H_\theta \right) \right| \\
& \leq \sup_{\theta \in B_r(\theta_0)} \left\| \nabla^2 f_n(\theta) - H_\theta \right\|_2 \\
& =  \sup_{\theta \in B_r(\theta_0)} \left\|\nabla^2 f_n(\theta) - \EE_{\theta_0}\left[ -\nabla^2 \log p_\theta(X) \right] + \EE_{\theta_0}\left[ -\nabla^2 \log p_\theta(X) \right] -  H_\theta   \right\|_2\\
& \leq \! \sup_{\theta \in B_r(\theta_0)} \!\left(\left\|\nabla^2 f_n(\theta) \!-\! \EE_{\theta_0}\!\left[ -\nabla^2 \log p_\theta(X) \right]\right\|_2 \!+\! \left\| \EE_{\theta_0}\!\left[ -\nabla^2 \log p_\theta(X) \right] \!-\!  H_\theta   \right\|_2  \right).
\]
Now we split $f_n$ into prior and likelihood, yielding that 
\[\label{eq:boundonfn}
& \leq \sup_{\theta \in B_r(\theta_0)} \left\|-n^{-1}\sum_{i = 1}^n \nabla^2\log p_\theta(X_i) - \EE_{\theta_0}\left[ -\nabla^2 \log p_\theta(X) \right]\right\|_2 \\
& \quad \! + \sup_{\theta \in B_r(\theta_0)}\! \|-n^{-1}\nabla^2\log\pi_0(\theta) \|_2 + \sup_{\theta \in B_r(\theta_0)}\!\left\| \EE_{\theta_0}\left[ -\nabla^2 \log p_\theta(X) \right] -  H_\theta   \right\|_2 .
\]
Given Condition $2$ of \cref{assump:regularity}, for all $\theta$, $\pi_0(\theta)$ is positive and $\nabla^2 \pi_0(\theta)$ is continuous; and further due to the compactness of  $B_r(\theta_0)$, we have that 
\[\label{eq:priorconv}
\forall r > 0, \quad	\sup_{\theta \in B_r(\theta_0)} \|-n^{-1}\nabla^2\log\pi_0(\theta) \|_2 \to 0, \quad\text{as } n \to \infty.	
\]
Then, it remains to bound the first term and the last term of \cref{eq:boundonfn}. For the first term, we aim to use the uniform weak law of large numbers to show its convergence to $0$. By Condition $5$ of \cref{assump:regularity}, there exists a $0 < r_1 \leq \kappa'$ and a measurable function $g$ such that for all $ \theta \in B_{r_1}(\theta_0)$ and for all $x$,
\[
\max_{i,j \in [d]}\left|\left(\nabla^2 \log p_{\theta}(x) \right)_{i,j} \right| < g(x), \quad \EE_{\theta_0}[g(X)] < \infty.
\]
Then, by the compactness of $B_{r_1}(\theta_0)$, we can apply the uniform weak law of large numbers \citep[Theorem 2]{jennrich1969asymptotic}, yielding that for all $i,j \in [d]$, 
\[
\sup_{\theta \in B_{r_1}(\theta_0)} \left| \left(-n^{-1}\sum_{i = 1}^n \nabla^2\log p_\theta(X_i)\right)_{i,j} - \left(\EE_{\theta_0}\left[ -\nabla^2 \log p_\theta(X) \right] \right)_{i,j}	\right| \stackrel{P_{\theta_0}}{\to } 0.
\]
Since the entrywise convergence of matrices implies the convergence in spectral norm,
\[\label{eq:term1lem}
	\sup_{\theta \in B_{r_1}(\theta_0)} \left\|-n^{-1}\sum_{i = 1}^n \nabla^2\log p_\theta(X_i) - \EE_{\theta_0}\left[ -\nabla^2 \log p_\theta(X) \right]\right\|_2 \stackrel{P_{\theta_0}}{ \to } 0.	
\]
For the last term of \cref{eq:boundonfn}, by Condition $4$ of \cref{assump:regularity}, 
\[
& \lim_{r\to 0} \sup_{\theta \in B_r(\theta_0)} \left\| \EE_{\theta_0}\left[ -\nabla^2 \log p_\theta(X) \right] -  H_\theta   \right\|_2\\
& = \lim_{r\to 0} \sup_{\theta \in B_r(\theta_0)} \left\| \EE_{\theta_0}\left[ -\nabla^2 \log p_\theta(X) \right] -  \EE_\theta\left[-\nabla^2 \log p_\theta (X)\right]   \right\|_2 \\
& \to 0.
\]
Thus, there exists a sufficiently small $r_2 > 0$ such that 
\[\label{eq:term3lem}
	\sup_{\theta \in B_{r_2}(\theta_0)} \left\| \EE_{\theta_0}\left[ -\nabla^2 \log p_\theta(X) \right] -  H_\theta   \right\|_2 \leq \frac{\eps}{8}.	
\]
Then, we combine \cref{eq:priorconv,eq:term1lem,eq:term3lem} and pick $r' = \min(r_1,r_2)\leq \kappa'$, yielding that
\[\label{eq:Hthetandfn}
\sP\left( \sup_{\theta \in B_{r'}(\theta_0)}\left|\lambda_{\min}\left( \nabla^2 f_n(\theta) - H_\theta \right) \right| \leq \frac{\eps}{4}  \right) \to 1,
\]
as $n \to \infty$. 
Then the local strong convexity is established. Note that we have already shown for all $\theta \in B_{\kappa'}(\theta_0), H_\theta \succeq \frac{\eps}{2} I$. By \cref{eq:Hthetandfn,eq:weylonfn},  we conclude that for all $\eps \leq \frac{\eps}{4}$,
\[
\lim_{n \to \infty} \sP\left( \forall \theta \in B_{r'}(\theta_0), \quad \nabla^2 f_n(\theta) \succeq \eps I\right) = 1.
\]

The smoothness argument follows from the same strategy. 
Weyl's inequality implies that 
\[
\nabla^2 f_n(\theta) =& 	\nabla^2 f_n(\theta) - H_\theta + H_\theta\\
\preceq& \lambda_{\max}\left( \nabla^2 f_n(\theta) - H_\theta \right) I +
\lambda_{\max}(H_\theta) I.
\]
By repeating the proof for local smoothness, we obtain that there exists
a sufficiently small $0 < r''$, such that $\forall \eps > 0$, 
\[
  \sP\left( \sup_{\theta \in B_{r''}(\theta_0)}\left| \|\nabla^2
f_n(\theta)\|_2 - \|H_\theta\|_2 \right| \leq \eps  \right) \to 1,
\]
as $n \to \infty$. Condition 4 and 5 of \cref{assump:regularity} yield that 
\[
  \sup_{\theta \in B_{r''}(\theta_0)}\|H_{\theta_0}\|_2 < \infty.
\]
Therefore, there exists a $\ell >0$ such that 
\[ \label{eq:fnsmothness}
  \lim_{n \to \infty} \sP\left( \forall \theta \in B_{r''}(\theta_0), \quad \nabla^2
f_n(\theta) \preceq \ell I\right) = 1.
\]
Then the proof is complete by defining $r \defined \min\{r', r''\}$.

\eprf

\subsection{Proof of \cref{cor:asymplocalconv}}

\bprf
We begin by verifying the conditions of \cref{thm:strongconvexsmooth} for $f_n$.
By \cref{assump:regularity} we know that $f_n$ is twice differentiable.
We also know that by \cref{lem:asymplocalconv}, under \cref{assump:regularity,assump:lsmooth}, there exist $\ell, r', \epsilon > 0$ such that
\[
\Pr\left(\sup_{\theta}\left\|- n^{-1}\nabla^2 \log\pi_n(\theta)\right\|_2 > \ell\right) &\to 0 \\
\Pr\left(\inf_{\|\theta - \theta_0\| < r'}\lambda_{\min}\left( -n^{-1}\nabla^2 \log\pi_n(\theta)\right) < \epsilon\right) &\to 0.
\]
By \cref{thm:optimum_inside} we know that $\mu_n^\star \stackrel{P_{\theta_0}}{\to} \theta_0$, so there exists an $r'> r>0$ such that 
\[
\Pr\left(\inf_{\|\theta - \mu_n^\star\| < r}\lambda_{\min}\left(- n^{-1}\nabla^2 \log\pi_n(\theta)\right) < \epsilon\right) &\to 0.
\]
Therefore by \cref{thm:strongconvexsmooth}, the probability that 
\[ \label{eq:FnGlobalsmooth}
\forall \mu, L, \quad -\ell D_n \preceq &n^{-1} \nabla^2\EE\left[-\log\pi_n(\mu+1/\sqrt{n}LZ)\right] \preceq \ell D_n,
\]
and 
\[\label{eq:F_nScvx}
\begin{aligned}
& \text{ for all } L \text{ and for } \|\mu - \mu_n^\star\|^2 < r^2/2,\\
& n^{-1} \nabla^2\EE\left[-\log\pi_n(\mu+n^{-1/2}LZ)\right] \succeq  D_n(\epsilon - \tau_n(\mu,L)\cdot(\epsilon + \ell)) ,
\end{aligned}
\]
hold converges to 1 as $n\to\infty$, where $D_n$ and $\tau_n(\mu,L)$ are as defined in \cref{eq:Ttau} 
and $x = \mu_n^\star$. Note that the gradient and Hessian in the above expression are taken with respect to a vector in $\reals^{d(d+1)}$ that stacks $\mu$ and each column of $L $ into a single vector.

Then for all $(\mu, L) \in \mcB_{r,n}$, we have
\[
& \|\mu-\mu_n^\star\|^2 \leq r^2/4\\
& \|L-L_n^\star\|_F^2 \leq 4\|I-L_n^\star\|_F^2 \implies\|L\|_F \leq 2\|I-L_n^\star\|_F + \|L_n^\star\|_F,	
\]
yielding
\[
\frac{r^2 - 2\|\mu-\mu_n^\star\|^2}{n^{-1}2\|L\|^2_F} \geq \frac{nr^2}{4\left(2\|I-L_n^\star\|_F + \|L_n^\star\|_F\right)^2}.
\]
Hence $\forall (\mu, L) \in \mcB_{r,n}$, $\tau_n(\mu,L ) \to 0$ as $n\to\infty$, yielding that under sufficiently large $n$,
\[
 \epsilon - \tau_n(\mu,L)\cdot(\epsilon + \ell) > \eps/2.	
\]
Therefore, the probability that for all $(\mu, L) \in \mcB_{r,n}$,
\[\label{eq:FnLocalcvx}
	\frac{1}{n} \nabla^2\EE\left[-\log\pi_n(\mu+1/\sqrt{n}LZ)\right] \succeq  \frac{\eps}{2} D_n 
\]
converges in $P_{\theta_0}$ to 1 as $n \to \infty$.

Combining \cref{eq:FnGlobalsmooth,eq:F_nScvx,eq:FnLocalcvx}, the proof is completed.
\eprf

\subsection{Proof of \cref{thm:convexsmoothedMAP} }

\subsubsection{Gradient and Hessian derivation}

The gradient for smoothed posterior is as follows,
\[
	\nabla \log \hat{\pi}_n(\theta) =& \nabla \log \left\{ \EE\left[\exp\left(-\frac{1}{2\alpha_n}\| \theta - W\|^2\right)\right] \right\} \\
	=& \frac{ \EE\left[\exp\left(-\frac{1}{2\alpha_n}\| \theta - W\|^2\right) \left(- \frac{1}{\alpha_n} \right) \left( \theta- W \right)\right]}{\EE\left[\exp\left(-\frac{1}{2\alpha_n}\| \theta - W\|^2\right)\right]}	,
\]
and the Hessian matrix is given by

\[ \label{eq:smpostHessian}
	\nabla^2 \log  \hpi_n(\theta) &= \frac{1}{\alpha_n^2} \frac{\EE\left[e^{\frac{-\|\theta- W\|^2-\|\theta- W'\|^2}{2\alpha_n}}W(W  -W')^T\right]}{\EE\left[e^{-\frac{\|\theta- W\|^2}{2\alpha_n}}\right]^2} - \frac{1}{\alpha_n} I,
\]
where $W,W' \distiid \Pi_{ n}$.

\subsubsection{Proof of $1^{\text{st}}$ statement of \cref{thm:convexsmoothedMAP}}

\begin{proof}[Proof of $1^{\text{st}}$ statement of \cref{thm:convexsmoothedMAP}]

To show the MAP estimation for smoothed posterior is asymptotically strictly
convex, we will show that 
\[\label{eq:logccvforM}
	\lim_{n \to \infty} \sP\left( \sup_{\|\theta-\theta_0\|\leq M} \lambda_{\text{max}}\left(\nabla^2 \log  \hpi_n(\theta) \right) < 0\right) = 1.	
\]

We focus on the first term of \cref{eq:smpostHessian}, and show that
asymptotically it is uniformly smaller than $\alpha_n^{-1}$ so that the overall
Hessian is negative definite.  For the denominator of \cref{eq:smpostHessian},
define  $B_n \defined \left\{W, W' : \max\{\|W'-\theta_0\|, \|W - \theta_0\|\} \leq \beta_n\right\}$ 
for any sequence $\beta_n = o(\alpha_n)$. Then we have
\[
\EE\left[e^{-\frac{\|\theta- W\|^2}{2\alpha_n}}\right]^2 &=
\EE\left[e^{\frac{-\|\theta- W\|^2-\|\theta- W'\|^2}{2\alpha_n}}1_{B_n}\right] + 
\EE\left[e^{\frac{-\|\theta- W\|^2-\|\theta- W'\|^2}{2\alpha_n}}1_{B^c_n}\right]\\ 
&\geq \EE\left[e^{\frac{-\|\theta- W\|^2-\|\theta- W'\|^2}{2\alpha_n}}1_{B_n}\right]\\
&\geq \EE\left[\left(\inf_{v,v' \in B_n} e^{\frac{-\|\theta-v\|^2-\|\theta-v'\|^2}{2\alpha_n}}\right)1_{B_n}\right]\\
& = \left(\inf_{v,v' \in B_n} e^{\frac{-\|\theta-v\|^2-\|\theta-v'\|^2}{2\alpha_n}}\right)\sP (B_n).
\]
By minimizing over $ v, v' \in B_n$, the above leads to
\[ \label{eq:denombound}
	\EE\left[e^{-\frac{\|\theta- W\|^2}{2\alpha_n}}\right]^2 \geq e^{\frac{-2(\|\theta-\theta_0\|+\beta_n)^2}{2\alpha_n}}\sP (B_n).
\]
%
For the numerator of the first term of \cref{eq:smpostHessian}, since $W, W'$ are \iid,
\[
	&\EE\left[e^{\frac{-\|\theta- W\|^2-\|\theta- W'\|^2}{2\alpha_n}}W(W  -W')^T\right] \\
	& \quad = \frac{1}{2} \EE\left[e^{\frac{-\|\theta- W\|^2-\|\theta- W'\|^2}{2\alpha_n}}(W  -W')(W  -W')^T\right],
\]
and since $\lambda_{\text{max}} \left( (W  -W')(W  -W')^T\right) = \|W  -W'\|^2$, 
\[\label{eq:nombound}
& \lambda_{\text{max}} \left( \EE\left[e^{\frac{-\|\theta- W\|^2-\|\theta- W'\|^2}{2\alpha_n}}W(W  -W')^T\right] \right)	\\
& \quad \leq \frac{1}{2} \EE\left[e^{\frac{-\|\theta- W\|^2-\|\theta- W'\|^2}{2\alpha_n}}\|W  -W'\|^2\right].
\]
With \cref{eq:denombound,eq:nombound}, we can therefore bound the maximal
eigenvalue of the Hessian matrix,
\[ \label{eq:maxeigenvalfraction}
& \lambda_{\text{max}}\left(\nabla^2 \log  \hpi_n(\theta) \right) \\
& \quad \leq \frac{1}{2 \alpha_n^2 \sP (B_n)}\EE\left[e^{\frac{-\|\theta- W\|^2-\|\theta- W'\|^2}{2\alpha_n}}e^{\frac{2(\|\theta-\theta_0\|+\beta_n)^2}{2\alpha_n}}\|W  -W'\|^2\right] - \frac{1}{\alpha_n}.
\]
We now bound the supremum of this expression over $\{\theta \in \reals^d: \|\theta-\theta_0\|\leq M\}$.  
Focusing on the exponent within the expectation,
\[
&\sup_{\|\theta-\theta_0\|\leq M} \frac{1}{\alpha_n} \left[2(\|\theta-\theta_0\|+\beta_n)^2 -\|\theta- W\|^2-\|\theta- W'\|^2 \right]\\
=& \sup_{\|\theta-\theta_0\|\leq M} \frac{1}{\alpha_n} \left[2(\|\theta-\theta_0\|+\beta_n)^2 -\|\theta- \theta_0 + \theta_0 -W\|^2 \right.\\
& \quad \qquad \qquad \left. -\|\theta-\theta_0 + \theta_0-W'\|^2 \right]\\
\leq& \frac{1}{\alpha_n} \left[\left(2 \beta_n^2 + 4 M \beta_n \right) - \left( \|\theta_0 - W\|^2  
+ \|\theta_0 - W'\|^2 \right) \right.\\
& \quad \left.+ 2M \left( \|\theta_0 - W\|  + \|\theta_0 - W'\| \right) \right],
\]
where the inequality is obtained by expanding the quadratic terms and bounding
$\|\theta - \theta_0\|$ with $M$. We combine the above bound with
\cref{eq:maxeigenvalfraction} to show that 
$\alpha_n^2\lambda_{\text{max}}\left(\nabla^2 \log  \hpi_n(\theta) \right) + \alpha_n$
is bounded above by
\[ \label{eq:FurtherBdHesian}
\hspace{-0.3cm}\frac{\beta_n}{2\sP(B_n)}	\!e^{\frac{2\beta_n^2 + 4M\beta_n}{\alpha_n} } \EE\!\left[ \!e^{\frac{ 2M \left(\! \|\theta_0 - W\|  + \|\theta_0 - W'\|\! \right) - \left( \!\|\theta_0 - W\|^2  + \|\theta_0 - W'\|^2 \!\right)}{\alpha_n}}  \frac{\|W  -W'\|^2}{\beta_n} \!\right]\!.\!\!
\]
By multiplying and dividing by $ \exp\left(\frac{ \|W-W'\|}{\sqrt{\beta_n}}\right)$, one notices that 
\[
	\frac{\|W  -W'\|^2}{\beta_n} =& \exp\left(\frac{ \|W-W'\|}{\sqrt{\beta_n}}\right) \exp\left(- \frac{ \|W-W'\|}{\sqrt{\beta_n}}\right) \frac{\|W  -W'\|^2}{\beta_n}\\
	\leq & 4e^{-2} \exp\left(\frac{ \|W -\theta_0\| + \|W' - \theta_0\|}{\sqrt{\beta_n}}\right),
\]
where the inequality is by the fact that $x^2 e^{-x}$ maximized at $x=2$
with value $4e^{-2}$ and $\|W  -W'\| \leq \|W\| + \|W'\|$. If we combine this bound with
\cref{eq:FurtherBdHesian} and note that $W, W'$ are iid,
\cref{eq:FurtherBdHesian} is bounded above by
\[ \label{eq:FFbdHessian}
\frac{2 e^{-2} \beta_n }{\sP(B_n)}	e^{\frac{2\beta_n^2 + 4M\beta_n}{\alpha_n} } 	 \EE\left[ e^{  \left( \frac{1}{\alpha_n}M + \beta_n^{-1/2} \right)\|W - \theta_0\| - \frac{1}{2\alpha_n} \|W - \theta_0\|^2  } \right] ^2.
\]

To show that the Hessian is asymptotically negative definite, it suffices to show that   
\cref{eq:FFbdHessian} is $o_{P_{\theta_0}}(\alpha_n)$. For the terms outside
the expectation, $\beta_n = o(\alpha_n)$ implies that
$2 e^{-2} \beta_n e^{\frac{2\beta_n^2 + 4M\beta_n}{\alpha_n} }  = o(\alpha_n)$,
and 
 \cref{assump:regularity} and \cref{lemma:posttailbound} together imply that
\[
	\sP(B_n) = \Pi_{n}\left( \left\{ W :  \|W - \theta_0 \| \leq \beta_n\right\} \right)^2 \stackrel{P_{\theta_0}}{\to}	1,
\]
so
\[
	\frac{2 e^{-2} \beta_n }{\sP(B_n)}	e^{\frac{2\beta_n^2 + 4M\beta_n}{\alpha_n} }  = o_{P_{\theta_0}}(\alpha_n).	
\]
Therefore, in order to show \cref{eq:FFbdHessian} is $o_{P_{\theta_0}}(\alpha_n)$, it is sufficient to show  that
\[
	\EE\left[ e^{  \left( \frac{1}{\alpha_n}M + \beta_n^{-1/2} \right)\|W - \theta_0\| - \frac{1}{2\alpha_n} \|W - \theta_0\|^2  } \right] = O_{P_{\theta_0}}(1).	
\]
The next step is to split the expectation into two 
regions---$\|W - \theta_0\|\leq \beta_n$ and $\|W - \theta_0\| > \beta_n$---and bound 
its value within them separately.

\benum
\item When $\|W - \theta_0\|\leq \beta_n$, the exponent inside the expectation is shrinking uniformly since $\beta_n = o(\alpha_n)$:
\[
	& \EE\left[ 1_{\{ \|W - \theta_0\| \leq \beta_n\}} e^{  \left( \frac{1}{\alpha_n}M + \beta_n^{-1/2} \right)\|W - \theta_0\| - \frac{1}{2\alpha_n} \|W - \theta_0\|^2  } \right] \\
	&\leq  \EE\left[ 1_{\{ \|W - \theta_0\| \leq \beta_n\}} \right] e^{  \left( \frac{1}{\alpha_n}M + \beta_n^{-1/2} \right)\beta_n} \\
	& = O_{P_{\theta_0}}(1).
\]

\item 
When $\|W - \theta_0\| > \beta_n$,  we take the supremum over the exponent (a quadratic function), yielding $\|W -
\theta_0\| = M + \alpha_n \beta_n^{-1/2}$ and the following bound,
\[\label{eq:threeterms}
&\left( \frac{1}{\alpha_n}M + \beta_n^{-1/2} \right)\|W - \theta_0\| - \frac{1}{2\alpha_n} \|W - \theta_0\|^2   \\
& \leq \sup_{\|v - \theta_0\|} \left( \left( \frac{1}{\alpha_n}M + \beta_n^{-1/2} \right)\|v - \theta_0\| - \frac{1}{2\alpha_n} \|v - \theta_0\|^2  \right)\\
&= \left( \frac{1}{\alpha_n}M + \beta_n^{-1/2} \right) \left(  M + \alpha_n \beta_n^{-1/2} \right) - \frac{1}{2\alpha_n} \left(  M + \alpha_n \beta_n^{-1/2} \right)^2 \\
& = \frac{M^2}{2\alpha_n} + \frac{M}{\beta_n^{1/2} }+ \frac{\alpha_n}{2\beta_n}.
\]
This yields
\[
&\EE\left[ 1_{\{ \|W - \theta_0\| > \beta_n\}} e^{  \left( \frac{1}{\alpha_n}M + \beta_n^{-1/2} \right)\|W - \theta_0\| - \frac{1}{2\alpha_n} \|W - \theta_0\|^2  } \right]\\
&\leq \Pi_{n}\left( \left\{ W :  \|W - \theta_0 \| > \beta_n\right\} \right)\exp\left( \frac{M^2}{2\alpha_n} + \frac{M}{\beta_n^{1/2} }+ \frac{\alpha_n}{2\beta_n} \right) .
\]

Note that  it is always possible to choose $\beta_n = o(\alpha_n)$ with $\beta_n = \omega(\alpha_n^2)$. With this choice of $\beta_n$, the dominating term among the three of \cref{eq:threeterms} is $\frac{M^2}{2\alpha_n} $. 

Then by \cref{lemma:posttailbound}, there exists a sequence $\beta_n = o(\alpha_n)$ with $\beta_n = \omega(\alpha_n^2)$ such that the following holds,
\[ 
\Pi_n( \left\{ W :  \|W - \theta_0 \| > \beta_n\right\} ) = o_{P_{\theta_0}}\left( \exp \left\{ -\frac{M^2}{2\alpha_n}\right\}\right),	
\] 
which implies \[
	\EE\left[ 1_{\{ \|W - \theta_0\| > \beta_n\}} e^{  \left( \frac{1}{\alpha_n}M + \beta_n^{-1/2} \right)\|W - \theta_0\| - \frac{1}{2\alpha_n} \|W - \theta_0\|^2  } \right] = o_{P_{\theta_0}}(1)	.
\]
\eenum
This finishes the proof.
\end{proof}

In the last step of the above proof, we require an exponential tail bound for
the posterior $\Pi_n$. We provide this in the following lemma, following
the general proof strategy of \citet[Thm 10.3]{van2000asymptotic}. 
The proof of \cref{lemma:posttailbound} involves many
probability distributions; thus, for mathematical convenience and explicitness,
in the proof of \cref{lemma:posttailbound} we use square bracket---$P\left[ X
\right]$---to denote the expectation of random variable $X$ with respect to a probability
distribution $P$. When taking expectation to a function of $n$ data points $f(X_1, \dots, X_n)$ , where $(X_i)_{ i =1}^n \distiid P_{\theta}$, we still write $P_\theta[f]$; and $P_\theta$ here represents the product measure.

\bnlem \label{lemma:posttailbound}
Under \cref{assump:regularity}, $\alpha_n^3 n \to \infty$, 
there exists a sequence $\beta_n$ satisfying $\beta_n = o(\alpha_n)$, $\beta_n = \omega(\alpha_n^2)$ and $\beta_n = \omega(n^{-1/2})$ such that for any fixed constant $M$,
\[
\Pi_n( \left\{ W :  \|W - \theta_0 \| > \beta_n\right\} ) = o_{P_{\theta_0}}\left( \exp \left\{ -\frac{M^2}{2\alpha_n}\right\}\right).	
\]
\enlem
\begin{proof}[Proof of \cref{lemma:posttailbound}]
In order to show that $\beta_n$ satisfies the tail probability bound,
it suffices to prove that
\[
e^{\frac{1}{\alpha_n}} P_{\theta_0}\left[ \Pi_n( \left\{ W :  \|W - \theta_0 \| > \beta_n\right\} ) \right] \to 0	,
\]
due to Markov's inequality (we absorb the $M^2/2$ constant into $\alpha_n$ because it does not affect the proof). 
To achieve this, we take advantage of the
existence of a test sequence applied from \cref{assump:regularity}. By
\citet[Lemma 10.6]{van2000asymptotic}, given the $1^\text{st}$ and the
$2^\text{nd}$ conditions of \cref{assump:regularity} and the fact that the
parameter space $\reals^d$ is $\sigma$-compact, there exists a sequence of
tests $\phi_n : \mcX^n \to[0,1]$, where $\mcX^n$ is the space of $(X_1, \dots, X_n)$, such that as $n \to \infty$, 
\[
	P_{\theta_0} [\phi_n]  \to 0,\quad \sup _{\left\|\theta-\theta_{0}\right\|>\varepsilon} P_{\theta}\left[1-\phi_{n}\right] \to 0.	
\]
Further, by \citet[Lemma 1.2]{kleijn2004bayesian} and \citet[Lemma 10.3]{van2000asymptotic}, 
under \cref{assump:regularity} and the existence of
the above test sequence $\phi_n$, for every $M_n \to \infty$, there exists a
constant $C > 0$ and another sequence of tests $\psi_n : \mcX^n\to [0,1]$ such that for all $\|\theta - \theta_0\| > M_n/\sqrt{n}$ and sufficiently large $n$,
\begin{equation} \label{eq:exponential_testability}
	P_{\theta_0} [\psi_n ]\leq \exp\{- Cn\},\quad  P_{\theta}\left[1-\psi_{n}\right] \leq \exp\{- Cn (\|\theta - \theta_0 \|^2 \wedge 1)\}.
\end{equation} 
Using $\psi_n$, we split the expectation as following,
\[
&  e^{\frac{1}{\alpha_n}} \cdot P_{\theta_0} [\Pi_n( \left\{ W :  \|W - \theta_0 \| > \beta_n\right\} )] \\
& =\underbrace{e^{\frac{1}{\alpha_n}} \cdot P_{\theta_0} [\Pi_n( \left\{ W :  \|W - \theta_0 \| > \beta_n\right\} ) \psi_n]}_{(\text{I})} \\
& \quad + \underbrace{e^{\frac{1}{\alpha_n}} \cdot P_{\theta_0} [\Pi_n( \left\{ W :  \|W - \theta_0 \| > \beta_n\right\} )(1- \psi_n)]}_{(\text{II})},
\]
and we aim to show both parts converging to $0$. 

For term (I), the first statement of \cref{eq:exponential_testability} implies that $\exists C >0$ such that
\[
	e^{\frac{1}{\alpha_n}} \cdot P_{\theta_0} [ \Pi_n( \left\{ W :  \|W - \theta_0 \| > \beta_n\right\} ) \psi_n] \leq e^{\frac{1}{\alpha_n}} P_{\theta_0}[ \psi_n ] \leq e^{\frac{1}{\alpha_n}} e^{-nC}	.
\]
where the first inequality follows by $ \Pi_n( \left\{ W :  \|W - \theta_0 \| > \beta_n\right\} ) \leq 1$. 
Since $n \alpha_n^3 \to \infty$, the last bound in the above expression converges to $0$. 

For term (II), we work with the shifted and scaled posterior distribution. Define $Z_n  = \sqrt{n}(W - \theta_0)$ and $B_n = \{Z_n: \|Z_n\| >  \sqrt{n\beta_n^2} \}$, and let $\tPi_{0}$ be the corresponding prior distribution on $Z_n$ and $\tPi_n$ be the shifted and scaled posterior distribution, which yields  
\[ \label{eq:testPU}
\begin{aligned}
	& e^{\frac{1}{\alpha_n}} \cdot  P_{\theta_0} [\Pi_n( \left\{ W :  \|W - \theta_0 \| > \beta_n\right\} )(1- \psi_n)] \\
	&\qquad =  e^{\frac{1}{\alpha_n}} \cdot  P_{\theta_0} \left[\tPi_n \left( B_n \right)(1- \psi_n) \right]. 
\end{aligned}
\]
Let $U $ be a closed ball around $0$ with a fixed radius $r$, then restricting $\tPi_0$ on $U$ defines a probability measure $\tPi_0^U$, i.e., for all measurable set $B$, $\tPi_0^U(B) = \tPi_0(B \cap U)/ \tPi_0(U)$.  Write $P_{n, z}$ for the joint distribution of $n$ data points $(X_1, \dots, X_n)$ parameterized under $\theta_0 + z/\sqrt{n}$ and hence write the marginal distribution of $(X_1, \dots, X_n)$ under $\tPi_0^U$ for $P_{n,U} = \int P_{n, z} \mbox{d} \tPi_0^U(z)$. The densities of these distributions will be represented using lower case, e.g., $p_{n,U}(x) = \int p_{n,z}(x) \tpi_0^U(z) \mbox{d} z$ is the PDF of $P_{n,U}$. Here we abuse the notation that $x$ represents $(x_1, \dots, x_n)$.

We replace $P_{\theta_0}$ in \cref{eq:testPU} with $P_{n,U}$. Under \cref{assump:regularity}, by \citet[P. 141]{van2000asymptotic}, $P_{n,U} $ is \textit{mutually contiguous} to $P_{\theta_0}$ \citep{lecam1960locally}, that is, for any statistics $T_n$ (a Borel function of $X^n$), $T_n \stackrel{P_{\theta_0}}{\to }0$ iff $T_n \stackrel{P_{n,U}}{\to }0$.  Thus, considering $\tPi_0 \left( B_n \right)(1- \psi_n)$ as the statistics $T_n$, the convergence to 0 of the expression in \cref{eq:testPU} is equivalent to 
\[
e^{\frac{1}{\alpha_n}}  \cdot P_{n, U} \left[\tPi_n \left( B_n \right)(1- \psi_n) \right]\to 0	.
\]
Manipulating the expression of $P_{n,U}$ and $\tPi_n \left( B_n \right)$ (we write $ \tPi_n \left( B_n, x\right) $ in the integral and write $\tPi_n \left( B_n, (X_i)_{i = 1}^n\right)$ in the expectation to make the dependence of posterior on the data explicit), 
\[
	P_{n, U}\left[ \tPi_n \left( B_n,(X_i)_{i = 1}^n\right)(1- \psi_n) \right] &=  \int \tPi_n \left( B_n, x \right)(1- \psi_n) \mbox{d} P_{n, U}(x)\\
	& = \int \tPi_n \left( B_n, x\right)(1- \psi_n) p_{n,U}(x) \mbox{d} x.
\]
Note that $p_{n,U}(x ) = \int p_{n,z}(x) \mbox{d} \tPi_0^U(z)$,
\[
& = \int \tPi_n \left( B_n,x \right)(1- \psi_n) \left( \int p_{n,z}(x) \mbox{d} \tPi_0^U(z) \right) \dee x	.
\]
Recall that for all measurable set $B$, $\tPi_0^U(B) = \tPi_0(B \cap U)/ \tPi_0(U)$, thus
\[
& = \frac{1}{\tPi_0(U)}\int \tPi_n \left( B_n, x \right)(1- \psi_n) \left( \int_U p_{n,z}(x) \mbox{d} \tPi_{0}(z) \right) \dee x.
\]
By using Bayes rule, we expand $\tPi_{n} \left( B_n, x \right) = \frac{\int \ind[B_n] p_{n,z}(x) \mbox{d} \tPi_0 (z)}{\int p_{n,z}(x)\mbox{d}\tPi_0(z)}$,
\[
& =	\frac{\int(1- \psi_n) \left(\int \ind[B_n] p_{n,z}(x) \mbox{d} \tPi_{0}(z) \right) \left( \int_U p_{n,z}(x) \mbox{d} \tPi_0(z) \right) \dee x}{\tPi_{0}(U) \int p_{n,z}(x)\mbox{d}\tPi_{0}(z)}.
\]
Note that $\tPi_n \left( U, x \right) = \frac{\int_U p_{n,z}(x) \mbox{d} \tPi_{0}(z)}{\int p_{n,z}(x)\mbox{d}\tPi_{0}(z)}$, 
\[
&= \frac{1}{\tPi_0(U)} \int  \left( \int \ind[B_n] p_{n,z}(x) \mbox{d} \tPi_{0}(z) \right) (1- \psi_n) \tPi_n \left( U,x \right) \dee x.
\]
By Fubini Theorem and $ \tPi_n \left( U,x \right) \leq 1$, 
\[
& \leq \frac{1}{\tPi_0(U)} \int_{B_n} \left( \int (1 - \psi_n) p_{n,z}(x) \dee x\right) \dee \tPi_0(z)\\
& =  \frac{1}{\tPi_{0}(U)} \int_{\{ \|z\| > \sqrt{n\beta_n^2}\}} P_{n,z} [1- \psi_n] \mbox{d} \tPi_0(z).
\]
Note that $P_{n,z} [1- \psi_n] \equiv P_{\theta}[1 - \psi_n]$ for $\theta = \theta_0 + z/\sqrt{n}$ and that $\sqrt{n\beta_n^2} \to \infty$ due to $\beta_n = \omega(n^{-1/2})$. Thus, we can use the second statement of \cref{eq:exponential_testability} to bound $P_{n,z} [1- \psi_n]$, yielding
\[
& \frac{1}{\tPi_0(U)} \int_{\{ \|z\| > \sqrt{n\beta_n^2}\}} P_{n,z} [1- \psi_n] \mbox{d} \tPi_0(z)\\
& \leq \frac{1}{\tPi_0(U)}  \int_{\{ \|z\| > \sqrt{n\beta_n^2}\}}  \exp\{- C (\|z\|^2 \wedge n)\} \dee \tPi_0(z).
\]
We then derive upper bounds for both the fraction and the integral to show the above is $o\left(e^{-\frac{1}{\alpha_n}}\right)$. For the fraction, we define $U_n \defined \left\{w\in \reals^d: \sqrt{n}(w - \theta_0) \in U  \right\}$, then
\[
	\tPi_0(U) = & \Pi_0(U_n) \geq \frac{\pi^{\frac{d}{2}}}{\gamma(\frac{d}{2}  +1)} \left(n^{-1/2} r\right)^d\inf_{w\in U_n}\pi_0(w).
\]
By \cref{assump:regularity}, for all $ w \in \reals^d, \pi_0(w)$ is positive and continuous, and hence
$\inf_{w\in U_n}\pi_0(w)$ is an increasing sequence that converges to $\pi_0(\theta_0) > 0$ as $n \to \infty$. Thus, there is a constant $D > 0$ such that for sufficiently large $n$,
\[\label{eq:boundforPi(U)}
\tPi_0(U) \geq D  n^{-d/2},	
\]
yielding that
\[
\exists C > 0, \quad \text{ s.t. } \quad \frac{1}{\tPi_{0}(U)} \leq C n^{d/2}.
\]
For the integral, by splitting $B_n$ into $\{\sqrt{n\beta_n^2} < \|z_n\| \leq k \sqrt{n} \}$ and $ \{ \|z_n\| > k \sqrt{n} \}$ for some positive $ k < 1$, 
	\[ 
		&\int_{\{ \|z\| > \sqrt{n\beta_n^2}\}}  \exp\{- C (\|z\|^2 \wedge n)\} \dee \tPi_{0}(z)\\
		& \leq  \int_{\{k \sqrt{n} \geq \|z\| >  \sqrt{n\beta_n^2} \}} \exp\{- C \|z\|^2 \} \dee \tPi_{0}(z)  + e^{-C k^2 n}.
	\]
Then by change of variable to $w = \frac{1}{\sqrt{n}} z + \theta_0$,
	\[
		&= \int_{\{k  \geq \|w- \theta_0\| >  \beta_n \}} \exp\{- C n \|w -\theta_0\|^2 \} \pi_0(w ) \mbox{d} w  + e^{-C k^2 n}.
	\]
Note that by \cref{assump:regularity}, $\pi_0(w)$ is continuous for all $w \in \reals^d$, we can choose a sufficiently small $k$ such that $\pi_0(\theta)$ is uniformly bounded by a constant $C$ over the region $\{k  \geq \|w- \theta_0\| >  \beta_n \}$. Thus, the above can be bounded above by 
\[& C \int_{\{k  \geq \|w- \theta_0\| >  \beta_n \}} \exp\{- C n \|w -\theta_0\|^2 \} \mbox{d} w  + e^{-C k^2 n}\\
& = C n^{-d/2} \int_{\{k \sqrt{n} \geq \|z\| > \sqrt{n \beta_n^2} \}} \exp\left\{ -C \|z\|^2\right\} \dee z + e^{-C k^2 n}\\
& \leq C n^{-d/2} \int_{\{\|z\| > \sqrt{n \beta_n^2} \}} \exp\left\{ -C \|z\|^2\right\} \dee z + e^{-C k^2 n},
\]
where the equality is by change of variable back to $z = \sqrt{n}(w - \theta_0) $. Then, consider the integral on RHS. Using spherical coordinates, there exists a fixed constant $ D > 0$ such that
\[
	\int_{\{\|z\| > \sqrt{n \beta_n^2} \}} \exp\left\{ -C \|z\|^2\right\} \dee z & = D \int_{\{r > \sqrt{n \beta_n^2}\}} e^{-Cr^2} r^{d-1} \dee r\\
	& = DC^{-d/2} \int_{\{s> C n\beta_n^2 \}}	e^{-s} s^{\frac{d}{2} -1}  \dee s,
\]
where the second equality is by setting $s = Cr^2$. Note that the integrand of RHS is proportional to the PDF of $\Gamma(\frac{d}{2}, 1)$. Using the tail properties of the Gamma random variable \citep[p.~28]{boucheron2013concentration}, we have that for some generic constant $D >0$,
\[
	\int_{\{\|z\| > \sqrt{n \beta_n^2} \}} \exp\left\{ -C \|z\|^2\right\} \dee z \leq D e^{ - C n\beta_n^2}.
\]
Therefore, for some generic constants $C , D >0$,
\[\label{eq:integbound}
&\int_{\{ \|z\| > \sqrt{n\beta_n^2}\}}  \exp\{- C (\|z\|^2 \wedge n)\} \dee \tPi_{0}(z)\\
& \leq	D n^{-d/2} e^{- C n\beta_n^2}  + e^{-C k^2 n}.
\]
Then we combine \cref{eq:boundforPi(U),eq:integbound}, yielding that for some constants $C, D > 0$ independent to $n$,
\[
& e^{\frac{1}{\alpha_n}}  \cdot P_{n, U} \left[\tPi_{0} \left( B_n\right)(1- \psi_n) \right]\\	
& \leq e^{\frac{1}{\alpha_n}} \frac{1}{\Pi_{n,0}(U)}  \int_{\{ \|z\| > \sqrt{n\beta_n^2}\}}  \exp\{- C (\|z\|^2 \wedge n)\} \dee \tPi_{0}(z)\\
& \leq e^{\frac{1}{\alpha_n}} C \sqrt{n}^d e^{-C n} + e^{\frac{1}{\alpha_n}} D e^{- C n\beta_n^2} .
\]
Lastly, it remains to show that there exists a positive sequence $\beta_n$ satisfying both $ \beta_n = o(\alpha_n)$ and $\beta_n = \omega(\alpha_n^2)$ such that the RHS converges to $0$. The first term always converges to $0$ due to $\alpha_n^3 n \to \infty$. For the second term, we consider two different cases. If $\alpha_n = o(n^{-1/6})$, we pick $\beta_n = n^{-1/3}$, which is both $ o(\alpha_n)$ and $\omega(\alpha_n^2)$. Then
\[
	e^{\frac{1}{\alpha_n}} D e^{- C n\beta_n^2} & = D \exp\left\{ \alpha_n^{-1} - C n^{1/3}\right\} \\
		&\longrightarrow 0,
\]
where the convergence in the last line is by $n \alpha_n^3 \to \infty \Leftrightarrow \frac{1}{\alpha_n} = o(n^{1/3})$. If $\alpha_n = \omega(n^{-1/6})$, we pick $\beta_n = \alpha_n^2$. Then 
\[
	e^{\frac{1}{\alpha_n}} D e^{- C n\beta_n^2} =  D \exp\left\{ \alpha_n^{-1} - C n \alpha_n^4 \right\}.
\]
Since $\alpha_n = \omega(n^{-1/6})$, $\frac{1}{\alpha_n} = o(n^{1/6})$ and $ n\alpha_n^4 = \omega(n^{1/3})$, yielding that the above converges to $0$ as $n \to \infty$.

This completes the proof.

\end{proof}

\subsubsection{Proof of $2^\text{nd}$ statement of \cref{thm:convexsmoothedMAP}}

In this section, we show that the smoothed MAP estimator 
($\htheta_n^\star$) is also a consistent estimate of $\theta_0$, but with a convergence rate 
that is slower than the traditional $\sqrt{n}$. This is the case because the variance of
the smoothing kernel satisfies $\alpha_n = \omega(n^{-1/3})$, and the convergence rate of $\htheta_n^\star $ is
determined by $\alpha_n$ via
\[
	\label{eq:smoothedMAPrate}
	\|\htheta_n^\star - \theta_{0}\| = O_{P_{\theta_0}}(\sqrt{\alpha_n}).	
\]
Recall that $\theta_{\text{MLE},n }$ is a $\sqrt{n}$-consistent estimate of
$\theta_0$. Thus, it is sufficient to show  $\left\|  \htheta_n^\star -
\theta_{\text{MLE},n} \right\| = O_{P_{\theta_0}}\left(\sqrt{\alpha_n}\right)$. 

Note that $\htheta_n^\star$ and $\theta_n^\star$ are maximals of stochastic process $\hpi_n(\theta)$ and $\pi_n(\theta)$ respectively, which can be studied in the framework of M-estimator \citep{van2000asymptotic,wellner2013weak}. A useful tool in establishing the asymptotics of M-estimators is the Argmax Continuous Mapping theorem ~\citep[Lemma 3.2.1]{wellner2013weak}, which is introduced as follows.
\bnlem[Argmax Continuous Mapping \citep{wellner2013weak}] \label{lemma:ArgmaxCont}
Let $\{f_n(\theta) \}$ and $f(\theta)$ be stochastic processes indexed by $\theta$, where $\theta \in \Theta$. Let $\htheta$ be a random element such that almost surely, for every open sets $G$ containing $\htheta$, 
\[
f(\htheta) >  \sup_{\theta \notin G} f(\theta).
\]
and $\htheta_n$ be a random sequence  such that almost surely
\[
f_n(\htheta_n) =  \sup_{\theta \in \Theta} f_n(\theta).
\]
If $\sup_{\theta\in \Theta} |f_n(\theta) - f(\theta)| = o_P(1)$ as $n \to \infty$, then 
\[
	\htheta_n \convd \htheta.	
\]	

\enlem

\!The proof strategy of the $2^\text{nd}$ statement of
\cref{thm:convexsmoothedMAP} is to apply \cref{lemma:ArgmaxCont}
in a setting where $f_n$ is $\hpi_n(\theta)$ and $f$ is a Gaussian density. Using
the Bernstein-von Mises Theorem \ref{thm_bvm}, we show that $\hpi_n(\theta)$ converges
uniformly to this Gaussian density, which implies that the MAP of
$\hpi_n(\theta)$ converges in distribution to the MAP of this Gaussian
distribution by the Argmax Continuous Mapping theorem. The detailed proof is as
follows.

\begin{proof}[Proof of $2^\text{nd}$ statement of \cref{thm:convexsmoothedMAP}]

	Note that 
	\[
		\|\htheta_n^\star - \theta_{0}\| \leq \|\htheta_n^\star - \theta_{\text{MLE},n}\|+ \| \theta_{\text{MLE},n} - \theta_0 \|.
	\] 
	By \cref{thm_bvm}, we have $\|\theta_{\text{MLE},n} - \theta_{0}\| \!= \!O_{P_{\theta_0}}(1/\sqrt{n})$. And in addition, given that $\sqrt{\alpha_n } = \omega (1/ \sqrt{n})$, in order to get \cref{eq:smoothedMAPrate}, it suffices to show
	\[
		\left\|  \hat{\theta}_n - \theta_{\text{MLE},n} \right\| = O_{P_{\theta_0}}\left(\sqrt{\alpha_n}\right). 	
	\]
	Thus, in this proof, we aim to show $\left\|   \htheta_n^\star - \theta_{\text{MLE},n} \right\| = O_{P_{\theta_0}}(\sqrt{\alpha_n}) $ and it is sufficient to prove
	\[
	\frac{1}{\sqrt{\alpha_n}}  \left(   \hat{\theta}_n - \theta_{\text{MLE},n}\right) \stackrel{P_{\theta_0}}{\to } 0.
	\]
	Let $\xi = \frac{1}{\sqrt{\alpha_n}}  \left(\theta - \theta_{\text{MLE},n}\right)$, $\xi^*_n = \frac{1}{\sqrt{\alpha_n}}  \left(   \hat{\theta}_n - \theta_{\text{MLE},n} \right) $ and $t = \frac{1}{\sqrt{\alpha_n}} \left(w - \theta_{\text{MLE},n}\right)$. By expressing $\hpi_n(\theta)$, which is defined in \cref{eq:smootheddensity},
	\[
	\xi_n^* &= \argmax_{\xi} \hpi\left(\sqrt{\alpha_n} \xi + \theta_{\text{MLE},n}\right)\\
	& = \argmax_{\xi} \int  \pi_n\left(\sqrt{\alpha_n} t+ \theta_{\text{MLE},n}\right)  \exp\left( -\frac{1}{2\alpha_n}\|\sqrt{\alpha_n}\xi - \sqrt{\alpha_n}t \|^2\right)  \mbox{d} t \\
	& = \argmax_{\xi} \int  \alpha_n^{d/2} \pi_n\left(\sqrt{\alpha_n} t+ \theta_{\text{MLE},n}\right)   \exp\left( -\frac{1}{2}\|\xi - t \|^2\right)  \mbox{d} t.
	\]
	Define \[ 
	f_n(\xi) =& 	\int  \alpha_n^{d/2} \pi_n\left(\sqrt{\alpha_n} t+ \theta_{\text{MLE},n}\right)   \exp\left( -\frac{1}{2}\|\xi - t \|^2\right)  \mbox{d} t, \\ 
	g_n(\xi) =& \int \phi\left(t; 0, \frac{1}{n \alpha_n}H_{\theta_0}^{-1}\right) \exp\left( -\frac{1}{2}\|\xi - t \|^2\right)  \mbox{d} t,\\
	f(\xi) =&  (2\pi )^{d/2} \phi\left(\xi; 0, I \right),
	\]
	where $\phi(\cdot; \mu, \Sigma)$ denotes the PDF of $\distNorm(\mu, \Sigma)$. 

	By adding and subtracting $f(\xi)$,   
	\[
		\xi_n^* =& \argmax_{\xi} f_n(\xi)\\
		=& \argmax_{\xi} \left\{f_n(\xi) - f(\xi) + f(\xi)\right\}.
 	\]
	We then apply  \cref{lemma:ArgmaxCont} to show $\xi_n^* \convd \argmax_{\xi} f(\xi)$. We start by verifying a condition of the argmax continuous mapping theorem that 
	\[ \label{eq:Argcond1}
	\lim_{n \to \infty} \sup_{\xi }	|f_n(\xi) - f(\xi)| = 0.
	\]
	By triangle inequality, for all $n$,
	\[ \label{eq:twotermsforuniformconv}
		\sup_{\xi }	|f_n(\xi) - f(\xi)| \leq \sup_{\xi }	|f_n(\xi) - g_n(\xi)| + \sup_{\xi }	|g_n(\xi) - f(\xi)|.
	\]
	Later we show both two terms on the RHS converging to $0$.

	For the first term. Note that $\alpha_n^{d/2} \pi_n(\sqrt{\alpha_n} t+ \theta_{\text{MLE},n})$ is the probability density function of $\Pi_{\sqrt{\alpha_n} t+ \theta_{\text{MLE},n}}$, which is the posterior distribution parameterized on $t$. Thus, for all n,
	\[
	& \sup_{\xi}|f_n(\xi) - g_n(\xi)| \\
	& = \sup_{\xi} \left\{\int \! \left| \alpha_n^{d/2} \pi_n(\sqrt{\alpha_n} t \!+ \!\theta_{\text{MLE},n}) \!-\! \phi(t; 0, \frac{1}{n\alpha_n}H_{\theta_0}^{-1})  \right| \times \right. \\
	&\qquad \qquad \left.  \! \exp \!\left( \!-\frac{1}{2}\|\xi - t \|^2 \!\right)  \mbox{d} t \!\right\}\\
	& \leq \tvd{\Pi_{\sqrt{\alpha_n} t+ \theta_{\text{MLE},n}}}{\mcN\left( 0, \frac{1}{n\alpha_n}H_{\theta_0}^{-1}\right)} ,
	\]
	where the inequality is by $ \sup_{\xi,t} \exp( -\frac{1}{2}\|\xi - t \|^2) \leq 1$. Under \cref{assump:regularity}, the posterior distribution admits Bernstein-von Mises theorem (\cref{thm_bvm}) that 
	\[\label{eq:bvmMLE}
		\tvd{\Pi_n}{ \distNorm\left(\theta_{\text{MLE}, n}, \frac{1}{n} H_{\theta_0}^{-1}\right)} = o_{P_{\theta_0}}(1).
	\]
	With the invariance of total variation under reparametrization, we have 
	\[\label{eq: W_n -M_n = op(1)}
	\tvd{\Pi_{\sqrt{\alpha_n} t+ \theta_{\text{MLE},n}}}{\mcN\left( 0, \frac{1}{n\alpha_n}H_{\theta_0}^{-1}\right)} =  o_{P_{\theta_0}}(1).
	\]
	This shows the uniform convergence from $f_n(\xi)$ to $g_n(\xi)$. Note that in \cref{eq:bvmMLE}, we states the Bernstein-von Mises theorem with a different centering process. It is allowed when $\theta_{\text{MLE},n}$ is asymptotically normal \citep[p.~144]{van2000asymptotic}, which is ensured by \cref{thm_bvm}.
	
	For the second term in \cref{eq:twotermsforuniformconv}.  Note that we can evaluate $g_n(\xi)$ since it is a convolution of two Gaussian PDFs, that is
	\[
	g_n(\xi)	 = (2\pi )^{d/2} \phi\left(\xi; 0, \frac{1}{n \alpha_n}H_{\theta_0}^{-1} + I \right).
	\]
	Comparing this to $f(\xi) =  (2\pi )^{d/2} \phi\left(\xi; 0, I \right)$, one notices that $ \frac{1}{n \alpha_n}H_{\theta_0}^{-1} + I \to I$ due to $\alpha_n^3 n \to \infty$. And further for Gaussian distributions, the convergence of parameters implies the uniform convergence of PDFs, yielding that 
	\[
	\lim_{n \to \infty} \sup_{\xi} |g_n(\xi) - f(\xi) | = 0.	
	\]
	Thus, we have \cref{eq:twotermsforuniformconv} converging to $0$ as $n \to \infty$.

	Now we look at $f(\xi)$ with the goal to apply \cref{lemma:ArgmaxCont} and to obtain 
	$\xi_n^* \convd \argmax_{\xi} f(\xi)$. Note that
	\[
		\argmax_{\xi} f(\xi) = 0 \quad \text{and} \quad  \sup_{\xi}f(\xi) =\det \left( I  \right)^{-1/2} = 1.
	\]
	To apply \cref{lemma:ArgmaxCont}, we need to ensure that for any open set $G$ that contains $0$,
	\[\label{eq:Argmaxcond2}
		f(0)> \sup_{\xi \in G} f(\xi).
	\]
	This holds by the unimodality of standard Gaussian distirbution.

	Therefore, with both conditioins \cref{eq:Argcond1} and \cref{eq:Argmaxcond2}, we can apply \cref{lemma:ArgmaxCont} to conclude that 
	\[
		\frac{1}{\sqrt{\alpha_n}}  \left(  \htheta_n^\star - \theta_{\text{MLE},n} \right) \stackrel{P_{\theta_0}}{\to } 0.	
	\]
	This completes the proof.
\end{proof}


\subsection{Proof of \cref{thm:mapasymp}}

\bprf 
\cref{lem:asymplocalconv} implies the existence of a locally asymptotically
convex and smooth region $B_r(\theta_0)$ of $f_n$; and \cref{thm_deterbvm,thm:convexsmoothedMAP}
show the consistency of $\theta_{n}^\star$ and $\htheta_n^\star$ to $\theta_0$ in
data-asymptotics.  
Let $\theta_\text{init}$ be the initial value of \cref{alg:csl} satisfying $\|\theta_\text{init} - \htheta^\star_n\| \leq \frac{r}{4(\ell + 1)}$.
If $\| \htheta_n^\star - \theta_0\| \leq r/4$ and $\|\theta_n^\star - \theta_0\| \leq r/4$,
\[
& \|\theta_\text{init}- \theta_n^\star\| \\
& \leq \|\theta_\text{init} - \htheta_n^\star\| + \|\htheta_n^\star - \theta_0\| + \|\theta_n^\star - \theta_0\| \\
& \leq (\frac{1}{2} + \frac{1}{4(\ell + 1)})r,
\]
and $B_{3r/4}(\theta_n^\star) \subseteq B_r(\theta_0)$.
Combining above shows that there exists $0 < r':= \frac{3r}{4}$ such that 
\[\label{eq:fnlocalcvxsmooth}
  \lim_{n \to \infty} \sP\left( \forall \theta \in B_{r'}(\theta_{n}^\star), \quad
    \eps I\preceq \nabla^2 f_n(\theta) \preceq \ell I\right) = 1.
\]
Then, it remains to show that the iterates produced by \cref{alg:csl} will be confined in
$B_{r'}(\theta_n^\star)$ as we get more and more data.
Since $\theta_n^\star$ is the global optimum of $f_n$,  
as long as backtracking line search ensures the decay of objective value within
the locally strongly convex $B_{r'}(\theta_n^\star)$\citep[Page 465]{boyd2004convex}, 
the gradient norm will decay in each iteration and hence the iterates converge to the
optimum. Therefore, it is sufficient to show the first iteration
stays inside.

By  $\ell$-smoothness of $f_n$ inside $B_{r'}(\theta_n^\star)$, 
we have $\|\nabla f_n(\theta_\text{init})\| \leq \frac{\ell r}{4(\ell +1)}$  and hence
\[
  \lim_{n \to \infty} \sP\left( \theta_\text{init}- \nabla f_n(\theta_\text{init})\in
  B_{r'}(\theta_n^\star) \right) = 1.
\]
This shows the confinement. The convergence rate of iterates is $\eta^k$, where
\[
 0 \leq \eta = \sqrt{\max\left\{|1 - \frac{2\eps}{\eps + \ell}|, |1 - \frac{2\ell}{\eps + \ell}|\right\}} < 1,
\]
which follows from the standard theory of gradient descent 
under $\eps$-strong convexity and $\ell$-Lipschitz smoothness \citep[Page 15]{ryu2016primer}.
This completes the proof.

\eprf

\subsection{Proof of \cref{thm:csvi}}
\bprf[Proof of \cref{thm:csvi}]

In this proof, we aim to apply \cref{thm:sgdconvergence} with 
\[
    \mcX = \{\mu \in \reals^p, L\text{ lower triangular with non-negative diagonals}\},    
\]
which is closed and convex.
Note that in the notation of this theorem, $x = (\mu^T, L_1^T, \dots, L_p^T)^T \in \reals^{(d+1)d}$ and $V \in \reals^{(d+1)d \times (d+1)d}$ is set to be a diagonal matrix with entries
$2$ for the $\mu$ components and $r/(2\|I - L_n^\star\|_F)$ for the $L$ components.
Therefore
\[
J(x) = J(\mu, L) = 4\|\mu-\mu_n^\star\|^2 + \frac{r^2}{4\|I-L_n^\star\|^2_F}\|L-L_n^\star\|^2_F.
\]
This setting yields two important facts. First, by \cref{thm:convexsmoothedMAP,thm:optimum_inside},
\[
    \htheta_n^\star \stackrel{P_{\theta_0}}{\to} \theta_0  \quad \text{and} \quad   \mu_n^\star \stackrel{P_{\theta_0}}{\to} \theta_0,
\]
yielding that 
\[
\sP\left( \|\htheta_n^\star -\theta_0\| + \|\mu_n^\star - \theta_0\| \leq \frac{r}{4\sqrt{2}} \right) \to 1, \quad \text{ as } n \to \infty.   
\]
For $\|\mu_0 - \htheta_n^\star \|^2 \leq \frac{r^2}{32}$, by triangle inequality, the probability 
that the following inequalities hold converges to $1$ in $P_{\theta_0}$ as $n \to \infty$,
\[
\|\mu_0 - \mu_n^\star\| \leq \| \mu_0 - \htheta_n^\star\| + \|\htheta_n^\star - \theta_0 \| + \|\mu_n^\star - \theta_0 \|  \leq \frac{r}{2\sqrt{2}}. 
\] 
Further with $L_0 = I$, $J(\mu_0, L_0) \leq \frac{3 r^2}{4} \leq r^2$. 
Hence, if we initialize $L_0 =  I $ and $\mu_0$ such that 
$\|\mu_0  - \htheta_n^\star\|_2^2 \leq \frac{r^2}{32}$, 
\[\label{eq:x0inJball}
\sP\left( x_0 \in \{x: J(x) \leq r^2 \} \right) \to 1, \quad \text{ as } n \to \infty.
\]
Second, if $J \leq r^2$ then $\mu$ is close to the optimal and $\|L\|_F$ is not too large, i.e.,
\[ \label{eq:Jball}
J(\mu,L) \leq r^2 &\implies \|\mu-\mu_n^\star\|^2 \leq r^2/4\\
J(\mu,L) \leq r^2 &\implies \|L-L_n^\star\|_F^2 \leq 4\|I-L_n^\star\|_F^2 \\
& \implies\|L\|_F \leq 2\|I-L_n^\star\|_F + \|L_n^\star\|_F,
\]
yielding that $\{J(\mu, L) \leq r^2 \} \subseteq \mcB_{r,n}$. Recall that 
\[
\mcB_{r,n} = \left\{\mu\in\reals^d, L\in\reals^{d\times d} : \| \mu - \mu_n^\star\|^2 \leq \frac{r^2}{4} \text{and}  \|L-L_n^\star\|_F^2 \leq 4\|I-L_n^\star\|_F^2\right\}.
\]
Then by \cref{cor:asymplocalconv}, under \cref{assump:regularity,assump:lsmooth}, the probability of the event that  
\[ \label{eq:cvxsmoothinJball}
F_n\text{ is }& \frac{\epsilon}{2} D_n\text{-strongly convex in $\{J(\mu, L) \leq r^2 \}$}\\
& \text{ and globally }\ell D_n\text{-Lipschitz smooth}
\]
converges to $1$ in $P_{\theta_0}$ as $n \to \infty$.

For brevity, we make the following definitions for the rest of this proof: 
recall  the definition of $f_n, F_n$ in \cref{eq:fn,eq:Fn} (we state here again):
\[
& I_n: \mcX \to \reals, \qquad I_n(x) := -\frac{1}{n}\log\det L \\
& f_n: \reals^d \to \reals, \qquad f_n( \theta ) := -\frac{1}{n}\log \pi_n(\theta)\\
& \tf_n: (\mcX, \reals^d) \to \reals, \qquad \tf_n(x, Z) :=  -\frac{1}{n}\log \pi_n\left(\mu + \frac{1}{\sqrt{n}} L Z \right)\\
&  F_n: \mcX \to \reals , \qquad F_n(x) := \EE\left[-\frac{1}{n}\log \pi_n\left(\mu + \frac{1}{\sqrt{n}} L Z \right) \right]\\
& \phi_n: = I_n + \tf_n ,\qquad \Phi_n: = I_n + F_n.
\]
Here $\phi_n(x, z)$ is the KL cost function with no expectation, and $\Phi_n(x)$
is the cost function with the expectation. To match the notation of
\cref{thm:sgdconvergence}, we reformulate the scaled gradient estimator defined
in \cref{eq:scaledLgradient} as $g_n$,
\[
g_n(x,Z) = \left\{\begin{array}{ll}
R_n(x)\nabla \phi_n(x, Z) & x \in \mcX^\mathrm{o}\\
\lim_{y\to x} R_n(y)\nabla \phi_n(y, Z) & x \in \partial \mcX
\end{array}\right.
,
\]
for a diagonal scaling matrix $R_n(x) \in \reals^{d(d+1)\times d(d+1)}$. Define
that $R_n(x)$ has entries $1$ for the $\mu$ components, 1 for the off-diagonal
$L$ components, and $1/(1+(n L_{ii})^{-1})$ for the diagonal $L$ components.
Note that $x \to \partial \mcX$ means that $L_{ii} \to 0$, ensuring that
$g_n(x,Z)$ has entries $-1$ for the $L_{ii}$. Since $Z$ is a standard normal
random variable, under the event that $-\frac{1}{n} \log \pi_n$ has Lipschitz
gradient, the gradient can be passed through the expectation so that the true
gradient is defined as below,
 \[
 G_n(x) := \EE\left[g_n(x,Z) \right]  = R_n(x)  \nabla \Phi_n(x).
 \]
Note that the projected stochastic iteration
\[
x_{k+1} = \Pi_\mcX\left(x_k - \gamma_k g_n(x_k, Z_k)\right), \quad k = \nats\cup\{0\},
\]
with $\Pi_\mcX(x) \defined \argmin_{y\in\mcX} \|V(x - y)\|^2$ is equivalent to the iteration described in \cref{alg:csvi}.
Note that the differentiability of $\phi_n$ only holds for $x \in
\mcX^\mathrm{o}$. For the case where $L_{ii} = 0$ for some $i \in [d]$, we can
use continuation via the limit 
$\lim_{L_{ii} \to 0}  -\frac{(n L_{ii})^{-1}}{1+ (n L_{ii})^{-1}} = -1$ 
to evaluate even though the gradient is not defined. For
the following proof, we do not make special treatments to those boundary points
when applying Taylor expansion and taking derivative.

Next we apply \cref{thm:sgdconvergence} to carry out the proof. The rest of
the proof consists two parts: \textbf{to show the confinement result (statement
2. of \cref{thm:sgdconvergence}) }  and \textbf{ to show the convergence result
(statement 3. of \cref{thm:sgdconvergence}) )}. We prove these two results
under the event that \cref{eq:x0inJball,eq:cvxsmoothinJball} hold; 
since the probability that these events hold converges in $P_{ \theta_0}$ to $1$ as $n
\to \infty$, the final result holds with the same convergent probability.

\textbf{We first show the confinement result by analyzing} $\epsilon(x)$,
$\ell^2(x)$, and $\sigma^2(r)$, which are defined in
\cref{eq:epsl2defs,eq:sigmadef} respectively. We aim to obtain that 
\benum
\item[i.] We can find sufficiently small $\gamma_k > 0$ such that 
\[
    \alpha_{k} = 1 + \ind\left[J(x_k) \leq r^2\right](-2\gamma_k\epsilon(r) + 2\gamma_k^2\ell^2(r)) \in (0,1]    
\]
holds for all $x\in \mcX$, i.e,
\[\label{eq:C1alphacond}
\forall x\in\mcX : J(x)\leq r^2, \quad 0 \leq 2\gamma_k\epsilon(x) - 2\gamma_k^2\ell^2(x) \leq 1.   
\]

\item[ii.] $\sigma^2(r) \to 0$ as $n \to \infty$ to guarantee the SGD iterations are eventually locally confined as $n \to \infty$ (based on \cref{thm:sgdconvergence}).
\eenum
\textbf{To show the statement i., \cref{eq:C1alphacond},} we start by deriving a lower bound for $2\gamma_k\epsilon(x) - 2\gamma_k^2\ell^2(x)$.  Examine the expression,
\[
& 2\gamma_k\epsilon(x) - 2\gamma_k^2\ell^2(x) \\
& = 2\gamma_k J(x)^{-1}(x-x^\star)^TV^TV R_n(x)\left(\nabla \Phi_n(x)-\nabla \Phi_n(x^\star)\right) \\
 &  \!-\! 2 \gamma_k^2 J(x)^{-1} \left(\nabla \Phi_n(x)\!-\!\nabla \Phi_n(x^\star)\right)^T R^T(x) V^TV R_n(x)\left(\nabla \Phi_n(x)\!-\!\nabla \Phi_n(x^\star)\right)\\
& =  \frac{2 \gamma_k}{J(x)}\left(V(x-x^\star)\right)^TV R_n(x)\left(\int \cdots \right)V^{-1}\left(V(x-x^\star)\right)\\
&  - \frac{2 \gamma_k^2}{J(x)}(V(x-x^\star))^T V^{-T} \left(\int \cdots \right)^T \left(V R_n(x) \right)^2 \left(\int \cdots \right)V^{-1}\left(V(x-x^\star)\right),
\]
where $\left(\int \cdots \right) = \left(\int_0^1 \nabla^2 \Phi_n((1-t)x^\star + tx)\dee t \right)$. By splitting $\Phi_n$ into the regularization $I_n(x)$ and the expectation $F_n(x)$; and defining 
\[
A(x) &\defined  V R_n(x)\left(\int_0^1 \nabla^2 I_n((1-t)x^\star + tx) \dee t \right)V^{-1}\\
B(x) &\defined  V R_n(x)\left(\int_0^1 \nabla^2 F_n((1-t)x^\star + tx) \dee t \right)V^{-1}\\
v(x) &\defined V(x-x^\star),
\]
the above expression can be written as 
\[ 
 &2\gamma_k\epsilon(x) - 2\gamma_k^2\ell^2(x) \\
&= 2\gamma_k \frac{v(x)^T(A(x)+B(x))v(x)  -\gamma_k \|(A(x)+B(x))v(x)\|^2 }{\|v(x)\|^2}\\
&\geq 2\gamma_k \frac{v(x)^TA(x)v(x)\!+\!v(x)^TB(x)v(x) \!-\!2\gamma_k \|A(x)v(x)\|^2 \!-\! 2\gamma_k\|B(x)v(x)\|^2 }{\|v(x)\|^2}\\
& \geq 2\gamma_k \left\{ \frac{v(x)^TA(x)v(x)+v(x)^TB(x)v(x)  -2\gamma_k \|A(x)v(x)\|^2}{\|v(x)\|^2} \right.\\
& \qquad \left. \frac{ - 2\gamma_k\|B(x)(VR_n(x)\ell D_n V^{-1})^{-1}\|^2\|VR_n(x)\ell D_n V^{-1} v(x)\|^2 }{\|v(x)\|^2} \right\}\\
&= 2\gamma_k \left\{ \frac{v(x)^TA(x)v(x)+v(x)^T(B(x)- VR_n(x)\frac{\epsilon}{2}D_n V^{-1})v(x)  }{\|v(x)\|^2} \right.\\
& \qquad \left. + \frac{  v(x)^T\left(VR_n(x)\frac{\epsilon}{2}D_n V^{-1}\right)v(x)-2\gamma_k \|A(x)v(x)\|^2  }{\|v(x)\|^2} \right.\\
& \qquad \left. - \frac{ 2\gamma_k\|B(x)(VR_n(x)\ell D_n V^{-1})^{-1}\|^2\|DR_n(x)\ell D_n V^{-1} v(x)\|^2 }{\|v(x)\|^2} \right\}
\]
Note that by \cref{cor:asymplocalconv} that $\frac{\eps}{2} D_n \preceq \nabla^2 F_n(x) \preceq \ell D_n$ and all the $V$, $R_n(x)$ are positive diagonal matrices, leading to 
\[
& B(x)- VR_n(x)\frac{\epsilon}{2}D_n V^{-1} \succeq 0 I\\
& \|B(x)(VR_n(x)\ell D_n V^{-1})^{-1}\|^2 \leq 1.
\]
Thus, the above expression can be bounded below by 
\[
&2\gamma_k\epsilon(x) - 2\gamma_k^2\ell^2(x) \\
& \geq 2\gamma_k  \left\{ \frac{v(x)^TA(x)v(x)+ v(x)^T\left(VR_n(x)\frac{\epsilon}{2}D_n V^{-1}\right)v(x)  }{\|v(x)\|^2} \right.\\
&\qquad \qquad \left. \frac{-2\gamma_k \|A(x)v(x)\|^2 - 2\gamma_k\|VR_n(x)\ell D_n V^{-1} v(x)\|^2 }{\|v(x)\|^2} \right\}\\
& = \frac{2}{\|v(x)\|^2} v(x)^T \left\{ \left[\gamma_k A(x) - 2\gamma_k^2 A^2(x) \right]  \right.\\
 & \qquad \qquad \left.+ \frac{1}{2}\left[ \eps\gamma_k R_n(x) D_n - 4 \ell^2 \left(\gamma_k R_n(x) D_n \right)^2 \right]\right\} v(x)\\
& \geq 2\lambda_{\min} \left(  \left[\gamma_k A(x) - 2\gamma_k^2 A^2(x) \right]  + \frac{1}{2}\left[ \eps\gamma_k R_n(x) D_n - 4 \ell^2 \left(\gamma_k R_n(x) D_n \right)^2 \right]  \right).
\]
Now, notice that $A(x)$ ,$R_n(x) D_n$ are all diagonal matrices with non-negative entries, 
\[ \label{eq:quadformsofC1}
\gamma_k A(x) - 2\gamma_k^2 A^2(x) & = \gamma_k A(x) \left(I - 2\gamma_k A(x) \right)\\
\eps\gamma_k R_n(x) D_n - 4 \ell^2 \left(\gamma_k R_n(x) D_n \right)^2 & = \gamma_k R_n(x) D_n \left( \eps - 4 \ell^2 \gamma_k R_n(x) D_n \right).
\]
As long as the entries of $A(x)$, $R_n(x) D_n$ are bounded above by a constant for all $n$, there exists a sufficiently small constant $c$ such that for all $\gamma_k 
< c$, \cref{eq:quadformsofC1} are both non-negative.
Given that for all $n$ and $\forall x \in \mcX$,
\[
& A(x)  = \diag\left(0, \cdots, \frac{(n L_{ii})^{-1}}{1 + (n L_{ii})^{-1}}  \frac{1}{L_{ii}^{\star}} , \cdots, 0\right) \preceq \frac{1}{\min_{i \in [d]} L_{ii}^\star } I\\ & R_n(x) D_n  \preceq I,
\]
we obtain the boundedness of the entries of $A(x)$, $R_n(x) D_n$. Therefore, we conclude that 
\[
\forall x\in\mcX, \quad 0 \leq 2\gamma_k\epsilon(x) - 2\gamma_k^2\ell^2(x).
\]

It remains to show the second inequality of \cref{eq:C1alphacond}, i.e.,
\[
\sup_{x\in\mcX : J(x)\leq r^2} 2\gamma_k\epsilon(x) - 2\gamma_k^2\ell^2(x) \leq 1.       
\]
This is true if 
\[
    \sup_{x\in\mcX : J(x)\leq r^2} \eps(x) \leq \gamma_k^{-1}.    
\]
Since $\gamma_k \to 0$ as $k \to \infty$, the above holds if $\sup_{x\in\mcX : J(x)\leq r^2} \eps(x)$ is bounded above by a constant that is independent to $n$. Now we consider the upper bound for $\sup_{x\in\mcX : J(x)\leq r^2} \eps(x)$.
Expanding $\eps(x)$,
\[
\eps(x) &= J(x)^{-1}(x-x^\star)^TV^TV R_n(x)\left(\nabla \Phi_n(x)-\nabla \Phi_n(x^\star)\right)\\
&  = \frac{v(x)^T(A(x)+B(x))v(x)}{\|v(x)\|^2}\\
&\leq \lambda_{\max}(A(x)+B(x))\\
&= \lambda_{\max} R_n(x)^{1/2}\left(\int_0^1 \nabla^2 \Phi_n((1-t)x^\star + tx)\dee t \right) R_n(x)^{1/2}.
\]
Split $\Phi_n$ into the regularization $I_n(x)$ and the expectation $F_n(x)$.
For the expectation, by \cref{cor:asymplocalconv} that $\nabla^2 F_n(x) \preceq \ell D_n$ and entries of $R_n(x)$ are bounded by $1$, we have
\[
R_n(x)^{1/2} \nabla^2 F_n(x) R_n(x)^{1/2} \preceq  \ell I,
\]
and for the regularization, note that $\nabla^2 I_n$ is a diagonal matrix with $0$ for $\mu$ and off-diagonals of $L$ and $ L_{ii}^{-2}/n$ for diagonals of $L$, so 
\[ \label{eq:supepsInbound}
\begin{aligned}
    &R_n(x)^{1/2} \left( \int_0^1 \nabla^2 I_n((1 - t) x^\star  +t x) \dee t\right) R_n(x)^{1/2}\\
&=  \diag\left(0, \cdots, \frac{(n L_{ii})^{-1}}{1 + (n L_{ii})^{-1}}  \frac{1}{L_{ii}^{\star}} , \cdots, 0\right) \\
& \preceq \frac{1}{\min_{i \in [d]} L_{ii}^\star } I
\end{aligned}   
\]
By the fact that $\forall i\in [d], L_{ii}^\star > 0$, we have \cref{eq:supepsInbound} is bounded above by a constant $C$. Use the Weyl's inequality to bound the maximal eigenvalue of the summation of two Hermitian matrices, we conclude that 
\[
\sup_{x \in \mcX: J(x) \leq r^2} \eps(x) \leq \ell + C.
\] 
Therefore, we have completed the proof for statement i., \cref{eq:C1alphacond}.

\textbf{Then we show the statement ii. by getting upper bound on $\sigma^2(r)$.} Recall that $\sigma^2(r)$ is the upper bound of the fourth moment of 
\[
    \left\| V R_n(x)\left(\nabla \phi_n(x, Z) - \nabla \Phi_n(x) \right)\right\| .   
\] 
Since the regularizor is cancelled in this expression, we only consider the expectation part. Note that $VR_n(x)$ is a diagonal matrix with positive diagonals,
\[
& \EE\left[ \left\|VR_n(x)\left( \nabla \tf_n(x, Z) - \nabla F_n(x)\right) \right\|^4 \right]^{1/4}\\
& \leq \max_{i \in [d(d+1)]}(VR_n(x))_{ii} \EE\left[ \left\| \nabla \tf_n(x, Z) - \nabla F_n(x)\right\|^4\right]^{1/4}.
\]
Let $Z_1, Z_2 $ be independent copies, by tower property of conditional expectation,
\[
& = \max_{i \in [d(d+1)]}(VR_n(x))_{ii}\EE\left[ \left\| \EE\left[ \nabla \tf_n(x, Z_1) -  \nabla \tf_n(x, Z_2) | Z_1\right] \right\|^4 \right]^{1/4}.
\]
By the convexity of $\|\cdot\|^4$ and Jensen's inequality,
\[
& \leq \max_{i \in [d(d+1)]}(VR_n(x))_{ii} \EE\left[  \EE\left[ \left\| \nabla \tf_n(x, Z_1) -  \nabla \tf_n(x, Z_2) \right\|^4  | Z_1\right] \right]^{1/4}\\
& =  \max_{i \in [d(d+1)]}(VR_n(x))_{ii}\EE\left[ \left\| \nabla \tf_n(x, Z_1) -  \nabla \tf_n(x, Z_2) \right\|^4 \right]^{1/4}.
\]
By $\left\| \nabla \tf_n(x, Z_1) \!- \!\nabla \tf_n(x, Z_2)  \right\|\!\leq \! \left\| \nabla \tf_n(x, Z_1)\right\| \!+\!\left\| \nabla \tf_n(x, Z_2) \right\| $ and Minkowski's inequality,
\[
&\leq  \max_{i \in [d(d+1)]}(VR_n(x))_{ii} \left\{\EE\left[ \|\nabla \tf_n(x, Z_1)\|^4 \right]^{1/4}  + \EE\left[ \|\nabla \tf_n(x, Z_1)\|^4 \right]^{1/4} \right\}\\
& = 2 \max_{i \in [d(d+1)]}(VR_n(x))_{ii} \EE\left[ \|\nabla \tf_n(x, Z)\|^4 \right]^{1/4}.
\]
Now we focus on bounding $\EE\left[ \|\nabla \tf_n(x, Z)\|^4 \right]^{1/4}$. We examine $\|\nabla \tf_n(x, Z)\|$,
\[
    \nabla \tf_n(x, Z) =  \left( \begin{array}{c}
\nabla f_n \left(\mu + \frac{1}{\sqrt{n}}L Z \right)\\
\frac{Z_1}{\sqrt{n}} \nabla f_n \left(\mu + \frac{1}{\sqrt{n}}L Z \right)\\
\vdots\\
\frac{Z_p}{\sqrt{n}} \nabla f_n \left(\mu + \frac{1}{\sqrt{n}}L Z \right)
\end{array}
\right)   \quad \in \reals^{d(d+1)},
\]
yielding
\[
\|\nabla \tf_n(x, Z)\|^4  = \left\|\nabla f_n \left(\mu + \frac{1}{\sqrt{n}}L Z \right) \right\|^4 \left( 1 + \frac{Z^T Z}{n}\right)^2.   
\]
By Cauchy-Schiwartz inequality,
\[
\EE\left[ \|\nabla \tf_n(x, Z)\|^4\right]^{1/4} \!\leq \EE\!\left[ \left\|\nabla f_n \left(\mu + \frac{1}{\sqrt{n}}L Z \right) \right\|^8 \right]^{1/8} \!\EE\!\left[ \left( 1 + \frac{Z^T Z}{n}\right)^4\right]^{1/8}.
\]
We then bounds these two terms on RHS separately. We use the sub-Gaussian property of $\left\|\nabla f_n \left(\mu + \frac{1}{\sqrt{n}}L Z \right)\right\|$ to bound its $8^{\text{th}}$ moment. First notice that $\left\|\nabla f_n \left(\mu + \frac{1}{\sqrt{n}}L Z \right)\right\|$ is a $\max_{i \in [d]} L_{ii} \frac{\ell}{\sqrt{n}}$-Lipschitz function of $Z$, 
\[
&\left| \left\|\nabla f_n \left(\mu + \frac{1}{\sqrt{n}}L Z_1 \right)\right\| - \left\|\nabla f_n \left(\mu + \frac{1}{\sqrt{n}}L Z_2 \right)\right\| \right|\\  
&\leq \left\|\nabla f_n \left(\mu + \frac{1}{\sqrt{n}}L Z_1 \right) - \nabla f_n \left(\mu + \frac{1}{\sqrt{n}}L Z_2 \right)\right\| \\
& = \left\|\int_0^1 \nabla^2 f_n\left(\mu+ (1-t)L Z_2/ \sqrt{n} + t Z_1/ \sqrt{n} \right) \dee t \frac{L}{\sqrt{n}}(Z_1 - Z_2) \right\|.
\]
Given that $\nabla^2 f_n \preceq \ell I$, the above is bounded by
\[
 \frac{\ell}{\sqrt{n}} \sqrt{(Z_1 - Z_2 )^T L^T L (Z_1 - Z_2 )}
& \leq \frac{\ell}{\sqrt{n}}  \lambda_{\max}(L^T L) \|Z_1- Z_2 \|\\
& = \frac{\ell}{\sqrt{n}}  \max_{i \in [d]} L_{ii}^2 \|Z_1- Z_2 \|.
\] 
Since a Lipschitz function of Gaussian noise is sub-Gaussian \citep[Thm 1]{kontorovich2014concentration}, i.e., let $Z \sim \distNorm(0, I_d)$, $\psi: \reals^d \to \reals$ be $L$-Lipschitz, then
\[
\Pr\left(|\psi(Z) - \EE [\psi(Z)] |> \eps \right) \leq 2\exp\left( -\frac{\eps^2}{4L^2} \right) .
\]
Thus, $\left\|\nabla f_n \left(\mu + \frac{1}{\sqrt{n}}L Z \right)\right\|$ is $\frac{4 \ell^2}{n} \max_{i \in [d]} L_{ii}^2$-sub-Gaussian. Then note that for a $\sigma^2$-sub-Gaussian random variable $X \in \reals$, for any positive integer $k \geq 2$, $\EE\left[|X|^k\right]^{1/k} \leq \sigma e^{1/e} \sqrt{k}$. Hence
we obtain
\[\label{eq: 8th moment of ...}
\EE\left[ \left\| \nabla f_n \left(\mu + \frac{1}{\sqrt{n}}L Z \right)\right\|^8\right]^{1/8} \leq   \frac{2\ell}{\sqrt{n}} e^{1/e} \sqrt{8} \max_{i \in [d]} L_{ii}.
\]
Along with the fact that Gaussian random variable has arbitrary order moments, 
\[
 \EE\left[ \left( 1 +\frac{Z^TZ}{n}\right)^4 \right] \leq C,
\]
for some constant $C$, we obtain 
\[
\EE\left[ \|\nabla_x f_n\|^4\right]^{1/4} \leq  \frac{2 C^{1/4} \ell}{\sqrt{n}} e^{1/e} \sqrt{8} \max_{i \in [d]} L_{ii},
\]
and hence 
\[
&\EE\left[ \left\|VR_n(x)\left( \nabla_x f_n - \nabla_x F_n\right) \right\|^4
\right]^{1/4} \\
&\leq \max_{i \in [d(d+1)]}(VR_n(x))_{ii} \frac{2 C^{1/4} \ell}{\sqrt{n}} e^{1/e} \sqrt{8} \max_{i \in [d]} L_{ii} .
\]
Taking supremum over $J(x) \leq r^2$, the RHS is bounded above by a universal constant, we therefore conclude that 
\[
\sigma^2(r) = \sup_{x\in \mcX: J(x)\leq r^2}    \EE\left[ \left\|VR_n(x)\left( \nabla \phi_n(x, Z) - \nabla \Phi_n(x) \right) \right\|^4 \right]^{1/4} \to 0, \; n \to \infty.
\]
Therefore, with $\forall x\in\mcX : J(x)\leq r^2, 0 \leq 2\gamma_k\epsilon(x) - 2\gamma_k^2\ell^2(x) \leq 1$ and $\sigma^2(r) \to 0$ as $n \to \infty$, applying \cref{thm:sgdconvergence} yields the confinement result, i.e.,  
\[\label{eq: c1 confinement}
\Pr\left(\sup_{k\in\nats} J(x_k) \leq r^2\right) \to 1
\]
in $P_{ \theta_0}$ as $n \to \infty$.

\textbf{Lastly, by statement 3. of \cref{thm:sgdconvergence}, we prove the convergence result by checking} 
\[
    \inf_{x\in \mcX, J(x) \leq r^2}\eps(x) > 0.     
\]
We use the similar way to expand the expression,
\[
\eps(x) &= J(x)^{-1}(x-x^\star)^TV^TV R_n(x)\left(\nabla \Phi_n(x)-\nabla \Phi_n(x^\star)\right)\\
&  = \frac{v(x)^T(A(x)+B(x))v(x)}{\|v(x)\|^2}\\
&\geq \lambda_{\min}(A(x)+B(x))\\
&= \lambda_{\min} R_n(x)^{1/2}\left(\int_0^1 \nabla^2 \Phi_n((1-t)x^\star + tx)\dee t \right) R_n(x)^{1/2}.
\]
By splitting $\Phi_n$ into the regularization and the expectation, we have
\[ \label{eq:eps(x)Ibound}
\begin{aligned}
    & R_n^{1/2}(x) \left( \int_0^1 \nabla^2 I_n((1 - t) x^\star  +t x) \dee t\right) R_n(x)^{1/2}\\
&=  \diag\left(0, \cdots, \frac{(n L_{ii})^{-1}}{1 + (n L_{ii})^{-1}}  \frac{1}{L_{ii}^{\star}} , \cdots, 0\right) \\
& \succeq 0 I,
\end{aligned}
\]
and 
\[ \label{eq:eps(x)Fbound}
\begin{aligned}
    &R_n(x)^{1/2}\left(\int_0^1 \nabla^2 F_n((1-t)x^\star + tx)\dee t \right) R_n(x)^{1/2}  \\
& \geq  R_n(x)^{1/2}  \frac{ D_n \eps}{2}  R_n(x)^{1/2} \\
& \succeq \eps/2n >0 .
\end{aligned}
\]
We then combine \cref{eq:eps(x)Ibound,eq:eps(x)Fbound} and use Weyl's inequality to bound the minimal eigenvalue of the summation of two Hermitian matrices,
yielding
\[
    \inf_{x\in \mcX, J(x) \leq r^2}\eps(x) > \eps/n >0.
\]
This gives the convergence result.

Then the proof is complete by applying \cref{thm:sgdconvergence}. We know that
$\xi_k$ is strictly positive. Since $\eps(r) > 0$ and $\ell(r) $ is bounded
above, there exists $\gamma_k = \Theta(k^{-\rho}), \rho \in (0.5, 1) $ so that
it satisfies the condition of the theorem. We have that $\sigma \to 0$, which
makes 3. in the statement of \cref{thm:sgdconvergence} become
\[
  \Pr\left( \| V(x_k - x^\star)\|^2 = O_{P_n}(k^{-\rho'})
  \right)\stackrel{P_{\theta_0}}{\to} 1,\, \rho' \in (0, \rho -0.5)\quad n\to\infty
\]
Even though $D$ is a function of $n$, $n$ is fixed as \cref{alg:csvi} runs.
Since $D$ is invertible, 
\[
  \Pr\left( \|x_k - x^\star\|^2 = O_{P_n}(k^{-\rho'})
  \right)\stackrel{P_{\theta_0}}{\to} 1, \,\,\rho' \in (0, \rho -0.5)\quad n\to\infty
\]
which is exactly our desired result: as the number of data $n\to\infty$, 
the probability that \cref{alg:csvi} finds the optimum in a rate of $k^{-\rho'}$ 
(as we take more
iterations, $k\to\infty$) converges to 1.
In other words, \emph{variational inference gets solved asymptotically} in a
rate of $k^{-\rho'}$  .

\eprf

\bnthm\label{thm:sgdconvergence}
Let $\mcX \subseteq \reals^p$ be closed and convex,
$g : \mcX \times \mcZ \to \reals^p$ be a function, 
$G(x) \defined \EE\left[g(x,Z)\right]$ for a random element $Z\in\mcZ$,
$x^\star\in\mcX$ be a point in $\mcX$ such that $G(x^\star) = 0$, 
$V \in \reals^{p\times p}$ be invertible,
$J(x) \defined \|V(x - x^\star)\|^2$, and $r \geq 0$.
Consider the projected stochastic iteration
\[
x_0 \in \mcX, \quad x_{k+1} = \Pi_\mcX\left(x_k - \gamma_k g(x_k, Z_k)\right), 
\quad k = \nats\cup\{0\},
\]
with independent copies $Z_k$ of $Z$, $\gamma_k \geq 0$, and
$\Pi_\mcX(x) \defined \argmin_{y\in\mcX} \|V(x - y)\|^2$.
If
\benum
\item For all $k\in\nats\cup\{0\}$, the step sizes satisfy
\[ \label{eq:epsl2defs}
 \forall x\in\mcX : J(x)\leq r^2, \quad 0 &\leq 2\gamma_k\epsilon(x) - 2\gamma_k^2\ell^2(x) \leq 1\\
\epsilon(x) &\defined \frac{1}{J(x)}(x-x^\star)^TV^TV\left(G(x)- G(x^\star)\right)\\
\ell^2(x) &\defined  \frac{1}{J(x)}\left\|V\left(G(x)-G(x^\star)\right)\right\|^2,
\]
\item For all $x\in\mcX$,  $\left(\EE\|V(g(x,Z) - G(x))\|^4\right)^{1/4} \leq\tsigma(x)$ for 
  $\tsigma : \mcX \to \reals_{\geq 0}$, and
\[ \label{eq:sigmadef}
\sigma(r) &\defined \sup_{x\in\mcX \,:\, J(x) \leq r^2} \tsigma(x),
\]
\eenum
then
\benum
\item The iterate $x_k$ is locally confined with high probability:
\[
\Pr\left(J(x_k) \leq r^2\right) &\geq\frac{\xi^2_k}{\xi_k^2 + 8\sigma(r)^2\zeta_k}\\
\xi_{k}(r) &\defined \max\{0, r^2 - J(x_0) - 2\sigma^2(r)\sum_{j<k}\gamma^2_j\}\\
\zeta_{k}(r) &\defined r^2\sum_{j<k}\gamma_j^2 + \sigma^2(r)\sum_{j<k}\gamma_j^4.
\]
\item The iterate $x_k$ stays locally confined for all $k\in\nats$ with high probability: 
\[
\Pr\left(\sup_{k\in\nats} J(x_k) \leq r^2\right) &\geq \frac{\xi^2}{\xi^2 + 8\sigma^2(r)\zeta}\\
\xi(r) &\defined \lim_{k\to\infty}\xi_{k}(r)
\quad
\zeta(r) \defined \lim_{k\to\infty} \zeta_{k}(r).
\]
\item If additionally
\[
\inf_{x\in \mcX : J(x) \leq r^2} \epsilon(x) > 0 \quad\text{and}\quad  \gamma_k
= \Theta(k^{-\rho}), \,\, \rho\in (0.5, 1],
\]
the iterate $x_k$ converges to $x^\star$ with high probability:
\[
 \lim_{k \to \infty} \Pr\left( J(x_k) \leq k^{-\rho'} \right) \geq \Pr\left(\sup_{k\in\nats} J(x_k)
  \leq r^2\right), \quad \forall \rho' \in (0, \rho - 0.5).
\]
\eenum

\enthm
\bprf
To begin, we show $\Pi_\mcX$ is non-expansive,
\[
    \|V(\Pi_\mcX(x) - \Pi_\mcX(y))\|^2 &\leq \|V(x - y)\|^2.
\]
For all $x, y \in \reals^p$, define $\langle x, y\rangle_V = x^T V^TV y$. Since $V$ is invertible, $V^TV$ is symmetric and positive definite, and hence $(\reals^p, \langle \cdot, \cdot\rangle_V)$ forms a Hilbert space. Any projection operator of a Hilbert space is non-expansive \citep[Prop. 4.4]{bauschke2011convex}.

Note that $x^\star = \Pi_{\mcX}(x^\star)$ and the projection operation is non-expansive, expanding the squared norm yields
\[
\|V(x_{k+1}-x^\star)\|^2 &\leq 
\|V(x_k - x^\star)\|^2 \\
&- 2\gamma_k (x_k-x^\star)^TV^TVg(x_k, Z_k) + \gamma_k^2\left\|Vg(x_k, Z_k)\right\|^2.
\]
Adding and subtracting $G(x_k)$ in the second and third terms, 
using the elementary bound $\left\|a+b\right\|^2 \leq 2\|a\|^2 + 2\|b\|^2$, and defining
\[
\beta_k(x) &\defined -2\gamma_k (x-x^\star)^TV^TV(g(x, Z_k) - G(x)) \\
& \quad + 2\gamma_k^2\left\|V(g(x, Z_k) - G(x))\right\|^2 - 2\gamma_k^2\EE\left[\|\cdot\|^2\right]\\
\epsilon(x) &\defined \frac{1}{J(x)}(x-x^\star)^TV^TV\left(G(x)-G(x^\star)\right)\\
\ell^2(x) &\defined \frac{1}{J(x)}\left\|V\left(G(x)-G(x^\star)\right)\right\|_2^2,
\]
 we have that
\[
J(x_{k+1})\leq & J(x_k)\left(1 - 2\gamma_k \epsilon(x_k) + 2\gamma_k^2\ell^2(x_k)\right)
+\beta_k(x_k) + 2\gamma_k^2 \tilde\sigma^2(x_k).
\]
We now define the filtration of $\sigma$-algebras
\[
\mcF_k = \sigma(x_1, \dots, x_k, Z_1, \dots, Z_{k-1}),
\]
 and the stopped process for $r > 0$,
\[
Y_{0} &= J(x_0)\\
Y_{k+1} &= \left\{ \begin{array}{ll}
Y_{k} & Y_{k}  > r^2\\
J(x_{k+1}) & \text{o.w.}
\end{array}\right.
\]
Note that $Y_k$ is $\mcF_k$-measurable,
and that 
 $Y_{k}$ ``freezes in place'' 
if $J(x_k)$  ever jumps larger than $r^2$; so for all $t^2 \leq r^2$,
\[
\Pr\left(J(x_k) > t^2\right)  
&= \Pr\left(J(x_k) > t^2 , Y_{k-1} > r^2 \right) + \Pr\left(J(x_k) > t^2 , Y_{k-1} \leq r^2 \right) \\
&=  \Pr\left(J(x_k) > t^2 , Y_k > r^2, Y_{k-1} > r^2 \right)\! +\! \Pr\left(Y_k > t^2, Y_{k-1} \leq r^2  \right)\\
& \leq \Pr\left(Y_k > r^2, Y_{k-1} > r^2  \right) + \Pr\left(Y_k > t^2, Y_{k-1} \leq r^2  \right)\\
& \leq \Pr\left(Y_k > t^2, Y_{k-1} > r^2  \right) + \Pr\left(Y_k > t^2, Y_{k-1} \leq r^2  \right)\\
& = \Pr\left(Y_k > t^2 \right).
\]
Therefore if we obtain a tail bound on $Y_k$, it provides the same bound on $J(x_k)$.
Now substituting the stopped process into the original recursion and collecting terms,
\[
&Y_{k+1}  \\
\leq &
 Y_{k}\left(1\!+ \!\ind\!\left[Y_k \!\leq\! r^2\right]\!(- 2\gamma_k\epsilon(x_k) \!+\! \gamma_k^2\ell^2(x_k))\right)\!+ \!\ind\!\left[Y_k \!\leq\! r^2\right]\!\left(\beta_k(x_k) \!+ \!2\gamma_k^2 \tsigma^2(x_k)\right)\\
\leq &
Y_{k}\left(1\!+ \!\ind\!\left[Y_k \!\leq\! r^2\right]\!(- 2\gamma_k\epsilon(x_k) \!+\! \gamma_k^2\ell^2(x_k))\right)\!+ \!\ind\!\left[Y_k \!\leq\! r^2\right]\!\beta_k(x_k) \!+\! 2\gamma_k^2 \sigma^2(r).
\]
Using the notation of \cref{lem:descent}, set
\[
\alpha_{k} &= 1 + \ind\left[Y_k \leq r^2\right](-2\gamma_k\epsilon(x_k) + 2\gamma_k^2\ell^2(x_k))\\ 
\balpha_{k} &= 1\\
\beta_{k} &= \ind\left[Y_k \leq r^2\right]\beta_k(x_k)\\
c_{k} &= 2\gamma_k^2 \sigma^2(r).
\]
By the fourth moment assumption,  $\beta_k$ has variance bounded above by $\tau_k^2$ conditioned on $\mcF_k$, where
\[
\tau_{k}^2 &= 8\gamma_k^2\ind\left[Y_k\leq r^2\right]\|V(x_k-x^\star)\|^2\tilde\sigma(x_k)^2 +
8\gamma_k^4\ind\left[Y_k\leq r^2\right]\tilde\sigma^4(x_k) \\
&\leq 8\gamma_k^2r^2\sigma^2(r) +
8\gamma_k^4\sigma^4(r).
\]
Therefore, using the descent \cref{lem:descent},
\[
&\Pr\left(Y_{k} > r^2\right) \leq \frac{\zeta_k}{\max\{r^2-\xi_k,0\}^2 + \zeta_k}\\
\xi_{k} = J(x_0) + &2\sigma^2(r)\sum_{j<k}\gamma^2_j
\quad \zeta_{k} = 8\sigma^2(r)\left(r^2\sum_{j<k}\gamma_j^2 + \sigma^2(r)\sum_{j<k}\gamma_j^4\right).
\]
yielding the first result.
Now since $Y_{k+1} \leq r^2 \implies Y_{k} \leq r^2$ for all $k\geq 0$,
the sequence of events $\left\{Y_k \leq r^2\right\}$ is decreasing. Therefore the second result follows from
\[
\Pr\left(\bigcap_{k=0}^\infty \left\{Y_k \leq r^2\right\}\right) &= \lim_{k\to\infty}\Pr\left(Y_k \leq r^2\right)\\
&\geq\lim_{k\to\infty} 1-\frac{\zeta_k}{\max\{r^2-\xi_k,0\}^2 + \zeta_k}\\
&= \frac{\max\{r^2-\xi,0\}^2}{\max\{r^2-\xi,0\}^2 + \zeta},
\]
where $\xi \defined\lim_{k\to\infty}\xi_{k}$ and $\zeta\defined\lim_{k\to\infty}\zeta_{k}$.
Finally, we analyze the conditional tail distribution of $Y_k$ given that it stays confined, i.e., $\forall k\geq 0$,  $Y_k\leq r^2$.
In the notation of \cref{lem:descent}, redefine
\[
0 \leq \balpha_{k} \defined \sup_{x \in \mcX : J(x)\leq r^2} 1 - 2\gamma_k\epsilon(x) + 2\gamma_k^2\ell^2(x) \leq 1,
\]
i.e., $\balpha_{k}$ is the largest possible value of $\alpha_{k}$ when $Y_k \leq r^2$. So again applying \cref{lem:descent},
\[
&\Pr\left(Y_{k}> t_k\given \forall k \,\,Y_k \leq r^2 \right) \\
&= \frac{\Pr\left(Y_{k}  >t_k, \forall k\,\, Y_k \leq r^2 \right)}{\Pr\left(\forall k\,\, Y_k \leq r^2 \right)}\\
&\leq \frac{\Pr\left(Y_{k} >t_k, \forall k\,\, Y_k \leq r^2 \right)}{ \frac{\max\{r^2-\xi,0\}^2}{\max\{r^2-\xi,0\}^2 + \zeta}}\\
\label{eq:sequencetailbound}
&\leq \frac{\left(\frac{\zeta'_k}{\max\{t_k-\xi'_k,0\}^2 +
\zeta'_k}\right)}{\left( \frac{\max\{r^2-\xi,0\}^2}{\max\{r^2-\xi,0\}^2 +
\zeta}\right)}\\
&\xi'_{k} = J(x_0)\prod_{i=0}^{k-1}\balpha_i + 2\sigma^2(r)\sum_{i=0}^{k-1}\gamma^2_i\prod_{j=i+1}^{k-1}\balpha_j\\
&\zeta'_{k} = 8\sigma^2(r)\sum_{i=0}^{k-1}(r^2\gamma_i^2 + \sigma^2(r)\gamma_i^4)\prod_{j=i+1}^{k-1}\balpha_j^2.
%
%
\]
Here $t_k  = \Theta(k^{-\rho'}), \rho' \in (0, \rho - 0.5)$ is a decreasing sequence, whose shrinking rate---such that
\cref{eq:sequencetailbound} still converges to $0$---will determine
the convergence rate of $Y_k$.


To understand the rate of \cref{eq:sequencetailbound}, the 
key is to characterizing the order of $\prod_{j< k}\balpha_{j}$ and
$\sum_{i = 0}^{k}\gamma_i^2 \prod_{j = i+1}^k \balpha_j^2$.
Since $\gamma_k = \Theta(k^{-\rho})$, $\rho\in (0.5, 1]$, 
and $\eps' \defined \inf_{x\in\mcX : J(x)\leq r^2} \epsilon(x) > 0$,
we know that
\[\label{eq:prodalphas}
  \prod_{j<k}\balpha_{j} 
  &=  \prod_{j<k} \left\{\sup_{x\in \mcX:J(x) \leq r^2} (1 -2\gamma_j \eps(x) +
  2\gamma_j^2\ell^2(x))\right\}\\
  &=  \prod_{j<k} \left(1 -2\gamma_j \inf_{x\in \mcX:J(x) \leq r^2} \{ \eps(x) -
  \gamma_j\ell^2(x) \} \right\}\\
  &\leq \prod_{j < k} ( 1- 2 c \gamma_j)\\
  & = \exp\left(\sum_{j < k}\log (1- 2c \gamma_j )\right)\\
  & \leq \exp\left(2c\sum_{j<k}\gamma_j  \right)
\]
for some $c > 0$, yielding that 
$\prod_{j<k}a_{j}= \Theta\left(\exp\left(-Ck^{1-\rho} \right)\right)$ 
for some $C > 0$. 
For the second term, since $\balpha \in (0,1)$,  
\[ \label{eq:zetarate}
\sum_{i = 0}^{k}\gamma_i^2 \prod_{j = i+1}^k \balpha_j^2
\leq \sum_{i  = 0 }^{k} \gamma_i^2  =  \Theta(k^{1- 2\rho }).
\]
Similarly, $\sum_{i = 0}^{k}\gamma_i^2 \prod_{j = i+1}^k \balpha_j = \Theta(k^{1
- 2\rho})$.
\footnote{Although the bound in \cref{eq:zetarate} is loose, the order
derscribed by the bound is actually tight. A more detailed analysis can be
obtained by approximating the summation with an integral, which yields the same
order.} 
Therefore, 
\[
  \xi'_k = \Theta\left( k^{1- 2\rho} \right), \quad 
  \zeta'_k = \Theta\left( k^{1- 2\rho} \right).
\]
Combined with the fact that $t_k = \Theta(k^{- \rho'}), \rho' \in (0, \rho -
0.5)$, this implies that \cref{eq:sequencetailbound} is $o(1)$ and hence for all
$\eps > 0, \rho' \in (0, \rho - 0.5)$, 
\[
  \lim_{k \to \infty}\Pr\left(k^{\rho'} Y_k >\eps \given \forall k \,\,Y_k \leq
  r^2 \right) = 0.
\]
Therefore, $\forall \rho' \in (0, \rho - 0.5)$,
\[
  \lim_{k \to \infty} \Pr\left(Y_k \leq k^{-\rho'} \right) 
 & \geq \lim_{k \to \infty} \Pr\left(Y_k \leq k^{-\rho'} \given \forall k \,\,Y_k\leq r^2\right)\Pr\left(\forall k \,\,Y_k\leq r^2\right) \\
& = \Pr\left(\forall k \,\,Y_k\leq r^2\right),
\]
and the result follows.
\eprf

\bnlem[Descent]\label{lem:descent}
Suppose we are given a filtration $\mcF_k \subseteq \mcF_{k+1}$, $k\geq 0$. Let
\[
Y_{k+1} \leq \alpha_k Y_k + \beta_k + c_k, \quad k\geq 0,
\]
where $Y_k \geq 0$ and
$0 \leq \alpha_k\leq \balpha_k$ are $\mcF_k$-measurable,
$\beta_k$ is $\mcF_{k+1}$-measurable and has mean 0 and variance conditioned on $\mcF_k$ bounded above by $\tau_k^2$,
and $Y_0, \tau^2_k, c_k, \balpha_k \geq 0$ are $\mcF_0$ measurable.  Then
\[
&\Pr\left(Y_k \geq t\right) \leq \frac{\zeta_k}{\max\{t-\xi_k, 0\}^2 + \zeta_k}\\
\xi_k = Y_0\prod_{i=0}^{k-1}\balpha_i + &\sum_{i=0}^{k-1}c_i\left(\prod_{j=i+1}^{k-1}\balpha_j\right) \quad \zeta_k = \sum_{i=0}^{k-1}\tau_i^2\left(\prod_{j=i+1}^{k-1}\balpha^2_j\right).
\]
\enlem

\bprf
Solving the recursion,
\[
Y_k &\leq \alpha_{k-1}Y_{k-1} + \beta_{k-1} + c_{k-1}\\
&\leq \alpha_{k-1}\left(\alpha_{k-2}Y_{k-2} +\beta_{k-2} + c_{k-2}\right)+ \beta_{k-1} + c_{k-1}\\
&\leq \dots\\
&\leq Y_0 \prod_{i=0}^{k-1}\alpha_i + \sum_{i=0}^{k-1}\beta_i\prod_{j=i+1}^{k-1}\alpha_j + \sum_{i=0}^{k-1}c_i\prod_{j=i+1}^{k-1}\alpha_j\\
&\leq Y_0 \prod_{i=0}^{k-1}\balpha_i + \sum_{i=0}^{k-1}\beta_i\prod_{j=i+1}^{k-1}\balpha_j + \sum_{i=0}^{k-1}c_i\prod_{j=i+1}^{k-1}\balpha_j.
\]
So
\[
\Pr\left(Y_k \geq t\right)
&\leq\Pr\left(\sum_{i=0}^{k-1}\beta_i\prod_{j=i+1}^{k-1}\balpha_j \geq t - \xi_k\right).
\]
By Cantelli's inequality and the fact that the $i^\text{th}$ term in the sum is $\mcF_{i+1}$-measurable,
\[
\Pr\left(Y_k \geq t\right) 
&\leq \frac{\sum_{i=1}^{k-1}\EE\left[\beta_i^2 \prod_{j=i+1}^{k-1} \balpha_j^2\right] }{\max\{t-\xi_k,0\}^2 + \sum_{i=1}^{k-1}\EE\left[\beta_i^2 \prod_{j=i+1}^{k-1} \balpha_j^2\right]}\\
&\leq \frac{\sum_{i=0}^{k-1}\tau_i^2 \prod_{j=i+1}^{k-1} \balpha_j^2}{\max\{t-\xi_k,0\}^2 + \sum_{i=0}^{k-1}\tau_i^2 \prod_{j=i+1}^{k-1} \balpha_j^2}\\
&= \frac{\zeta_k}{\max\{t-\xi_k,0\}^2 + \zeta_k}.
\]
\eprf

\end{document}